\documentclass[10pt]{iopart}

\usepackage{cite}

\expandafter\let\csname equation*\endcsname\relax
\expandafter\let\csname endequation*\endcsname\relax

\usepackage{amsmath}
\usepackage{amssymb}
\usepackage{graphicx}
\usepackage{color}
\usepackage{caption}
\usepackage{soul}

\usepackage{bm}
\usepackage{bbold}

\usepackage{mathtools}
\usepackage{colortbl}

\usepackage{nicefrac}
\usepackage{hhline}
\usepackage{subfig}

\usepackage{multirow} 
\usepackage{tabularx} 
\usepackage{array}
\usepackage{units}

\usepackage{tensor} 
\usepackage{braket}

\usepackage{hyperref}
\usepackage{comment}

\newcommand{\grafe}[1]{\left\{ #1 \right\}}
\newcommand{\tonde}[1]{\left( #1 \right)}
\newcommand{\quadre}[1]{\left[ #1 \right]}

\usepackage{array}
\newcolumntype{L}{>{$}l<{$}}

\begin{document}

\title{Distribution of rare saddles in the $p$-spin energy landscape}

\author{Valentina Ros}

\address{Laboratoire de Physique de l’Ecole Normale Superieure,
ENS, Universit\'{e} PSL, CNRS, Sorbonne Universit\'{e},
Universit\'{e} Paris-Diderot, Sorbonne Paris Cit\'{e}, Paris, France}
\ead{valentina.ros@lps.ens.fr}
\vspace{10pt}
\begin{indented}
\item[]December 2019
\end{indented}

\begin{abstract}
We compute the statistical distribution of index-1 saddles surrounding a given local minimum of the $p$-spin energy landscape, as a function of their distance to the minimum in configuration space and of the energy of the latter. We identify the saddles also in the region of configuration space in which they are subdominant in number (\emph{i.e.}, rare) with respect to local minima, by computing large deviation probabilities of the extremal eigenvalues of their Hessian. As an independent result, we determine the joint large deviation probability of the smallest eigenvalue and eigenvector of a GOE matrix perturbed with both an additive and multiplicative finite-rank perturbation. 
\end{abstract}

%
%
%
%
%

\section{Introduction}
High-dimensional systems are typically associated to complex, highly non-convex energy landscapes, in which the  number of stationary points (local minima, maxima or saddles) increases steeply with the dimensionality. Classifying these points in terms of their energy, of their stability and of their location in the underlying configuration space is a topic that is of interest in a large variety of fields, including disordered systems~\cite{crisom95,giardinacavagnaparisi,Monasson,BrayDean,CLR2, Rizzo1, Aspelmeier0, fyodorov, FyodorovWilliams, FyodNadal, auffingerbenaouscerny, eliran,Majumdar, FyodorovTexier, FyodLeDoussal}, ecology and biology~\cite{MayNature, FyodorovKhoru, remprotein, Fitness1}, neural networks~\cite{sompolinsky1,touboul}, inference~\cite{tengyuma, MontanariBenArous, FyodLinearly, SpikedRepKacRice, Passed}, game theory~\cite{Galla}, string theory and cosmology~\cite{Douglas, Aazami}. 
In many of these contexts, a crucial motivation for determining the distribution of stationary points is to understand how the energy functional is explored dynamically, through algorithms that proceed via local moves in configuration space, biased towards lower-energy configurations. When metastable local minima proliferate, indeed, the dynamical search of the global minimum (or optimal state) is likely hampered by the ruggedness and glassiness of the landscape.
In high-dimensional glassy systems, several features of the resulting slow dynamics (such as aging~\cite{CuKu, ReviewDyn, Folena}) have been characterized in detail. However, it is still to large extent an open question~\cite{FranzFirstSteps,BerthierBiroli2011} how the system escapes dynamically from the metastable, trapping local minima via 
activated crossings of the surrounding energy barriers. 

Addressing this question is notoriously challenging, as it requires to determine the energetic cost of the paths in the landscape connecting different local minima. It is clear that a pivotal role in fixing such cost is played by the critical points lying along the path, in particular by the saddles: characterizing how the saddles are arranged with respect to local minima and how they are connected   in configuration space is therefore crucial. Key questions in this respect are: given a local minimum, what is the number and what is the energy distribution of the saddles that lie at a fixed distance from it in configuration space? Which among these saddles are \emph{geometrically connected} to the minimum, meaning that there exist descending paths in the landscape that connect the saddle to the minimum? Do these saddles represent potential escape states for the system that is dynamically trapped in a metastable local minimum?

For random landscapes, these questions can be approached within a statistical framework. The so called spherical $p$-spin model~\cite{Gross,DerridaR,crisom92,crisom95} gives one of the simplest incarnations of a random landscape: the energy functional is in this case a monomial of degree $p$ with random coefficients and Gaussian statistics, defined on a sphere of large dimension $N \gg 1$. In this model the random fluctuations give rise to a rugged landscape, with an exponentially-large (in the dimension $N$) number $\mathcal{N}  \sim {\rm exp}\quadre{N \Sigma + o(N)} $ of stationary points, $\Sigma$ being their `complexity'. These points are non-trivially distributed in terms of their energy and stability: local minima are typically confined below a certain energy level called the \emph{threshold}, above which saddles of extensive index $k=O(N)$ dominate (the index being the number of unstable directions in configuration space). More precisely, at any value of energy below the threshold one typically finds an exponentially-large number $\mathcal{N}_k \sim {\rm exp}\quadre{N \Sigma_k + o(N)}$ of saddles of arbitrary non-extensive index $k=o(N)$. These saddles are distributed hierarchically, with complexities $\Sigma_k$ that are strictly decreasing with $k$:  the dominant (at the exponential scale in $N$) stationary points below the threshold are minima with $k=0$, followed by index-1 saddles, index-2 saddles and so on~\cite{giardinacavagnaparisi,auffingerbenaouscerny}. 

Because of the large-dimensionality of configuration space, for any given local minimum of the $p$-spin landscape the saddles lie in overwhelming majority at very large distance from it in configuration space, and are geometrically disconnected to it. Those saddles that are close and connected to the minimum are \emph{atypical} in the sense that they constitute an exponentially-small (in $N$) fraction of the whole population: computing their complexity requires to condition explicitly to be nearby the reference minimum in configuration space. 
A calculation of this type was first performed in Ref.~\cite{AnInvestigation} (see also \cite{CGPConstrainedComplexity, ThreeReplicas}), where the \emph{constrained} complexity of stationary points at fixed distance from a reference minimum was obtained through the replica formalism and within the so called annealed approximation. More recently, the same results have been recovered within a quenched formalism exploiting the Kac-Rice formalism~\cite{RBCBarriers}, and supplemented with the statistical analysis of the Hessian of the counted stationary points, that allowed to determine their stability. For the $p$-spin model, it is found that the stationary points that are closer to the minimum are typically saddles of index-1 connected geometrically to it, while those at larger distance are other local minima. As a consequence, information on the statistics of the energy barriers surrounding the minimum can be extracted from the energy distribution of the nearby index-1 saddles. 

The information obtained in this way is however not fully complete, as it corresponds only to the saddles that are closest to the reference minimum. In other words, the calculation performed in \cite{RBCBarriers} allows to identify only the saddles that lie in the region of configuration space where they are the typical stationary points (\emph{i.e.}, those having larger complexity). At larger distance from the reference minimum, it is likely that other index-1 saddles connected to the minimum are present, but are not traced as they have smaller complexity with respect to minima. The purpose of this work is to identify these saddles and determine their energy distribution and complexity. 

To target the saddles in the regions of configuration space dominated by minima, we need to impose explicit constraints on the Hessian matrices of the stationary points we are counting. These matrices have the statistics of a GOE matrix deformed with some finite-rank perturbations, that are generated by conditioning the stationary point to be at fixed distance from the reference minimum. In particular, computing the complexity requires to determine the joint probability distribution of the smallest eigenvalue of such deformed GOE matrix, and of the corresponding eigenvector. 
Random matrix ensembles deformed with low-rank perturbations have been widely investigated in the literature: extensive effort has been devoted in particular to the characterization of the eigenvalues transitions (by now generically known as \emph{BBP transitions}~\cite{BBPpaper}) occurring when outliers (or isolated eigenvalues) appear in the spectrum. For deformed Wigner matrices (in particular in the case of Gaussian entries), several results have been derived on the typical value of the isolated eigenvalues \cite{EdwardsJones, Kosterlitz, Peche, Peche2, Frankel}, on their fluctuations \cite{ FluctuationsFeral, Pizzo, BenaychFluctuations} and on the typical value of the eigenvector projection along the direction of the perturbation \cite{Benaych-Georges}. The large deviations of the isolated eigenvalue in the case of a deterministic additive perturbation have been determined in~\cite{Maida}. This result has been recently pushed forward in~\cite{BiroliGuinnet}, by computing the \emph{joint} large deviations of the isolated eigenvalue and of the projection of the corresponding eigenvector along the direction of the additive perturbation. This work builds on~\cite{BiroliGuinnet} to extend the large deviation results to the case in which the GOE matrix is deformed with a combination of both an additive and multiplicative perturbations, which is relevant to characterize the statistics of the $p$-spin Hessian matrices at a critical point.

This paper is split into three parts: in the the first part (Section \eqref{sec:PartOne}) we present the results on the $p$-spin energy landscape. In the second part (Section \eqref{sec:PartTwo}) we state the large deviation functions of the smallest eigenvalue and eigenvectors of a deformed GOE matrix in general form, and summarize the main steps of the derivation. The third part (Section \eqref{sec:PartThree}) is devoted to the derivation of these large deviation principles. The second and third parts of the paper are formulated in general terms, and can be read independently from the first. A more detailed summary of the structure of the paper is given at the beginning of each part. The conclusions are given in Section \eqref{sec:Conclusions}.

\section{Part I: rare saddles in the landscape of the spherical  $p$-spin model}\label{sec:PartOne}
In this first part of the work, we discuss how the complexity of index-1 saddles of the spherical $p$-spin model is obtained, and present the results of the calculation. In Section \ref{sec:Known} we summarize the general formalism for the computation of the complexity and we recall the statistical properties of the Hessian of the energy landscape, evaluated at the stationary points. In Section \ref{sec:AtypicalSaddles} we set up the calculation of the complexity of the \emph{atypical} saddles, and we state the expressions of the large deviation functions for the minimal eigenvalue and eigenvector of the Hessian matrices. In Section \ref{sec:Results} we present the resulting complexity of the index-1 saddles at fixed given overlap from a reference minimum of the landscape, and we comment on the implications for the dynamical exploration of the landscape.

\subsection{The $p$-spin energy landscape: total constrained complexity and Hessian statistics}\label{sec:Known}

\subsubsection{Constrained complexity and Kac-Rice formula.}
We consider the energy landscape of the spherical $p$-spin model with $p \geq 3$:
\begin{equation}\label{eq:HampSpin}
 E\quadre{{\bf s}}=-\sum_{i_1 < i_2 \dots < i_p}J_{i_1,i_2,\dots,i_p } s_{i_1} s_{i_2} \dots s_{i_p}, 
\end{equation}
where $i_k \in \grafe{1, \cdots, N}$, the configurations ${\bf s}=(s_1, \cdots, s_N)$ lie on the surface of a sphere and satisfy $\sum_{i=1}^{N} s_i^2=N$, and their closeness is measured in terms of the overlap $q({\bf s}, {\bf s'})={\bf s} \cdot {\bf s'}/N$.  The quenched random couplings $J_{i_1,i_2,\dots,i_p}$ are independent Gaussian variables with zero mean and variance $\langle J_{{\bf i} }^2 \rangle =p!/2N^{p-1}$. The random energy landscape \eqref{eq:HampSpin} is therefore itself Gaussian, with zero average $\langle E\quadre{{\bf s}}  \rangle =0 $ and  covariance
\begin{equation}\label{eq:Isotropy}
\langle  E\quadre{{\bf s}}   E\quadre{{\bf s'}} \rangle = \frac{N}{2} \tonde{\frac{{\bm s} \cdot{ \bm s'}}{N}}^p
\end{equation}
that is isotropic, meaning that it depends on ${\bm s}, {\bm s'}$ only through their overlap. In the following, we denote 
 the energy density of a configuration ${\bf s}$  by $\epsilon= \lim_{N \to \infty} E[{\bf s}]/N$. The threshold value of the energy reads $\epsilon_{\text{th}}=-[2 (p-1)/p]^{1/2}$, while $\epsilon_{\text{gs}}$ denotes the density of the ground state configurations. 
 
 At energy densities $\epsilon > \epsilon_{\rm th}$ the landscape is dominated by saddles with a huge index $k =O(N)$: this portion of the landscape is easily explored dynamically since stationary points have plenty of directions in configuration space in which the energy landscape is  descending~\cite{CuKu}, and it is not of interest in the light of activated dynamics. We therefore restrict to the energy regime $\epsilon_{\text{gs}} \leq \epsilon \leq \epsilon_{\text{th}}$, which is dominated by stationary points that are either trapping local minima or saddles with few negative directions $k \sim o(N)$. The complexities $\Sigma_k (\epsilon)$ count the number of such stationary points of energy density $\epsilon$ and index $k$, at the exponential scale in $N$. The \emph{total} complexity $\Sigma (\epsilon)$ is obtained as
 \begin{equation}
 \Sigma (\epsilon)= \max_{k} \Sigma_k (\epsilon).
 \end{equation}
For the spherical $p$-spin $\Sigma(\epsilon)=\Sigma_0 (\epsilon)$ for all $\epsilon_{\rm gs} \leq \epsilon \leq \epsilon_{\rm th}$: at each value of energy below the \emph{threshold} the typical (most numerous) stationary points are local minima. 
 
In the following we aim at characterizing stationary points ${\bf s}$ of energy density $\epsilon$ and index $k$ that are at overlap $q={\bf s} \cdot {\bf s}_0/N$ with respect to some fixed local minimum ${\bf s}_0$ of the landscape, extracted with uniform measure among those at energy $\epsilon_0$. We denote with $ \Sigma_k(\epsilon, q| \epsilon_0)$ the corresponding complexities, and with $ \Sigma(\epsilon, q| \epsilon_0)$ the total one, obtained maximizing over $k$. More precisely,  following the notation of Ref. \cite{RBCBarriers} we define rescaled spin configurations on the unit sphere, ${\bm \sigma}= {\bf s}/\sqrt{N}$, and introduce the rescaled energy 
$h[{\bm \sigma}] \equiv  \sqrt{{2}/{N}} E[\sqrt{N} {\bm \sigma}]$. Given a reference local minimum ${\bm \sigma}^0$ drawn at random from the population of minima with energy $\epsilon_0$ ($\epsilon_{\text{gs}} \leq \epsilon_0 \leq \epsilon_{\text{th}}$), we denote with $\mathcal{N}_{{\bm \sigma}^0}(\epsilon, q| \epsilon_0)$ the number of stationary points with energy $\epsilon$ that are at fixed overlap ${\bm \sigma}^0 \cdot {\bm \sigma}= q$ with the minimum, and define the associated total quenched complexity as:
\begin{equation}\label{eq:QuenchedConstrained}
 \Sigma(\epsilon, q| \epsilon_0)= \lim_{N \to \infty}\frac{1}{N}\Big\langle \,  \log \mathcal{N}_{{\bm \sigma}^0}(\epsilon, q| \epsilon_0)
 \Big\rangle_{0},
\end{equation}
where the average $\langle \cdot \rangle_0$ is over both the local minima of energy $\epsilon_0$ at fixed realization of the random energy field \eqref{eq:HampSpin}, and over the different realizations of the latter. Notice that for $q=0$, which is the typical value of the overlap between an arbitrary pair of stationary points, the constraint is ineffective and \eqref{eq:QuenchedConstrained} reproduces the well-known complexity curve of local minima $\Sigma(\epsilon)=\Sigma_0(\epsilon)$.

The total complexity \eqref{eq:QuenchedConstrained} has been computed in \cite{RBCBarriers} using the Kac-Rice formula and its generalizations~\cite{fyodorov, auffingerbenaouscerny, SpikedRepKacRice}. From that calculation it followed that the quenched complexity actually coincides with its annealed counterpart computed in \cite{CGPConstrainedComplexity}, obtained exchanging the average with the logarithm in \eqref{eq:QuenchedConstrained}. 
The latter be easily obtained as the large-$N$ asymptotic of the Kac-Rice formula for the first moment of $\mathcal{N}_{{\bm \sigma}^0}$. To state the formula, we introduce the gradient ${\bf g} \quadre{{\bm \sigma}}$ of the energy field $h[{\bm \sigma}]$: since the functional is restricted to the sphere, its gradient lies in the $M=(N-1)-$dimensional tangent plane to the sphere at the point ${\bm \sigma}$; similarly, the Hessian matrix $\mathcal{H}\quadre{{\bm \sigma}}$ collects the components of the second derivatives of $h[{\bm \sigma}]$ along the directions corresponding to some basis $\grafe{{\bf e}_i[{\bm \sigma}]}_{i=1}^M$ spanning the tangent plane. In terms of these quantities, the constrained complexity reads:
{\medmuskip=0mu
\begin{equation}\label{eq:FirstMomentKacRice}
\Sigma(\epsilon, q | \epsilon_0)=\lim_{N \to \infty} \frac{1}{N} \log \quadre{\int d{\bm \sigma} \, \delta\tonde{{\bm \sigma} \cdot {\bm \sigma}^0\hspace{-0.05 cm}-\hspace{-0.05 cm} q\hspace{-0.01 cm}}\mathcal{E}_{{\bm \sigma}|{\bm \sigma}_0}(\epsilon, q|\epsilon_0) \,
\, p_{{\bm \sigma}|{\bm \sigma}^0}({\bf 0}, \epsilon)},
 \end{equation}}
where the integration is over the configurations ${\bm \sigma}$ at fixed overlap $q$ with the reference minimum, $ p_{{\bm \sigma}|{\bm \sigma}^0}({\bf 0}, \epsilon)$ denotes the joint density function of the gradient and field $({\bf g} [{\bm \sigma}], h[{\bm \sigma}])$, conditioned to the values of gradient and field at ${\bm \sigma}^0$ and evaluated at $({\bf 0}, \sqrt{2 N} \epsilon)$, and $\mathcal{E}_{{\bm \sigma}|{\bm \sigma}_0}(\epsilon, q|\epsilon_0)$ is the following expectation value
\begin{equation}
\mathcal{E}_{{\bm \sigma}|{\bm \sigma}_0}(\epsilon, q|\epsilon_0)= \Big\langle  \left| \text{det} \mathcal{H}[{\bm \sigma}]\right| \Big|  \grafe{ \begin{subarray}{l}
 {\bf g}[{\bm \sigma}^0]=0, {\bf g}[{\bm \sigma}]=0\\
  h[{\bm \sigma}^0]=\sqrt{2 N} \epsilon_0,   h[{\bm \sigma}]=\sqrt{2 N} \epsilon
  \end{subarray}} \Big\rangle.
\end{equation}
Notice that, while in principle the quantity inside the logarithm in \eqref{eq:FirstMomentKacRice} depends on the particular local minimum ${\bf \sigma_0}$, as a consequence of the isotropy of the $p$-spin covariances the dependence is only on the overlap parameter $q$. Therefore the uniform average on the local minima at fixed value of $q$ yields a constant factor equal to one (see the Supplemental Material of Ref.  \cite{RBCBarriers}, in particular Sec. G.1, for the derivation of this formula). The asymptotic of \eqref{eq:FirstMomentKacRice} is determined by computing the conditional distribution of the energy field and of its derivatives, which can be determined explicitly due to Gaussianity. In particular, the average of the Hessian determinant in \eqref{eq:FirstMomentKacRice} is over the distribution of $\mathcal{H}[{\bm \sigma}]$ \emph{conditioned} to the fact that ${\bm \sigma}$ is a stationary point of energy density $\epsilon$, at fixed overlap $q$ from another stationary point (a minimum) of energy $\epsilon_0$, as we recall in the following section.

\subsubsection{Statistics of the Hessians at overlap $q$ from a reference minimum.}\label{sec:statHess}

In absence of conditioning (equivalently, for $q=0$) the Hessian at a stationary point ${\bm \sigma}$ has the statistical distribution of a GOE matrix, shifted by a constant diagonal matrix that depends only on the energy density $\epsilon$. This follows from the isotropy of the correlations \eqref{eq:Isotropy}, which translates into a matrix distribution that is itself invariant under basis rotations in the tangent plane. The energy-dependent shift follows from the spherical constraint imposed on the variables ${\bm \sigma}$, and it is such that for any $\epsilon< \epsilon_{\rm th}$ the typical configuration of the Hessian density of states (in the large-$N$ limit) is a semicircle which is entirely supported on the positive semi-axis, implying that typical stationary points are minima. Saddles are generated by large deviations of the smallest eigenvalues of the Hessian, that are pulled out of the bulk of the density of states and into the negative semi-axis: this happens with a large-deviation probability that is exponentially decaying in $N$ \cite{LargeDevGOE}, implying the exponential suppression of the complexity of saddles with respect to the one of minima \cite{giardinacavagnaparisi,auffingerbenaouscerny}.

 \begin{figure}[t!]
  \centering
       \includegraphics[width=.46\linewidth]{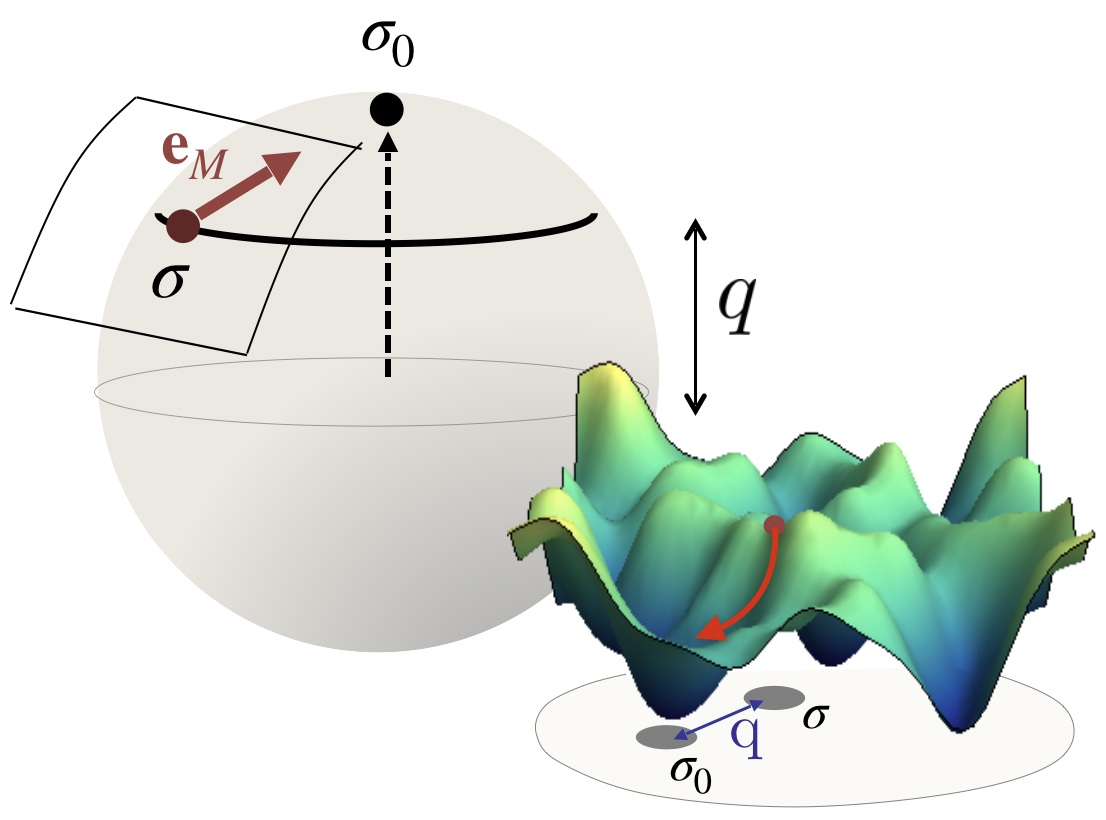} 
\caption{\small Schematic representation of configuration space, with the reference minimum ${\bm \sigma_0}$ and a saddle ${\bm \sigma}$ at overlap $q$, that is geometrically connected to the minimum. The vector ${\bf e}_M$ lies in the tangent plane to the sphere at ${\bm \sigma}$, along the direction connecting ${\bm \sigma}$ to the reference minimum ${\bm \sigma}_0$.  }\label{fig:disegnino}
  \end{figure}

When we enforce the point ${\bm \sigma}$ to be at finite overlap $q$ from another local minimum ${\bm \sigma}_0$, the isotropy is broken along the direction in configuration space that connects the two stationary points. At the level of the Hessian statistics, this translates into rank-1 perturbations (both additive and multiplicative) to an otherwise GOE distributed matrix, that depend explicitly on the parameters $\epsilon, \epsilon_0$ and $q$~\cite{RBCBarriers} (see also \cite{eliran}). 
To express it compactly, it is convenient to choose a basis $\grafe{{\bf e}_i[{\bm \sigma}]}_{i=1}^M$ in the tangent plane at ${\bm \sigma}$  in such a way that the last vector ${\bf e}_M=(q{\bm \sigma}-{\bm \sigma}_0)/\sqrt{1-q^2}$ is the only one having a projection on ${\bm \sigma}_0$, while all the remaining ones are arbitrary vectors spanning the space orthogonal to ${\bm \sigma}, {\bm \sigma}_0$, see Fig. \ref{fig:disegnino}.  With this choice of basis the conditioned Hessian is distributed as:
\begin{equation}\label{eq:Shift}
\mathcal{H}[{\bm \sigma}] \sim  \mathcal{M}- \sqrt{2 N} p \epsilon\, \mathbb{1},
\end{equation}
where $\mathbb{1}$ is the identity matrix and $\mathcal{M}$ is an $M$-dimensional matrix with the following properties: the $(M-1)-$dimensional block made of the entries $m_{ij (\neq M)}$ has GOE statistics with zero average and variance 
\begin{equation}\label{eq:VariancePspin}
\sigma^2(p)=p(p-1);
\end{equation}
the elements $m_{iM}$ for $i \neq M$ have a different variance $\Delta^2(q)< \sigma^2$ depending explicitly on the overlap parameter $q$, 
and the element $m_{MM}$ has a non-zero average $\mu(q, \epsilon, \epsilon_0)$ and yet another variance $ \Tilde{\Delta}^2 (q)<\Delta^2(q)$. These functions depend explicitly on $p$: for $p=3$, for instance, one finds $ \Tilde{\Delta}^2(q)=0$. Their explicit form is recalled in Appendix~\ref{app:ConstantsHessian}.

To further simplify the notation, we introduce an $M \times M$ deterministic matrix of the form:
\begin{equation}
 F(q) \equiv \mathbb{1}- \quadre{1- \frac{\Delta(q)}{\sigma}} \, {\bf e}_{M} {\bf e}^T_{M},
\end{equation}
and define a complex (purely imaginary) variable $\zeta(q)$ through the identity: 
\begin{equation}
\frac{\Delta^4(q) }{\sigma^2} + [\zeta(q)]^2= \tilde{\Delta}^2(q).
\end{equation}
The matrix $\mathcal{M}$ can then be re-written as: 
\begin{equation}\label{eq:DistrHess0}
 \mathcal{M}= F(q) \mathcal{X} F(q) + \tonde{\sqrt{N}\mu + \zeta(q) \,\xi} {\bf e}_{M} {\bf e}^T_{M}.
\end{equation}
Here  $\mathcal{X}$ is a GOE matrix with variance \eqref{eq:VariancePspin} and $\xi$ is a Gaussian random variable independent of $\mathcal{X}$, having zero average and unit variance. Notice that the variance of the $MM$ element of the perturbed matrix $F(q) \mathcal{X} F(q)$ equals to $\Delta^4/ \sigma^2$, which is different with respect to the variance of $m_{MM}$: the fluctuating variable $\xi$ is added to compensate for this difference.  As we recall in the next section, the main effect of the finite-rank perturbation in \eqref{eq:DistrHess0} is to modify the typical configuration of the density of states giving rise to an isolated eigenvalue.

 \subsubsection{The isolated eigenvalue of the Hessian and the saddles.}
When the  finite-rank perturbations to the GOE matrix $\mathcal{M}$ in \eqref{eq:Shift} are sufficiently strong, they generate a sub-leading correction to the density of states
\begin{equation}\label{eq:SemicircleP}
\rho_{\epsilon}(\lambda)=\frac{\sqrt{4 \sigma^2(p)-(\lambda+ \sqrt{2} p \epsilon)^2}}{2 \pi \sigma^2(p)}
\end{equation}
of the Hessian matrix,
in the form of a single eigenvalue  $ \lambda_0(q, \epsilon, \epsilon_0)$ that is isolated and detached from the support of \eqref{eq:SemicircleP}, meaning that  $\lambda_0(q, \epsilon, \epsilon_0) < -2 \sigma^2- \sqrt{2} \,p\, \epsilon$. The explicit expression of this eigenvalue has been determined in \cite{RBCBarriers}. It is more conveniently given in terms of the resolvent\footnote{ The resolvent is defined for $|z|>2 \sigma$ as the solution of the quadratic equation $\sigma^2 G^2_\sigma(z)-z G_\sigma(z)+1=0$ satisfying $G_\sigma(z) \to 0$ as $|z| \to \infty$.} of the unperturbed GOE matrix $\mathcal{X}$ with variance $\sigma$: 
\begin{equation}\label{eq:Resolvent0}
 G_\sigma(z)=\left \langle \frac{1}{M}\text{Tr} {\frac{1}{z-\mathcal{X}}} \right \rangle \stackrel{z \text{ real}}{=}\frac{1}{2 \sigma^2} \tonde{z- \text{sign}(z) \sqrt{z^2- 4 \sigma^2}} \in \quadre{-\frac{1}{\sigma}, \frac{1}{\sigma}}.
 \end{equation}
 Setting
 \begin{equation}
 \lambda_0(q, \epsilon, \epsilon_0)= \lambda_{\rm min}^{\rm typ}(q, \epsilon, \epsilon_0) - \sqrt{2} p \epsilon,
 \end{equation}
 it is found in \cite{RBCBarriers} that the typical value $\lambda_{\rm min}^{\rm typ}(q, \epsilon, \epsilon_0)$ of the smallest eigenvalue of $\mathcal{M}$ is the solution of $\lambda - \mu(q, \epsilon, \epsilon_0)-\Delta^2(q) G_\sigma(\lambda)=0$, and reads explicitly: 
 \begin{equation}
\lambda_{\rm min}^{\rm typ}(q, \epsilon, \epsilon_0)=\frac{1}{2 (\sigma^2-\Delta^2)} \tonde{2 \mu \sigma^2- \Delta^2 \mu+ \Delta^2 \sqrt{\mu^2-4 (\sigma^2- \Delta^2)}}= \mu + \Delta^2 G_{\sigma'}(\mu), 
\end{equation}
where $ G_{\sigma'}(\mu)$ has a modified variance 
\begin{equation}\label{eq:SigmaPrimePspin}
\sigma'(p,q)=\sqrt{\sigma^2(p)-\Delta^2(q)}.
\end{equation}
 Notice that this expression is independent of the Gaussian fluctuations with variance $\zeta(q)$ of the element $m_{MM}$.
Using the equation satisfied by $\lambda_{\rm min}^{\rm typ}$ we get:
 \begin{equation}\label{eq:TypFinal}
\lambda_{\rm min}^{\rm typ}(q, \epsilon, \epsilon_0)=G^{-1}_\sigma\tonde{G_{\sigma'}(\mu)}=\frac{1}{G_{\sigma'}(\mu)}+ \sigma^2 G_{\sigma'}(\mu),
 \end{equation}
 where 
 \begin{equation}\label{eq:InvResolvent}
G_\sigma^{-1}(z)= \frac{1}{z}+ \sigma^2 z
\end{equation}
 is the inverse of the resolvent operator, restricted to the domain $|z|<1/\sigma$. This expression is consistent provided that  condition is that the argument of $G_{\sigma}^{-1}$ belongs to $\quadre{-1/\sigma, 1/\sigma}$. Assuming that $G_{\sigma'}(\mu)<0$, this gives:
\begin{equation}\label{eq:CondBPP}
G_{\sigma'}(\mu)>- \frac{1}{\sigma} \longrightarrow \mu< - \sigma \quadre{1+ \frac{(\sigma')^2}{\sigma^2}},
\end{equation}
which identifies the regime of parameters for which the isolated eigenvalue exists. 
We denote the threshold value with:
\begin{equation}\label{eq:MuBBPpSpin}
\mu_{\rm c}(p,q) \equiv - \sigma(p) \quadre{1+  \tonde{\frac{\sigma'(p)}{\sigma(p)}}^2}=- \sqrt{p(p-1)} \quadre{2- \frac{\Delta^2(q)}{p(p-1)}}.
\end{equation}
In this regime, the eigenvector ${\bf v}_0$ associated to $\lambda_{\rm min}^{\rm typ}$ has a projection $ {\bf v}_0 \cdot {\bf e}_M[{\bm \sigma}]$ along the direction connecting the two stationary points which remains non-zero as $N \to \infty$. 
Notice that $\lim_{\sigma' \to 0}  G_{\sigma'}(\mu)= 1/\mu$, implying that when $\Delta(q) \to \sigma$ the eigenvalue exists for $\mu<-\sigma$ and reduces to $\lambda_{\rm min}^{\rm typ}= \mu + \sigma^2/\mu$, reproducing the well-known expression resulting from a purely additive perturbation \cite{EdwardsJones, Kosterlitz, Peche, Peche2}. In presence of a multiplicative perturbation given by the matrices $F(q)$, the same form holds with $1/\mu$ replaced with $G_{\sigma'}(\mu)$.

When the parameters are such that the shifted eigenvalue $\lambda_0(q, \epsilon, \epsilon_0)<0$, the associated stationary points are saddles of index-1. As found in \cite{RBCBarriers}, this happens when the overlap $q$ with the reference minimum is large enough (for any fixed $\epsilon$, larger than a given $q_{\rm ms}(\epsilon|\epsilon_0)$, see Fig. \ref{fig:energies}): the total complexity \eqref{eq:QuenchedConstrained} is therefore contributed by saddles for large enough $q$. These saddles are \emph{geometrically connected} to the reference minimum ${\bm \sigma_0}$, meaning that their unstable direction has an $O(1)$ projection along the direction pointing towards ${\bm \sigma_0}$ in configuration space. Notice that no large deviation calculation is necessary to find these saddles, as the \emph{typical} configurations of the Hessian have a negative mode: in other words, at these values of the overlap index-1 saddles are the typical, exponentially most numerous stationary points. At smaller values of $q$, the typical stationary points are instead minima with no isolated eigenvalue; in this regime the complexity of saddles has to be obtained with a large deviation calculation, by conditioning explicitly the Hessian to exhibit one negative isolated eigenvalue.

\subsection{Computing the complexity of atypical saddles}\label{sec:AtypicalSaddles}

\subsubsection{The constrained complexity of saddles.}\label{sec:ConstrComp1}
We now give a formula for the constrained complexity of saddles at overlap $q$ with the reference minimum, in the \emph{annealed} approximation. Let us denote with  $\mathcal{N}_{{\bm \sigma}^0}(\epsilon, q, \lambda, u| \epsilon_0)$ the number of stationary points ${\bm \sigma}$ having an Hessian with smallest eigenvalue taking a given value $\lambda_{\rm min}=\lambda$ and such that the corresponding eigenvector ${\bf v}_{\rm min}$ has a macroscopic projection $u_{\rm min}=|{\bf v}_{\rm min} \cdot {\bf e}_M[{\bm \sigma}]|^2=u>0$ along the direction connecting the two stationary points in configuration space. The complexity $ \Sigma(\epsilon, q, \lambda, u| \epsilon_0)$ of these points in the annealed approximation is given by: 
\begin{equation}\label{eq:QuenchedConstrained1}
 \Sigma(\epsilon, q, \lambda,u| \epsilon_0)= \lim_{N \to \infty}\frac{1}{N}  \log \Big\langle \, \mathcal{N}_{{\bm \sigma}^0}(\epsilon, q, \lambda,u| \epsilon_0)
 \Big\rangle_{0},
\end{equation}
where the average number can be written as:
\begin{equation}\label{eq:FirstMomentKacRiceS}
\begin{split}
 \Big\langle \mathcal{N}_{{\bm \sigma}^0}(\epsilon, q, \lambda, u| \epsilon_0)\Big\rangle_0= & \int d{\bm \sigma} \, \delta\hspace{-0.05 cm}\tonde{{\bm \sigma} \hspace{-0.05 cm}\cdot\hspace{-0.05 cm} {\bm \sigma}^0\hspace{-0.05 cm}-\hspace{-0.05 cm} q\hspace{-0.01 cm}} 
  \Big\langle  \left| \text{det} \mathcal{H}[{\bm \sigma}]\right| \Big|  \grafe{ \begin{subarray}{l}
 {\bf g}[{\bm \sigma}^0]=0, {\bf g}[{\bm \sigma}]=0\\
  h[{\bm \sigma}^0]=\sqrt{2 N} \epsilon_0,   h[{\bm \sigma}]=\sqrt{2 N} \epsilon\\
  \lambda_{\rm min}=\lambda,   \, \,  u_{\rm min}=u
  \end{subarray}} \Big\rangle \times \\
  & \times
\, p_{{\bm \sigma}|{\bm \sigma}^0}({\bf 0}, \epsilon) \, \mathbb{G}_{{\bm \sigma}|{\bm \sigma}^0}\tonde{\lambda,u}.
\end{split}
 \end{equation}
In this modified version of the Kac-Rice formula, the expectation value of the Hessian is conditioned also to the event $\lambda_{\rm min}=\lambda$ and $u_{\rm min}=u$.  The case $\lambda<0$ corresponds to saddles with \emph{at least} one unstable direction. The constraint on the overlap  $u_{\rm min}=u$ is added to track whether the saddles are geometrically connected to ${\bm \sigma_0}$ (when $u>0$), or whether the downhill direction is uncorrelated with the minimum ${\bm \sigma}_0$ (when $u=0$). The function $\mathbb{G}_{{\bm \sigma}|{\bm \sigma}^0}\tonde{\lambda,u}$ is the joint distribution of $(\lambda_{\rm min}, u_{\rm min})$ induced by the statistics of the conditioned Hessian described in Sec. \ref{sec:statHess}.  

In Appendix \ref{app:determinantHessian} we argue that conditioning on $\lambda$ and $u$ does not modify the typical density of states of the Hessian to leading order in $N$, which therefore remains equal to \eqref{eq:SemicircleP}. The effect of the conditioning is (at most) to generate isolated eigenvalues, that are sub-leading corrections to the density of states. As a consequence, to (exponential) order in $N$ the expectation value of the determinant in \eqref{eq:FirstMomentKacRiceS} is insensitive to the conditioning on the smallest eigenvalue. Additionally, the distribution $\mathbb{G}_{{\bm \sigma}|{\bm \sigma}^0}\tonde{\lambda,u}$  depends on ${\bm \sigma}$ and ${\bm \sigma}^0$ only through the parameters $q, \epsilon$ and $\epsilon_0$, because the full distribution of the Hessian does. We re-label it as $\mathbb{G}_{\epsilon, q| \epsilon_0} \tonde{\lambda,u}$ in the following. 
For values of $\lambda, u$ that are different with respect to the typical ones, $\mathbb{G}_{\epsilon, q| \epsilon_0} \tonde{\lambda,u}$ is a large deviation probability with a given rate function to be determined: 
\begin{equation}\label{eq:RateP1}
 \lim_{N \to \infty} \frac{\log \mathbb{G}_{\epsilon, q| \epsilon_0} \tonde{\lambda,u}}{N}= -{L}_{\epsilon, q| \epsilon_0}(\lambda,u).
\end{equation}
It follows from these considerations that we can re-write \eqref{eq:QuenchedConstrained1} as:
\begin{equation}\label{eq:QuenchedConstrained12}
 \Sigma(\epsilon, q, \lambda, u| \epsilon_0)=  \Sigma(\epsilon, q| \epsilon_0)- {L}_{\epsilon, q| \epsilon_0}(\lambda,u),
 \end{equation}
 where $\Sigma(\epsilon, q| \epsilon_0)$ is the total constrained complexity already computed in \cite{RBCBarriers}. In the following, we shall consider \emph{typical} values $u_{\rm typ}(\lambda)$ of the overlap $u$, defined as:
\begin{equation}
u_{\rm typ}(\lambda) \equiv \underset{u \in \quadre{0,1}}{\mathrm{argmin}} \, {L}_{\epsilon, q| \epsilon_0}(\lambda,u),
\end{equation}
and set
\begin{equation}\label{eq:Rat2}
F_{\epsilon, q| \epsilon_0}(\lambda) \equiv L_{\epsilon, q| \epsilon_0}(\lambda,u_{\rm typ}(\lambda)).
\end{equation}
The complexity of the most numerous stationary points with $\lambda_{\rm min}=\lambda$ is then:
\begin{equation}\label{eq:bah}
\Sigma(\epsilon, q, \lambda | \epsilon_0)= \Sigma(\epsilon, q| \epsilon_0)- F_{\epsilon, q| \epsilon_0}(\lambda),
\end{equation}
and thus it is readily obtained from the large deviation rate $F_{\epsilon, q| \epsilon_0}(\lambda)$ of the smallest eigenvalue of an Hessian. Saddles are obtained setting $\lambda<0$. 
The second and third parts of this work are devoted to the computation of the rate function $F_{\epsilon, q| \epsilon_0}(\lambda)$. In the following section, we adapt the general result to the case of the $p$-spin Hessians.

\subsubsection{Large deviations of the smallest eigenvalue of the Hessians. }\label{sec:LargeDevPspin}
In the third part of this work we derive the large deviation function of the smallest eigenvalue of matrices of the general form:
\begin{equation}
 \mathcal{Y}=   \tonde{ \mathbb{1}- \frac{\beta}{1+ \beta} \, {\bf e}_{M} {\bf e}^T_{M}} \mathcal{X} \tonde{ \mathbb{1}- \frac{\beta}{1+ \beta} \, {\bf e}_{M} {\bf e}^T_{M}}  + \theta \, {\bf e}_{M} {\bf e}^T_{M},
\end{equation}
where $\mathcal{X}$ is a GOE matrix with variance $\sigma^2$, $\beta$ is a non-negative constant and $\theta$ is a Gaussian random variable with mean $\overline{\theta}<0$ and variance $\sigma^2_\theta$.  The Hessian matrices \eqref{eq:DistrHess0} follow this distribution, with 
\begin{equation}
\begin{split}
\sigma \to \sigma(p) \equiv \sqrt{p(p-1)}, \quad 
\beta \to \frac{\sqrt{p (p-1)}}{\Delta(q)}-1, \quad 
\overline{\theta} \to \mu(q, \epsilon, \epsilon_0),
\end{split}
\end{equation}
and 
\begin{equation}
\sigma^2_\theta \to \zeta^2(q)= \tilde{\Delta}^2(q)-\frac{\Delta^4(q)}{\sigma^2},
\end{equation}
where the explicit expressions of these functions are given in Appendix \ref{app:ConstantsHessian}.  We let $\mathcal{F}_{\epsilon, q| \epsilon_0}(\lambda)$ be the corresponding rate function for the minimal eigenvalue. Given the diagonal shift in \eqref{eq:Shift}, we have that the rate in \eqref{eq:Rat2} is obtained as:
\begin{equation}
F_{\epsilon, q| \epsilon_0}(\lambda)=\mathcal{F}_{\epsilon, q| \epsilon_0}(\lambda+ \sqrt{2}\, p\, \epsilon).
\end{equation}
We not adapt the general result of section \ref{sec:OptimizedU} to this case. 
We introduce the threshold values:
\begin{equation}\label{eq:Goal00}
\lambda^{\pm}_p(\epsilon, q| \epsilon_0) \equiv x^{\pm}_{\sigma(p)}\tonde{\mu(q, \epsilon, \epsilon_0), \frac{\sqrt{p(p-1)}}{\Delta}-1} -\sqrt{2}\, p \,\epsilon,
\end{equation}
where the functions $x^{\pm}_\sigma$ are given in \eqref{eq:ExPIuMinus}. 
Given the shifted variance \eqref{eq:SigmaPrimePspin} and the critical value \eqref{eq:MuBBPpSpin}, we define the following three regimes:
\begin{itemize}
\item Regime A:  $- 2 \sigma'(p,q) < \mu(q, \epsilon, \epsilon_0)<0$ 
\item Regime B.1:  $\mu_{\rm c}(p,q)\leq \mu(q, \epsilon, \epsilon_0) \leq - 2 \sigma'(p,q)$
\item Regime B.2:  $\mu(q, \epsilon, \epsilon_0)<\mu_{\rm c}(p,q) $.
\end{itemize}
These three Regimes can be understood in terms of the typical value of the smallest eigenvalue $\lambda_{\rm min}^{\rm typ}(q, \epsilon, \epsilon_0)$ of the matrix \eqref{eq:DistrHess0}: in Regime B.2 the eigenvalue isolated from the bulk of the density of states, $\lambda_{\rm min}^{\rm typ}(q, \epsilon, \epsilon_0)< -2 \sigma(p)$, see \eqref{eq:CondBPP}. In regime B.1. it holds instead $\lambda_{\rm min}^{\rm typ}(q, \epsilon, \epsilon_0)= -2 \sigma(p)$, and the quantities \eqref{eq:Goal00}  are real. The Regime A corresponds to values of the parameters $q, \epsilon$ and $\epsilon_0$ for which the quantities \eqref{eq:Goal00} are complex. Notice that it always holds $\mu_{\rm c}(p,q)<- 2 \sigma'(p,q) \leq 0$. When $\Delta(q) \neq \sigma(p)$, we find  $\sigma'(p) \to 0$: therefore, Regime A is present only when the multiplicative perturbation to the Hessian is present.

To state the form of the large deviation function, we further introduce the function:
\begin{equation}
\mu^*_p(x| q, \epsilon, \epsilon_0)= \theta^*_0 \tonde{x \Big| \, \sigma(p), \frac{\sqrt{p(p-1)}}{\Delta}-1, \mu(q, \epsilon, \epsilon_0), \zeta^2(q)},
\end{equation}
where the function $\theta^*_0$ is defined in \eqref{eq:NewSP} \footnote{For generic $p$, this function has a lengthy expression in terms of the parameters $q, \epsilon, \epsilon_0$. In the special case $p=3$, however, some simplifications occur due to the fact that $\tilde{\Delta}(q) =0$, meaning that in this case the $MM$-element of the Hessian does not fluctuate. In this particular case we find:
{
\medmuskip=0mu
\thinmuskip=0mu
\thickmuskip=0mu
\begin{equation}
\mu^*_{p=3}=\mu -\frac{(1-q^2) x+\sqrt{4 \mu ^2 (1+q^2)^2-4 \mu  (1+3 q^4+4 q^2) x+(3 q^2 +1)^2x^2+24 (1-q^2)^2}}{2 \left(1+q^2\right)}.
\end{equation}
}}.
Given these quantities, the expression of the large deviation function $F_{\epsilon, q| \epsilon_0}(\lambda)$ reads as follows:
\begin{itemize}
\item In Regime A, 
\begin{equation}\label{eq:Ultimo2p}
F_{\epsilon, q| \epsilon_0}(\lambda) = \mathcal{G}_0(\lambda + \sqrt{2} \, p \, \epsilon) 
    \end{equation}

\item In Regime B.1,  
\begin{equation}\label{eq:Ultimo2p}
F_{\epsilon, q| \epsilon_0}(\lambda) = 
\begin{cases}
\mathcal{G}_{q, \epsilon| \epsilon_0}(\lambda + \sqrt{2} \, p \, \epsilon) &\lambda^{-}_p(\epsilon, q| \epsilon_0) <\lambda<\lambda^{+}_p(\epsilon, q| \epsilon_0)\\
\mathcal{G}_0(\lambda + \sqrt{2} \, p \, \epsilon) &\lambda<\lambda^{-}_p \, \text{or} \,\, \lambda^{+}_p <\lambda<- 2 \sigma(p)-\sqrt{2} p \epsilon \\
     \end{cases}
\end{equation}

\item In Regime B.2, 
\begin{equation}\label{eq:Ultimo1p}
F_{\epsilon, q| \epsilon_0}(\lambda) = 
\begin{cases}
\mathcal{G}_{q, \epsilon| \epsilon_0}(\lambda + \sqrt{2} \, p \, \epsilon) & \lambda^{-}_p(\epsilon, q| \epsilon_0) <\lambda<-2 \sigma(p)- \sqrt{2} p \epsilon\\
\mathcal{G}_0(\lambda + \sqrt{2} \, p \, \epsilon) & \lambda<\lambda^{-}_p(\epsilon, q| \epsilon_0).     \end{cases}
\end{equation}
\end{itemize}

Here the large deviation function $\mathcal{G}_0(x)$ is the one of an unperturbed GOE matrix, given by \cite{LargeDevGOE}:
\begin{equation}
\begin{split}
\mathcal{G}_{0}(x)&=\int_{x}^{-2 \sigma} \frac{\sqrt{z^2-4 \sigma^2}}{2 \sigma^2} dz=\frac{x^2}{4 \sigma^2}- \mathcal{I}(x)- \frac{1}{2}+ \log \sigma,
\end{split}
\end{equation}
where for $x<-2 \sigma$:
\begin{equation}\label{eq:GOEIs}
\begin{split}
\mathcal{I}(x)=
\log \tonde{ -\frac{x}{2}+\frac{1}{2}\sqrt{x^2- 4 \sigma^2}}- \frac{1}{2}+ \frac{x^2}{4 \sigma^2}+\frac{x}{4 \sigma^2}\sqrt{x^2-4 \sigma^2}.
\end{split}
\end{equation}
The other rate function is obtained as:
\begin{equation}
\mathcal{G}_{q, \epsilon| \epsilon_0}(x) \equiv  \mathcal{G}_{\theta, \beta}(x) \Big|_{ \; \theta \to \mu^*_p, \; \; 
\beta \to \frac{\sqrt{p(p-1)}}{\Delta(q)}-1}
\end{equation}
where the explicit form of $ \mathcal{G}_{\theta, \beta}(x)$ is given in \eqref{eq:FundamentalRates}. In the regimes in which the large deviation function equals to $\mathcal{G}_0(x)$ it holds $u_{\rm typ}(\lambda)=0$, while in the other regimes one finds $u_{\rm typ}(\lambda)>0$: therefore, the latter regimes correspond to the saddles that are geometrically connected to the reference minimum. In the following, these results are used to determine statistical distribution of index-1 saddles.

 \subsubsection{Quenched {\it vs} annealed complexity: a comment.}
Before discussing the results of the complexity calculation of saddles, it is necessary to comment on  ``annealed'' nature of the calculation we are performing. The complexity in \eqref{eq:QuenchedConstrained1} gives the asymptotic value of the {\it average} number of stationary points with the desired properties; this may in principle differ from the asymptotic value of the {\it typical} number of such stationary points, that is controlled by the so-called quenched complexity which is obtained exchanging the average and the logarithm in \eqref{eq:QuenchedConstrained1}. The calculation of the latter is in general more involved; it requires to 
resort to representation of the logarithm in terms of higher moments of the number of stationary points, \begin{equation}\label{eq:RepTrick}
 \Big\langle \, \log \mathcal{N}_{{\bm \sigma}^0}(\epsilon, q, \lambda,u| \epsilon_0) \Big\rangle_0= \lim_{n \to 0} \frac{\langle \mathcal{N}_{{\bm \sigma}^0}^n(\epsilon, q, \lambda,u| \epsilon_0)\rangle_0-1}{n},
 \end{equation}
 and to analytically continue the expression of these moments in order to values $n \to 0$. As shown explicitly in \cite{RBCBarriers}, when computing the {\it total} constrained complexity $\Sigma(\epsilon, q |\epsilon_0)$  the two procedures are equivalent. The computation of the quenched complexity through the replica trick, indeed, naturally leads to the emergence of an order parameter $q_1$ that can be interpreted as the typical overlap between the stationary points of energy $\epsilon$ that are at overlap $q$ from the reference minimum. The calculation shows that this overlap takes the particularly simple value $q_1=q^2$, indicating that the stationary points have the weakest possible correlation with each others. It is this feature that implies that (i) the quenched and annealed constrained total complexities  $\Sigma(\epsilon, q |\epsilon_0)$ coincide, (ii) the statistical properties of the Hessian described in Sec. \ref{sec:statHess} can be themselves determined in an annealed setting, computing the distribution of $\mathcal{H}[{\bm \sigma}]$ over the realizations of the random energy field only, and not over all the stationary points ${\bm \sigma}$ at fixed overlap from the reference minimum. In the calculation presented here, we are assuming that the same remains true when conditioning to the value of the smallest eigenvalues of the Hessian. As we discuss in Appendix~\ref{app:QuenchedKR}, this corresponds to assuming that the conditioning does not affect the value of the typical overlap $q_1$ between stationary points with those stability properties, introducing additional correlations between them. This is {\it a priori} not guaranteed, and it is therefore an approximation: in the same Appendix, we discuss what would be the steps required to perform a quenched calculation of  $\Sigma (\epsilon, q, \lambda,u| \epsilon_0)$ and comment further on the assumptions on which the annealed approximation relies.

\subsection{Complexity of saddles: the results}\label{sec:Results}

\subsubsection{Transitions in the population of saddles. }\label{sec:TransPop}
\begin{figure}[h!]
  \centering
              \includegraphics[width=.7\linewidth]{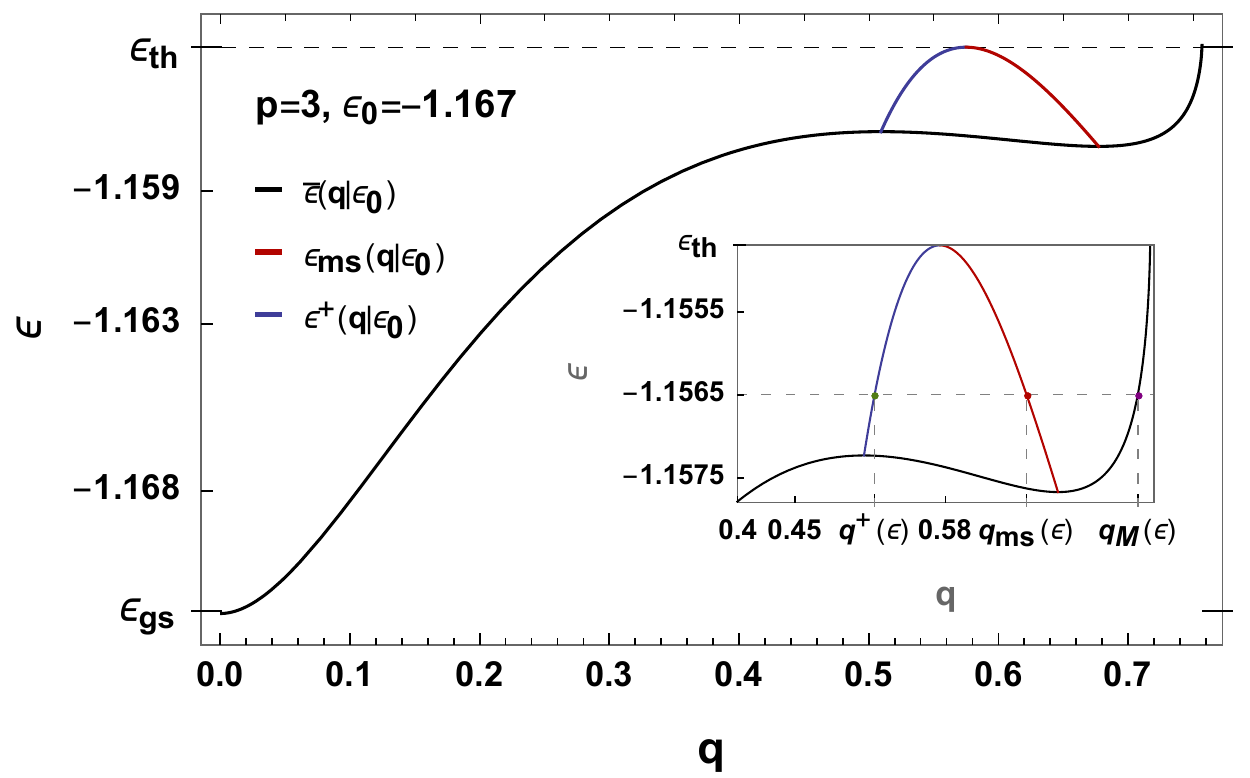} 
\caption{\small Plot of the energy curves $\overline{\epsilon}(q|\epsilon_0)$, $\epsilon_{\rm ms}(q|\epsilon_0)$ 
and $\epsilon^+(q|\epsilon_0)$. \emph{Inset. }Zoom of the main plot. The dashed lines identify the overlaps $q^+(\epsilon| \epsilon_0)$, 
$q_{\rm ms}(\epsilon| \epsilon_0)$ and $q_M(\epsilon| \epsilon_0)$ for $\epsilon=-1.1565$. }\label{fig:energies}
  \end{figure}

For fixed energy $\epsilon_0$ of the reference minimum ${\bm \sigma}_0$, we are interested in characterizing the properties of the \emph{dominant} saddles ({\it i.e.}, of those having higher complexity) as a function of their overlap $q$ with  ${\bm \sigma}_0$ and of their energy density $\epsilon$.  We anticipate that for the values of $q, \epsilon$ for which the complexity of saddles is non-zero, the dominant ones have always index $k=1$. Their properties however change as a function of $q, \epsilon$. To discuss this, it is convenient to introduce three special values of the overlap $q^+(\epsilon| \epsilon_0),  q_{\rm ms}(\epsilon|\epsilon_0), q_M(\epsilon| \epsilon_0)$ and of the energy density  $\epsilon^+(q|\epsilon_0), \overline{\epsilon}(q|\epsilon_0), \epsilon_{\rm ms}(q| \epsilon_0)$ defined in terms of the total constrained complexity $\Sigma(\epsilon, q|\epsilon_0)$ and of
$\lambda^{+}_p(\epsilon, q| \epsilon_0)$  in \eqref{eq:Goal00} in the following way (see Fig. \ref{fig:energies}):
\begin{itemize}
\item The overlap $q_M(\epsilon| \epsilon_0)$ is the one at which the total constrained complexity becomes non-negative, {\it i.e.},  for each $q> q_M(\epsilon| \epsilon_0)$ one finds  $\Sigma(\epsilon, q| \epsilon_0)<0$, implying that typically there are \emph{no} stationary points at those values of the overlap. Similarly, the energy curve $\overline{\epsilon}(q|\epsilon_0)$ gives the energy density of the deepest stationary points found at overlap $q$ with the reference minimum, and it is defined from $\Sigma(\overline{\epsilon}, q| \epsilon_0)=0$: for $\epsilon <\overline{\epsilon}(q|\epsilon_0)$, typically there are \emph{no} stationary points at overlap $q$ with the reference minimum.
\item The overlap $q_{\rm ms}(\epsilon| \epsilon_0)$ is the one at which the stationary points contributing to the total constrained complexity are marginal saddles, with an Hessian having an isolated eigenvalue that is exactly equal to zero: $\lambda_0(q_{\rm ms}, \epsilon, \epsilon_0)=0$. In the high-overlap regime $q_{\rm ms}(\epsilon| \epsilon_0) \leq q \leq q_M(\epsilon| \epsilon_0)$ the complexity $\Sigma(\epsilon, q|\epsilon_0)$ is contributed by index-1 saddles that are geometrically connected to the reference minimum, whereas for $0 \leq q \leq q_{\rm ms}(\epsilon| \epsilon_0)$ it is contributed by local minima. The energy curve $\epsilon_{\rm ms}(q|\epsilon_0)$ gives the energy at which the typical value of the isolated eigenvalue vanishes, and it is defined by $\lambda_0(q, \epsilon_{\rm ms}, \epsilon_0)=0$.
\item The overlap $q^+(\epsilon| \epsilon_0)$ and the energy density $\epsilon^+(q| \epsilon_0)$ are defined as the points where $\lambda^{+}_\sigma(\epsilon, q| \epsilon_0)$ is exactly equal to zero:  $$\lambda^{+}_\sigma(\epsilon, q| \epsilon_0) \Big|_{\epsilon=\epsilon^+(q| \epsilon_0)}=0=\lambda^{+}_\sigma(\epsilon, q| \epsilon_0) \Big|_{q=q^+(\epsilon| \epsilon_0)}$$. 
\end{itemize} 
A plot of the transition overlaps and energies is given in Fig. \ref{fig:energies} for $\epsilon_0=-1.167$ and $p=3$, and the notation is summarized in Table \ref{table:energies}.

\begin{figure}[h!] 
  \begin{minipage}[b]{0.5\linewidth}
    \centering
    \includegraphics[width=.95\linewidth]{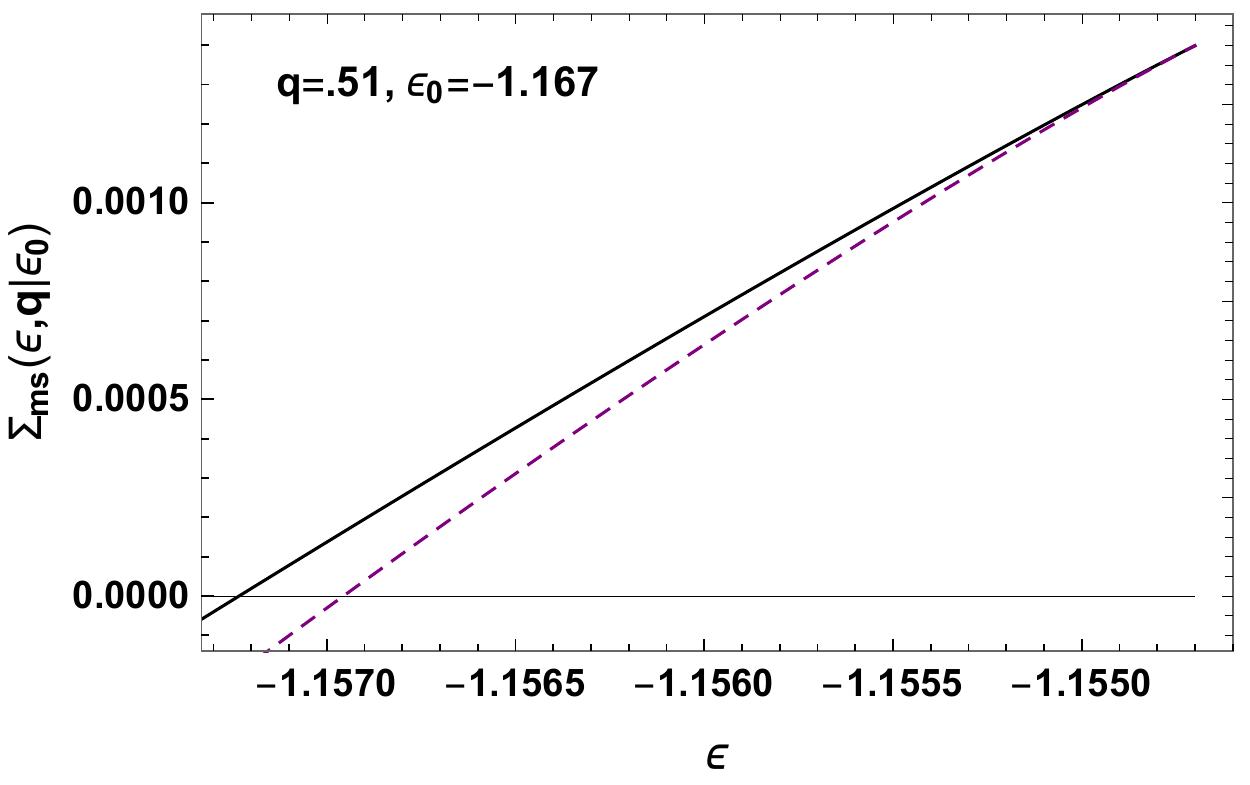}
        \vspace{4ex}
  \end{minipage}
  \begin{minipage}[b]{0.5\linewidth}
    \centering
    \includegraphics[width=.95\linewidth]{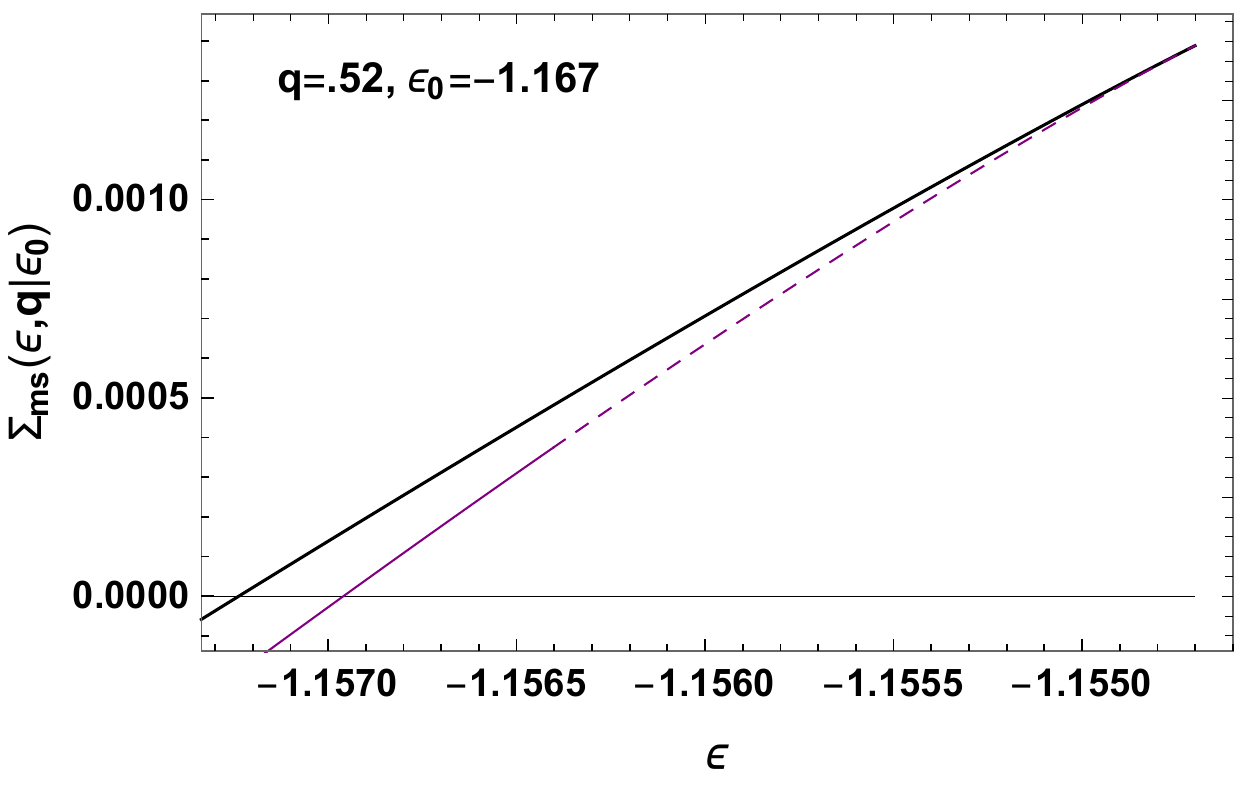} 
    \vspace{4ex}
  \end{minipage} 
  \begin{minipage}[b]{0.5\linewidth}
    \centering
    \includegraphics[width=.95\linewidth]{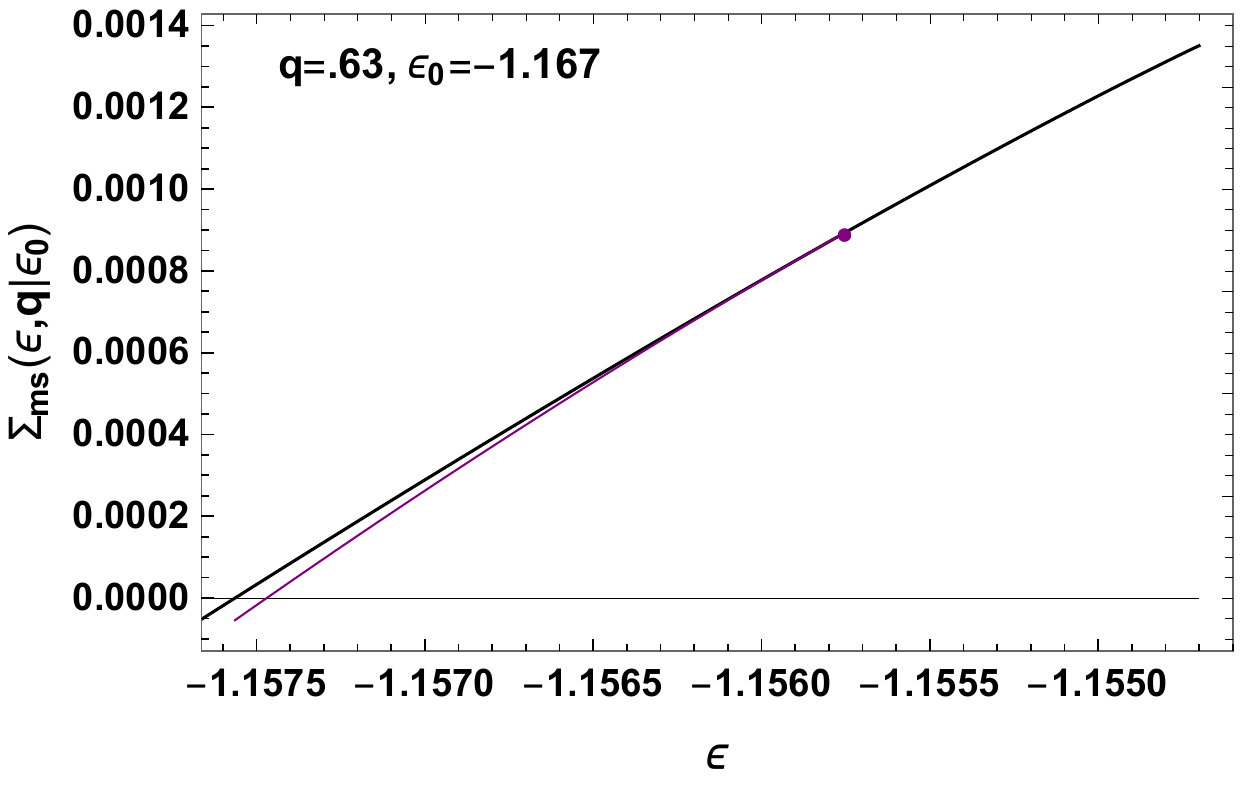} 
    \vspace{4ex}
  \end{minipage}
  \begin{minipage}[b]{0.5\linewidth}
    \centering
    \includegraphics[width=.9\linewidth]{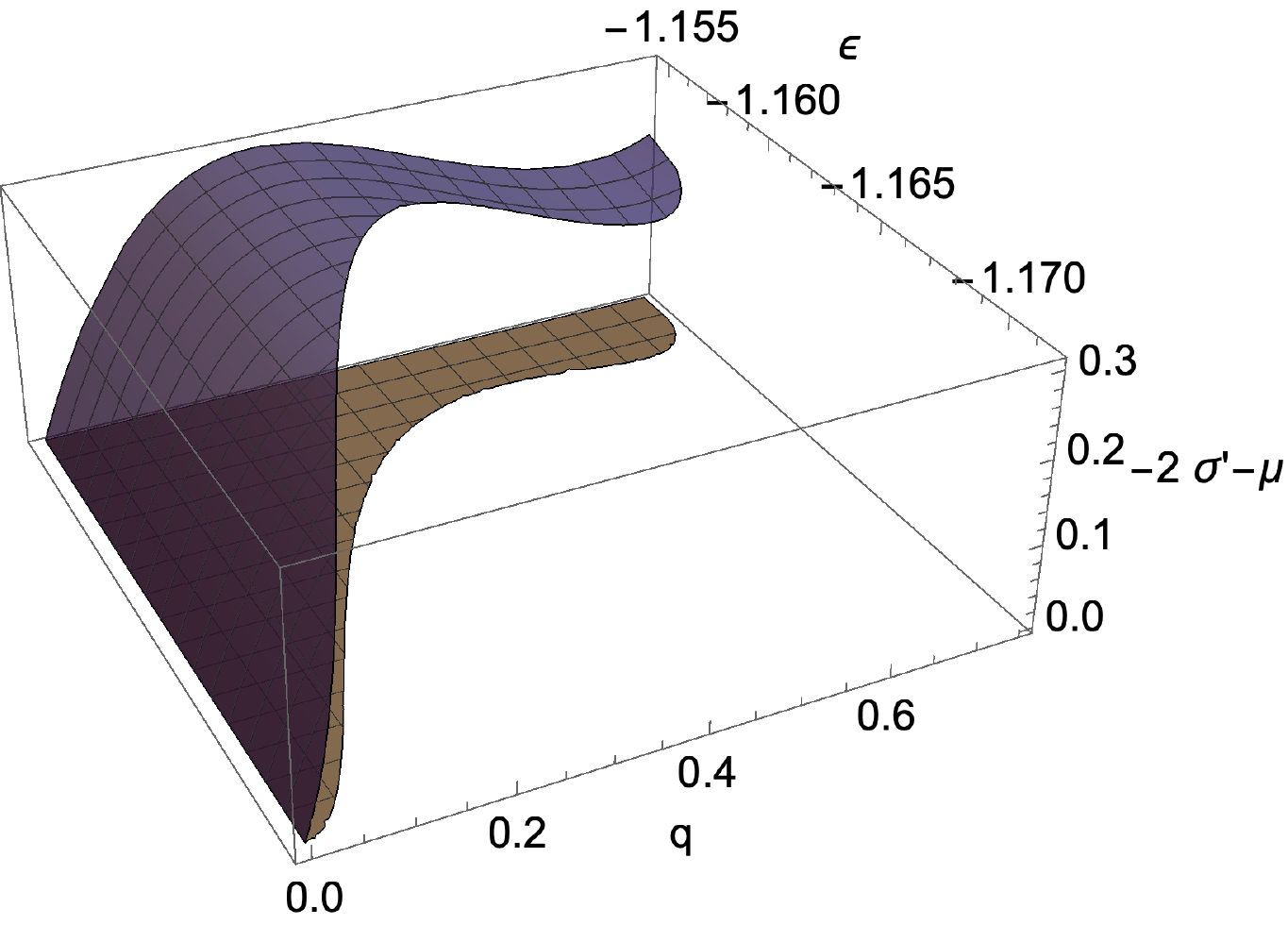} 
    \vspace{4ex}
  \end{minipage} 
   \caption{\small Total constrained entropy $\Sigma(\epsilon, q|\epsilon_0)$ (black) and entropy of the marginal saddles $\Sigma_{\rm ms}(\epsilon, q|\epsilon_0)$ (purple) for fixed $\epsilon_0=-1.167$ and different values of $q$. The continuous part of the purple lines corresponds to saddles that are geometrically connected to the reference minimum (meaning that $u_{\rm typ}>0$), while the dashed part corresponds to disconnected saddles. \emph{Top left. } For this value of $q$ none of the lines in Fig. \ref{fig:energies} is crossed: typically the Hessian has no isolated eigenvalue, and the index-1 saddles are not connected to the minimum as $\epsilon^+(q|\epsilon_0)< \overline{\epsilon}(q|\epsilon_0)$. \emph{Top right. } For this value of $q$ the line $\epsilon^+(q|\epsilon_0)$ in Fig. \ref{fig:energies} is crossed: the index-1 saddles at smaller energy have $u_{\rm typ}>0$ while those at higher energy have $u_{\rm typ}=0$.
   \emph{Bottom left. } For this value of $q$ the curve  $\epsilon_{\rm ms}(q|\epsilon_0)$ is crossed: above a given energy (point in the figure) the typical stationary points are index-1 saddles with a negative isolated eigenvalue, which vanishes at the point where $\Sigma(\epsilon, q|\epsilon_0)=\Sigma_{\rm ms}(\epsilon, q|\epsilon_0)$. \emph{Bottom right. }   Plot of the function $-2 \sigma'(p,q)-\mu(q, \epsilon, \epsilon_0)$ (blue surface) for $\epsilon_0=-1.167$, $p=3$ and $\epsilon \geq \overline{\epsilon}(q|\epsilon_0)$. The function is always larger than zero (gray surface), indicating that for these parameters Regime B holds.
   }\label{fig:EntropiesA}
\end{figure}

When $ q \leq q_{\rm ms}(\epsilon| \epsilon_0)$ and local minima are the dominant stationary points, 
a population of saddles with finite complexity exists, with a whole range of values of $\lambda<0$ and complexity  \eqref{eq:bah}. These saddles are  have at least one negative mode of the Hessian, but not extensively-many of them, {\it i.e.}, $k=o(N)$ \footnote{Indeed, the bulk of the density of states is not altered by the conditioning, and it is therefore equal to a semicircle law entirely supported on the positive semi-axis for all $\epsilon< \epsilon_{\rm th}$.}. The complexity of the dominant ones is obtained minimizing the large deviation function in \eqref{eq:bah} over $\lambda \leq 0$. It can be checked that in the relevant regime of parameters the following inequality is satisfied:
\begin{equation}\label{eq:Inequality}
-2 \sigma'(p,q)- \mu(q, \epsilon, \epsilon_0) \geq 0
\end{equation}
and therefore that Regime B holds (see Fig. \ref{fig:EntropiesA}). The large deviation function to optimize is therefore \eqref{eq:Ultimo2p}, which is a decreasing function of $\lambda$, minimal at the boundary value $\lambda=0$. It follows that for $ q \leq q_{\rm ms}(\epsilon| \epsilon_0)$ the dominant saddles are \emph{marginally stable}, with a single Hessian mode that is exactly equal to zero. 
We denote the complexity of these saddles with:
\begin{equation}
\Sigma_{\rm ms}(\epsilon, q| \epsilon_0)\equiv \Sigma(\epsilon, q| \epsilon_0) -
\begin{cases}
\mathcal{G}_{q, \epsilon| \epsilon_0}(\sqrt{2} \, p \, \epsilon) &\text{ if  } \lambda^{-}_p(\epsilon, q| \epsilon_0) <0<\lambda^{+}_p(\epsilon, q| \epsilon_0)\\
\mathcal{G}_0( \sqrt{2} \, p \, \epsilon) &\text{ if  }  0<\lambda^{-}_p(\epsilon, q| \epsilon_0)  \; \; \text {or} \; \; \lambda^{+}_p(\epsilon, q| \epsilon_0) <0\\
     \end{cases}
\end{equation}
where the subscript  stands for ``marginal saddles''. These saddles are geometrically connected to the reference minimum only whenever $\lambda^{-}_p(\epsilon, q| \epsilon_0) <0<\lambda^{+}_p(\epsilon, q| \epsilon_0)$. We find that, for the values of parameters we are interested in, $\lambda^{-}_p(\epsilon, q| \epsilon_0) <0$ always, and the relevant condition is $0<\lambda^{+}_p(\epsilon, q| \epsilon_0)$: for $\epsilon<\epsilon^+(q| \epsilon_0)$ defined above, it holds $\lambda^+_p(\epsilon, q| \epsilon_0)>0$ and thus the corresponding saddles satisfy $u>0$.

\begin{table}[t!]
\centering
\begin{tabular}{ |p{6.2cm}|p{6.2cm}|  }
\hline
\multicolumn{2}{|l|}{\quad \quad \quad \quad \quad \quad \quad \quad \quad  Special overlaps and energies at fixed $\epsilon_0$} \\
 \hline
  \noalign{\global\arrayrulewidth=.05mm}
  \arrayrulecolor{black}\hline
 \footnotesize{${\bf q_M(\epsilon|\epsilon_0)}$: overlap of stationary points at energy $\epsilon$ that are closer to the reference minimum }  &  \footnotesize{${ \overline{{\bm \epsilon}}(q|\epsilon_0)}$: energy of deepest stationary points at overlap $q$ with the reference minimum }  \\ 
\footnotesize{${\bf q_{\rm ms}(\epsilon|\epsilon_0)}$: transition between typical ($q>q_{\rm ms}$) and atypical ($q<q_{\rm ms}$) saddles}   & \footnotesize{${ {\bm \epsilon}_{\rm ms}(q|\epsilon_0)}$: transition between typical ($\epsilon>\epsilon_{\rm ms}$) and atypical ($\epsilon<\epsilon_{\rm ms}$) saddles}  \\ 
\footnotesize{${\bf q^+(\epsilon|\epsilon_0)}$: transition between connected  ($q>q^+$)  and disconnected ($q<q^+$) saddles}   & \footnotesize{${ {\bm \epsilon}^+(q|\epsilon_0)}$: transition between connected  ($\epsilon<\epsilon^+$) and disconnected ($\epsilon>\epsilon^+$) saddles}    \\
   \hline
  \hline
 \footnotesize{${\bf q^*}(\epsilon_0)$: overlap of the deepest saddle(s) connected to the reference minimum}   &  \footnotesize{${\bf {\bm \epsilon}^*}(\epsilon_0)$: energy of the deepest saddle(s) connected to the reference minimum}   \\ 
\footnotesize{${\bf q^{\rm mx}_{\rm ms}}(\epsilon_0)$: overlap of the farthest saddle(s) connected to the reference minimum}   & \footnotesize{${\bf{ \bm \epsilon}^{^*}_{1}}(\epsilon_0)$: energy of the farthest saddle(s) connected to the reference minimum} \\ 
 \hline
\end{tabular}
\caption{\small Summary of the special values of the overlaps/energy densities defined in the text. Each function depends on the energy density $\epsilon_0$ of the reference minimum. }\label{table:energies}
\end{table}

As a consequence, we find that the saddles dominating the energy landscape are always index index-1 saddles, with complexity:
\begin{equation}\label{eq:FinalCompIndex1}
\Sigma_{1}(\epsilon, q| \epsilon_0) =
\begin{cases}
0  &\text{ if  } q_M(\epsilon| \epsilon_0)<q\\
\Sigma(\epsilon, q| \epsilon_0) &\text{ if  } q_{\rm ms}(\epsilon| \epsilon_0)\leq q \leq q_M(\epsilon| \epsilon_0) \\
\Sigma_{\rm ms}(\epsilon, q| \epsilon_0) &\text{ if  }  q<  q_{\rm ms}(\epsilon| \epsilon_0).
     \end{cases}
\end{equation}
The population of dominating saddles displays three regimes, separated by two transitions: (i) at high-overlap with the reference minimum, the saddles have a single Hessian mode that is strictly negative, and are geometrically connected with the minimum; (ii) at intermediate overlaps, the saddles are marginal, and still geometrically connected to the minimum; (iii) at low overlaps, the dominant saddles are marginal, but uncorrelated to the reference minimum. 
 Plots of the total constrained complexity $\Sigma(\epsilon,q|\epsilon_0)$ and of the complexity $\Sigma_{\rm ms}(\epsilon,q|\epsilon_0)$ of marginally stable saddles are given in Fig. \ref{fig:EntropiesA}, in the different regimes.

\subsubsection{Iso-complexity curves and deepest saddles at fixed overlap. }
\begin{figure}[h!] 
    \centering
      \includegraphics[width=.7\linewidth]{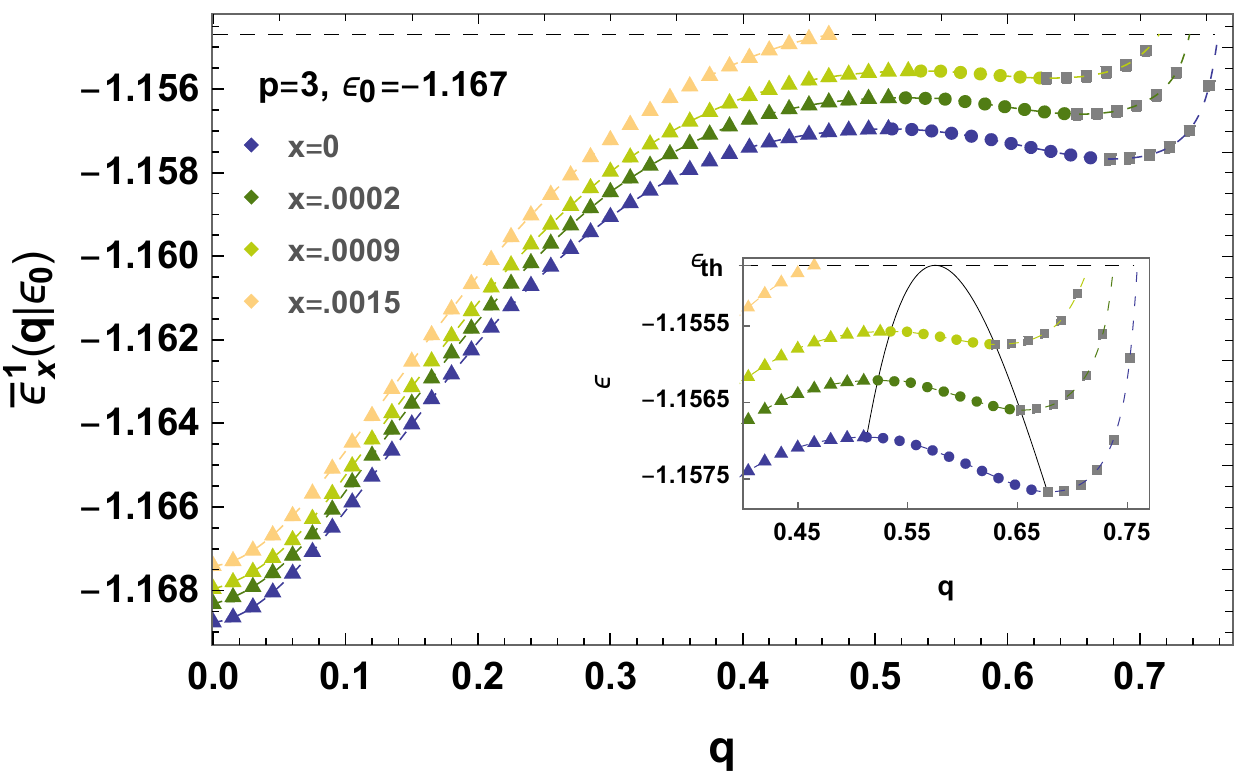} 
   \caption{\small Iso-complexity curves of index-1 saddles, see \eqref{eq:IC}. The different symbols correspond to saddles with one negative Hessian mode (squares), marginal saddles geometrically connected to the reference minimum (circles) and marginal saddles that are disconnected (triangles). The gray part of the curves correspond to the typical saddles already determined in Ref. \cite{RBCBarriers}. \emph{Inset.} Zoom of the iso-complexity curves. The black lines are the curves $\epsilon^+(q| \epsilon_0)$ and $\epsilon_{\rm ms}(q| \epsilon_0)$.}\label{fig:Isocomp}
\end{figure}

A convenient way to represent the saddles complexity is through iso-complexity curves $\overline{\epsilon}^{1}_x(q|\epsilon_0)$, see Fig.~\ref{fig:Isocomp}, which give the energies of the index-1 saddles having a fixed value of the complexity:
\begin{equation}\label{eq:IC}
\Sigma_1(\overline{\epsilon}^{1}_x, q| \epsilon_0)=x.
\end{equation}
The smallest of these curves $\overline{\epsilon}^{1}_{x=0}(q|\epsilon_0)$ corresponds to zero complexity and gives the energy of the deepest index-1 saddles found at overlap $q$ with the reference minimum. A comparison between this energy and the energy of the deepest stationary points $\overline{\epsilon}(q|\epsilon_0)$ at the same overlap is given in Fig.~\ref{fig:energies2}. The two curves coincide for overlap larger than $q^*(\epsilon_0) \equiv q_{\rm ms}(\overline{\epsilon}^{1}_{x=0}| \epsilon_0)$, which is also the local minimum of the two curves, as shown explicitly\footnote{Actually, it is shown in \cite{RBCBarriers}  that for arbitrary value of $x$, the iso-complexity curve have local minima at  overlaps $q_x$ which coincide with the overlaps $q_{\rm ms}$ at which the typical value of the isolated eigenvalue vanishes: the transition between the marginal saddles and the saddles with a negative eigenvalue occurs exactly at the minimum of these iso-complexity curves.} in \cite{RBCBarriers}. Following the notation of that work, we denote the corresponding energy with $\epsilon^*(\epsilon_0)$. It follows from Fig.~\ref{fig:Isocomp} that this is the energy of the deepest saddles that are geometrically connected to the reference minimum, and therefore it corresponds to the optimal (\emph{i.e.}, lowest) energy barrier.

For $q<q^*(\epsilon_0)$, the energy of the deepest marginal saddles $\overline{\epsilon}^{1}_{0}(q|\epsilon_0)$ is higher than the one of the deepest minima (as it follows naturally from the fact that their complexity is smaller). This curve has a local maximum at an overlap $q \equiv q^{\rm mx}_{\rm ms}(\epsilon_0)$, corresponding to an energy density $ \epsilon^*_1(\epsilon_0)$. We find that this overlap coincides with the point at which $\overline{\epsilon}^{1}_{0}(q|\epsilon_0)$ intersects $\epsilon^+(q| \epsilon_0)$,
\begin{equation}
 \epsilon^*_1(\epsilon_0)=\overline{\epsilon}^{1}_{0}( q^{\rm mx}_{\rm ms}|\epsilon_0) =\epsilon^+( q^{\rm mx}_{\rm ms}| \epsilon_0),
\end{equation}
meaning that exactly at these overlap the deepest saddles become geometrically disconnected from the reference minimum. 
Notice that also the curve $\overline{\epsilon}(q|\epsilon_0)$ is maximal at the point where $\epsilon^*(q| \epsilon_0)=\epsilon^+(q| \epsilon_0)$: this overlap is smaller than $q^{\rm mx}_{\rm ms}(\epsilon_0)$, and corresponds to saddles that are not geometrically connected to the minimum. 

\subsubsection{Distribution of escape states and dynamical barrier. }\label{sec:DiscussionDynamics}
From the analysis above it follows that local minima below the threshold are surrounded by an exponential multiplicity of index-1 saddles that are geometrically connected to the minima. The energy density of these saddles is distributed over an interval $\epsilon \in \quadre{\epsilon^*(\epsilon_0), \epsilon_{\rm th}}$ whose width depends on the energy $\epsilon_0$ of the local minimum. These connected saddles are distributed in a region of configuration space that corresponds to overlaps $q \in \quadre{q^{\rm mx}_{\rm ms}(\epsilon_0), q_M(\epsilon_0)}$: outside this interval, saddles are either absent, or the dominant ones are uncorrelated to the reference minimum, in the sense that the corresponding downhill direction in configuration space does not point towards the minimum.  


Each of the connected index-1 saddles represents a potential escape state for the system that is dynamically trapped in the reference minimum. However, it is not guaranteed that once the system escapes through a saddle, it is able to decorrelate from the initial minimum, \emph{i.e.}, to reach regions of configuration space that are orthogonal to it. It is indeed likely that the escape from a local minimum is a complicated dynamical process involving a sequence of jumps between minima that are sufficiently close to each others in configuration space, until decorrelation is achieved. The true ``dynamical barrier" would then correspond to the maximal energy barrier crossed in this composite process. 

\begin{figure}[h!] 
  \begin{minipage}[b]{0.5\linewidth}
    \centering
      \includegraphics[width=.95\linewidth]{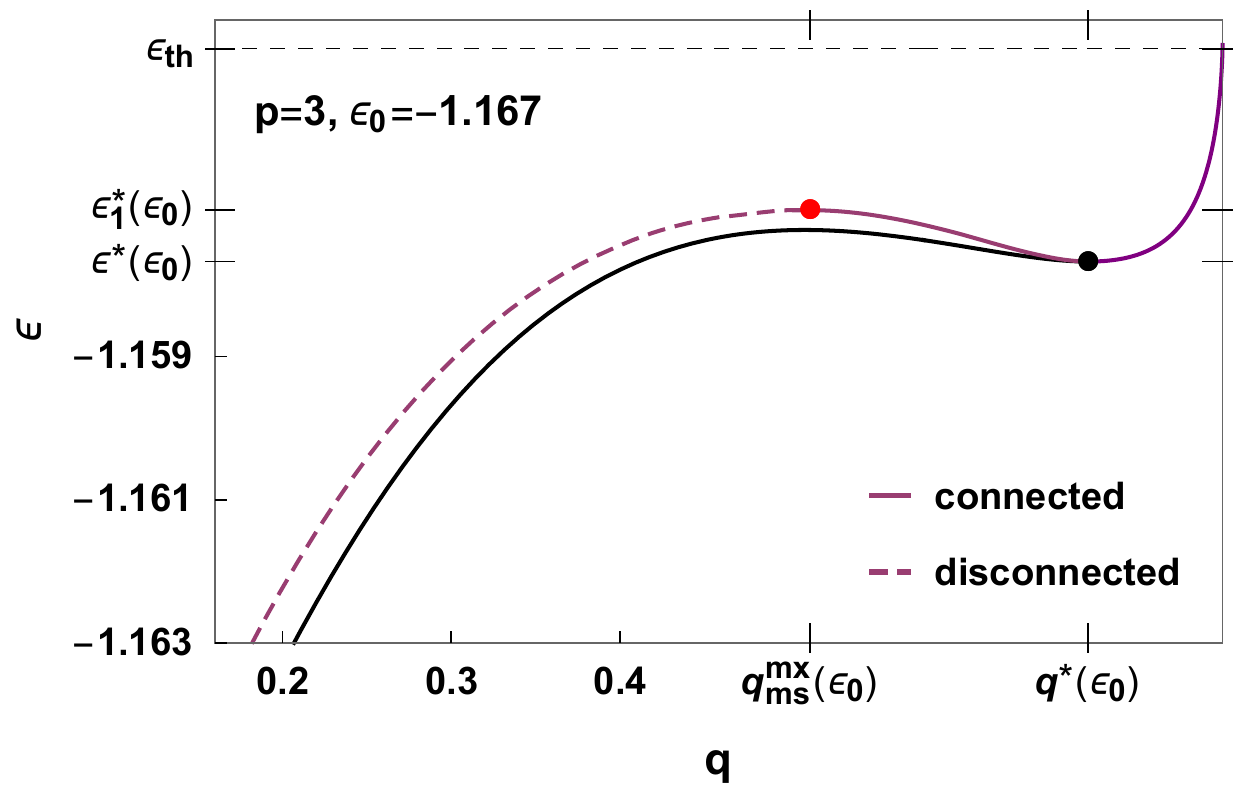} 
    \vspace{4ex}
  \end{minipage}
  \begin{minipage}[b]{0.5\linewidth}
    \centering
      \includegraphics[width=.95\linewidth]{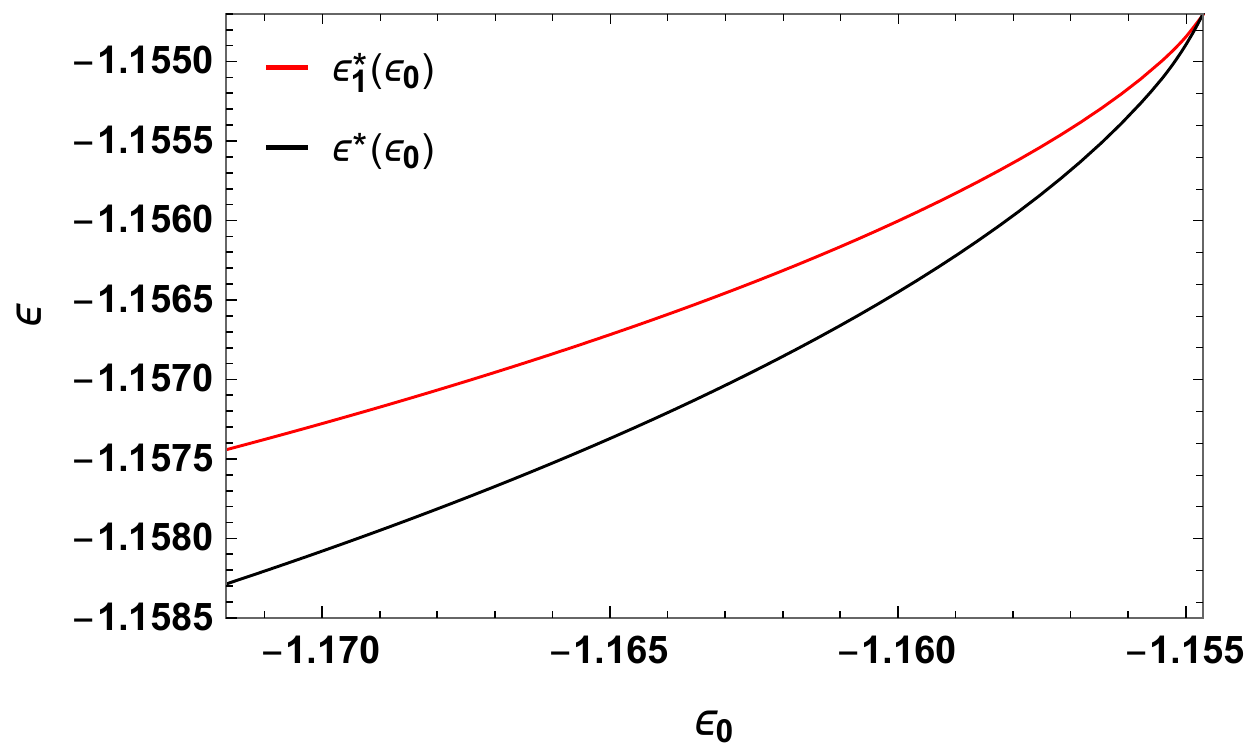} 
    \vspace{4ex}
  \end{minipage} 
   \caption{\small \emph{Left. } Comparison between the energy of the deepest minima (black) and of the deepest saddles (purple) either correlated (solid) or uncorrelated (dashed) with the minimum. The local maximum corresponds to $\epsilon^*_1(\epsilon_0)$ (red dot), the local minimum to $\epsilon^*(\epsilon_0)$ (black dot). \emph{Right.} Dependence of the energies $\epsilon^*_1(\epsilon_0)$ and $\epsilon^*(\epsilon_0)$ on the energy density of the reference minimum $\epsilon_0$. }\label{fig:energies2}
\end{figure}

A lower bound to the dynamical barrier can be obtained from the zero-temperature Franz-Parisi potential \cite{FP, FP2}, as the energy corresponding to the local maximum of the potential curve. As shown in \cite{RBCBarriers}, the local maximum of the Franz-Parisi potential coincides exactly with the local maximum of the curve $\overline{\epsilon}(q|\epsilon_0)$ (and it is thus contributed by local minima). The minimal-energy saddles at $q^*(\epsilon_0)$ correspond to a smaller energy barrier, indicating that the system escaping from those saddles does not fully decorrelate from the initial local minimum. Indeed, this is consistent with the study of the dynamics \cite{RBCProgress}. On the other hand, some of the marginal saddles at smaller overlap $q$ identified in this work satisfy the bound, see Fig. \ref{fig:energies2}. In particular, the local maximum of the curve $\overline{\epsilon}^{1}_{0}(q|\epsilon_0)$, where the transition occurs between saddles that are geometrical connected to the minimum and saddles that are not, corresponds to an energy barrier $\epsilon^*_1(\epsilon_0)- \epsilon_0$ satisfying the bound. The dependence of $\epsilon^*_1(\epsilon_0)$ on the depth $\epsilon_0$ of the reference minimum is shown in Fig. \ref{fig:energies2}. These saddles represent potential candidates for the dynamical barriers: checking whether this is the case through the study of the dynamics is an interesting open problem.

\section{Part II: general statements of the large deviation functions}\label{sec:PartTwo}
In this second part of the paper, we give the general expressions of the large deviation functions of the smallest eigenvalue and eigenvector of GOE matrices deformed with both and additive and multiplicative perturbation along a fixed direction in configuration space. In particular, in Section \ref{sec:GenStatMatr} we 
recall the general expression for the \emph{typical} value of the isolated eigenvalue of the Hessian, and define the various large deviation functions to be determined. In Sections \ref{sec:LDFfixedXU}, \ref{sec:OptimizedU} and \ref{sec:FlucThetaLDP} we report the general expressions of these large deviation functions, and discuss their interpretation in terms of a \emph{BBP-like} transition of the second smallest eigenvalue of the perturbed matrices.  In Section \ref{sec:SummaryConto} we give a summary of the main steps of the calculation, which is presented in detail in the third part of the paper.

\subsection{Perturbed GOE matrix:typical values and large deviations}\label{sec:GenStatMatr}
We let $\mathcal{X}$ be a $M$-dimensional GOE matrix with entries $x_{ij}$ with respect to some basis ${\bf e}_{i}$, and variances
$
 \langle x_{ij}^2 \rangle= (\sigma^2/M)[1+ \delta_{ij}].
$
This corresponds to the distribution:
\begin{equation}
 P(\mathcal{X})= \frac{1}{Z_M(\sigma)} e^{-\frac{M}{4 \sigma^2} \text{Tr} \mathcal{X}^2},
\end{equation}
where $Z_M(\sigma)$ is the normalization. For $\beta \geq 0$ we define the $M \times M$ matrix:
\begin{equation}
 F_{\beta} = \mathbb{1}- \frac{\beta}{1+\beta} \, {\bf e}_{M} {\bf e}^T_{M}
\end{equation}
where $\mathbb{1}$ is the identity matrix, ${\bf e}_{M}$ is a unit vector 
and we set
\begin{equation}\label{eq:BasicMf}
 \mathcal{Y}= F_\beta \mathcal{X} F_\beta + \theta \, {\bf e}_{M} {\bf e}^T_{M},
\end{equation}
where $\theta$ is the strength of the additive perturbation, which we take to be a fluctuating variable with distribution:
\begin{equation}\label{eq:MuGaussian}
f_{\overline{\theta}, \sigma_\theta}(\theta)= \frac{1}{\sqrt{2 \pi \sigma^2_\theta}}e^{- \frac{M}{2 \sigma^2_\theta} (\theta- \overline{\theta})^2}.
\end{equation}
We denote with $\mu_M \leq \mu_{M-1} \leq \cdots \leq \mu_1$ the eigenvalues of $\mathcal{Y}$. 
Notice that
the statistics of the rescaled, conditioned Hessian described in Sec. \ref{sec:statHess} (up to the shift by the $\epsilon$-dependent diagonal matrix) is the one of a matrix of the form \eqref{eq:BasicMf} with parameters $\sigma^2 \to p(p-1)$, $\beta \to \sigma/\Delta(q)-1$, $\overline{\theta} \to \mu(q, \epsilon, \epsilon_0)$ and $\sigma_\theta \to \zeta(q)$.

In the following, we restrict to the case $\overline{\theta}<0$, which is of interest for the $p$-spin landscape problem. We denote with $\rho_M^{\rm typ}(\mu)$ the \emph{typical} eigenvalue density of the matrix $\mathcal{Y}$. For certain values of the parameters $\overline{\theta}, \beta$, the latter exhibits a sub-leading correction with respect to the GOE semicircle:
\begin{equation}\label{eq:UsualGOE}
\rho_{\sigma}(\mu)=\frac{\sqrt{4 \sigma^2-\mu^2}}{2 \pi \sigma^2},
\end{equation}
that corresponds to the smallest eigenvalue being isolated from the bulk of the density of states. This happens whenever:
\begin{equation}\label{eq:BBP}
\overline{\theta} \leq \theta_{\rm c} \equiv -\sigma \tonde{1+ \frac{\sigma'^2}{\sigma^2}}=- \sigma \tonde{\frac{1+ 2 \beta^2+ 4 \beta }{[1+ \beta]^2}} 
\end{equation}
or equivalently
\begin{equation}\label{eq:BBP2}
 \sigma^2 G_{\sigma'}(\overline{\theta})\geq - \sigma,
\end{equation}
where
\begin{equation}
\sigma'= \sigma \sqrt{\frac{\beta(\beta+2)}{(1+\beta)^2}} < \sigma
\end{equation}
and where $G_{\sigma}$ is the GOE resolvent:
\begin{equation}\label{eq:Resolvent2}
 G_\sigma(z) \stackrel{z \text{ real}}{=}\frac{1}{2 \sigma^2} \tonde{z- \text{sign}(z) \sqrt{z^2- 4 \sigma^2}} \in \quadre{-\frac{1}{\sigma}, \frac{1}{\sigma}}.
 \end{equation}
In this case one has that the typical value of the smallest eigenvalue $\mu_M$ reads:
\begin{equation}\label{eq:Lambda0}
\mu_M^{\rm typ} = \mu_0(\overline{\theta}, \beta)  \equiv G^{-1}_\sigma(G_{\sigma'}(\overline{\theta}))=\frac{1}{G_{\sigma'}(\overline{\theta})}+ \sigma^2 G_{\sigma'} (\overline{\theta}) \leq -2 \sigma,
\end{equation}
and thus the typical density of eigenvalues is
\begin{equation}\label{eq:DOS}
\rho_M^{\rm typ}(\mu)=\frac{\sqrt{4 \sigma^2-\mu^2}}{2 \pi \sigma^2}+ \frac{1}{M}\delta \tonde{\mu-\mu_0(\overline{\theta}, \beta)}+ o\tonde{\frac{1}{M}}.
\end{equation}
When \eqref{eq:BBP} is not satisfied, the sub-leading contribution to  \eqref{eq:DOS} is absent and $ \mu_M^{\rm typ}=-2 \sigma$. In absence of the multiplicative perturbation (when $\beta=0$), we have $\sigma' \to 0$; using that 
\begin{equation}
\lim_{\sigma' \to 0} G_{\sigma'}(x)= \frac{1}{x}
\end{equation}
we recover the well known results for the minimal eigenvalue of a GOE matrix subject to an additive rank-1 perturbation \cite{EdwardsJones, Peche, Peche2},
\begin{equation}
\lim_{\beta \to 0}\mu_0(\overline{\theta}, \beta)=\overline{\theta}+\frac{\sigma^2}{\overline{\theta}}.
\end{equation}

Notice that the typical density of states \eqref{eq:DOS} does not depend on the fluctuations of $\theta$ but only on its average value. 
The fluctuations enter into play when looking at large deviations of $\mu_M$. We denote with ${\bf v}_{M}$ the corresponding eigenvector, and define $u_{\rm M}= |{\bf v}_{M} \cdot {\bf e}_M|^2$. We use the notation $\mathcal{P}_{ \overline{\theta},\sigma_\theta,\beta}(x)$ for the distribution of the smallest eigenvalue $\mu_M$, which is given by:
\begin{equation}
\mathcal{P}_{\overline{\theta}, \sigma_\theta, \beta}(x)= \int_{0}^1 du \int_{-\infty}^{\infty} \, d\theta\, f_{\overline{\theta}, \sigma_\theta}(\theta)\, \tilde{\mathcal{P}}_{\theta,\beta}(x, u),
\end{equation}
where $ \tilde{\mathcal{P}}_{\theta,\beta}(x, u)$ is the joint probability density of $\mu_{ M}$ and $u_{ M}$, \emph{conditioned} to a fixed value of the additive perturbation $\theta$. In the following we compute the large deviation function:
\begin{equation}\label{eq:LargeDUno}
\lim_{M \to \infty} \frac{\log \tilde{\mathcal{P}}_{\theta,\beta}(x, u)}{M} =- \mathcal{L}_{\theta, \beta}(x,u).
\end{equation}
For each $x$, we determine the typical value $u_{\rm typ}(x)$ maximizing the large deviation function,
\begin{equation}\label{eq:DefUTyp}
u_{\rm typ}(x) \equiv \underset{u \in \quadre{0,1}}{\mathrm{argmin}} \, \mathcal{L}_{\theta, \beta}(x,u) 
\end{equation}
and set
\begin{equation}\label{eq:LargeDDue}
\overline{\mathcal{L}}_{\theta,\beta}(x) \equiv  \mathcal{L}_{\theta, \beta}(x,u_{\rm typ}(x)).
\end{equation}
The large deviation function for fluctuating $\theta$ is then obtained as:
\begin{equation}\label{eq:FinalLDF}
\lim_{M \to \infty} \frac{\mathcal{P}_{\overline{\theta}, \sigma_\theta, \beta}(x)}{M}\equiv -\mathcal{F}_{\overline{\theta}, \sigma_\theta, \beta}(x)= -\min_{\theta} \quadre{\frac{(\theta- \overline{\theta})^2}{2 \sigma_\theta^2} + \overline{\mathcal{L}}_{\theta,\beta}(x)}.
\end{equation}
This large deviation function exhibits an explicit dependence on the variance $\sigma^2_\theta$; nevertheless, as we shall see, its minimum is always attained at the typical value $\mu_M^{\rm typ}$ of the smallest eigenvalue, that does not depend on $\sigma^2_\theta$ and it is given by  $\mu_0(\overline{\theta}, \beta)$ in \eqref{eq:Lambda0} when \eqref{eq:BBP} is satisfied, and by $-2 \sigma$ otherwise.

\subsection{Large deviation function at fixed $u$ and $\theta$}\label{sec:LDFfixedXU}
We begin by stating the form of the large deviation function \eqref{eq:LargeDUno}. 
We define the constants:
\begin{equation}\label{eq:Constants}
 \begin{split}
  & {C}_2=-2 \theta \, (1+ \beta)^4, \quad \quad 
  {C}_3= 2 \beta(2+ \beta), \quad \quad
  C_4(x,u)=C_2 + \frac{C_3^2}{2} x u,
 \end{split}
\end{equation}
and introduce:
\begin{equation}\label{eq:kappa}
\kappa_{\theta, \beta}(x,u)= \frac{\sigma^2 C_3 [2+ C_3(1-u)]^3}{C_4^2(x,u)(1-u)}=\frac{4 \sigma ^2 \beta  (2+\beta )  [1+\beta  (\beta +2) (1-u)]^3}{(1-u) [\beta ^2 (\beta +2)^2 u
   x-(\beta +1)^4 \theta ]^2}.
\end{equation}
We identify the following two regimes of parameters:
\begin{equation}\label{eq:Cases}
\begin{split}
&\text{Case A:} \quad \quad \kappa_{\theta, \beta}(x,u)>1\\
   &\text{Case B:} \quad \quad \kappa_{\theta, \beta}(x,u) \leq 1,
\end{split}
\end{equation}
and define the rate functions:
{
\medmuskip=0mu
\thinmuskip=0mu
\thickmuskip=0mu
\begin{equation}\label{eq:Case11}
\begin{split}
&\mathcal{L}^{(a)}_{\theta, \beta}(x,u)= \frac{1}{4 \sigma^2} \tonde{x^2+ C_2 x u + \frac{C_3^2}{4} x^2 u^2+ C_3 x^2 u}-\mathcal{I}(x) + \frac{1}{2}- \frac{1}{2} \log \tonde{\frac{2 \sigma^2 (1-u)}{C_3(1-u)+2 }}\\
&\hspace{1.8 cm}-\frac{C_4^2(x,u) (1-u)^2}{4 \sigma^2 \quadre{2 + C_3(1-u)}^2},\\
&\mathcal{L}^{(b)}_{\theta, \beta}(x)=\frac{1}{4 \sigma^2} \tonde{2 x^2+ C_3 x^2 + C_2 x + \frac{C_3^2}{4} x^2}- \frac{3}{2} \mathcal{I}(x)+ \frac{1}{2}+ \frac{1}{2}\log \quadre{\frac{C_3^2 x+ 2 C_3 x+ 2 C_2}{4 \sigma^2 }}
\end{split}
\end{equation}
}
where:
{
\medmuskip=0mu
\thinmuskip=0mu
\thickmuskip=0mu
\begin{equation}\label{eq:GOEI}
\begin{split}
\mathcal{I}(z)=& \int d\lambda \, \rho_\sigma(\lambda) \log |\lambda-z|=\\
=&
\begin{cases}
\log \tonde{ -\frac{z}{2}+\frac{1}{2}\sqrt{z^2- 4 \sigma^2}}- \frac{1}{2}+ \frac{z^2}{4 \sigma^2}+\frac{z}{4 \sigma^2}\sqrt{z^2-4 \sigma^2} &\text{   if  } z<-2 \sigma\\
\frac{z^2}{4 \sigma^2}- \frac{1}{2}+ \log \sigma &\text{   if  } -2 \sigma<z<0\\
\end{cases},  
\end{split}
\end{equation}
}
and 
\begin{equation}\label{eq:Constf}
l(\theta, \beta)= 1-\frac{1}{2}\log \tonde{\frac{2 \sigma^4}{C_3+2}} -\frac{C_2^2}{4 \sigma^2 (C_3+2)^2}=1-\log \tonde{\frac{\sigma^2}{1+ \beta}}-\frac{\theta^2}{2 \sigma^2 [1+\beta]^2}. 
\end{equation}
When Case B holds, we further define the following functions:
{
\medmuskip=0mu
\thinmuskip=0mu
\thickmuskip=0mu
\begin{equation}\label{eq:Varie}
\begin{split}
F(x,u)&=-\frac{C_4(x,u)(1-u)+ \sqrt{C_4^2(x,u)(1-u)^2- \sigma^2 C_3(1-u) \quadre{2+ C_3(1-u)}^3}}{ \sigma^2 \quadre{2 + C_3(1-u)}^2}\\
\mu_1(x,u)&=-\frac{2 \tonde{C_4(1-u)\quadre{1+C_3(1-u)}- \sqrt{(1-u) \quadre{C_4^2(1-u)- \sigma^2 C_3 \quadre{2+ C_3(1-u)}^3}}}}{C_3(1-u) \quadre{2 + C_3(1-u)}^2}.
\end{split}
\end{equation}
}
Notice that the functions \eqref{eq:Varie} are complex in Case A, when $ \kappa_{\theta, \beta}(x,u)>1$.
In terms of these quantities, the large deviation function \eqref{eq:LargeDUno} is given by the following expressions:
\begin{itemize}
\item  When Case A holds:
\begin{equation}\label{eq:LDP0}
\mathcal{L}_{\theta, \beta}(x,u) = \mathcal{L}^{(a)}_{\theta, \beta}(x,u)- l(\theta, \beta).
\end{equation}
\item When Case B holds:
\begin{equation}\label{eq:LDP1}
\mathcal{L}_{\theta, \beta}(x,u)  = \begin{cases}
 \mathcal{L}^{(a)}_{\theta, \beta}(x,u)  -l(\theta, \beta)&\text{ if  }  \sigma^2 F(x,u) \geq - \sigma\\
   \mathcal{L}^{(b)}_{\theta, \beta}(x)-l(\theta, \beta) &\text{ if  }  \sigma^2 F(x,u) < - \sigma \quad \text{ and} \quad x \geq \mu_1(x,u)\\
   \mathcal{L}^{(a)}_{\theta, \beta}(x,u)-l(\theta, \beta) &\text{ if  }   \sigma^2 F(x,u) < - \sigma \quad \text{ and} \quad x < \mu_1(x,u).
\end{cases}
\end{equation}
This expression is continuous at the point $x=\mu_1(x,u)$.
\end{itemize}
The constant shift equals to $l(\theta, \beta)= \mathcal{L}^{(a)}_{\theta, \beta}(x_{ \rm typ},u_{\rm typ})$, where $x_{\rm typ},u_{\rm typ}$ are the typical values for the given parameters; it is added to ensure that $\mathcal{L}_{\theta, \beta}(x_{\rm typ},u_{\rm typ})=0$.  In Appendix \eqref{app:LimitingCases}, we discuss how limiting cases known in the literature are recovered.

\subsubsection{Interpretation in terms of the second-smallest eigenvalue}\label{sec:InterpretationSmaller}
When $\theta$ is kept fixed, the typical value of the smallest eigenvalue $\mu_M$ undergoes a transition at $\theta=\theta_{\rm c}$ given in~\eqref{eq:BBP}: for $\theta \leq \theta_{\rm c}$, it equals to $\mu_0(\theta, \beta)$, which can be equivalently written as:
\begin{equation}\label{eq:Lambda02}
 \mu_0(\theta, \beta)  = G^{-1}_\sigma(G_{\sigma'}(\overline{\theta}))= m^+_\sigma[C_2, C_3]
 \end{equation}
in terms of the constants \eqref{eq:Constants}, where
\begin{equation}\label{eq:Emme}
m^\pm_\sigma [a,b]= \frac{2}{b(b+2)^2} \tonde{-a (b+1)\pm  \sqrt{a^2-\sigma^2 b(b+2)^3}}.
\end{equation}
The different regimes of the large deviation function \eqref{eq:LDP1} in Case B can be interpreted in terms of an analogous transition of the typical value $\mu_{M-1}^{\rm typ}$ of the \emph{second-smallest} eigenvalue of the matrix $\mathcal{Y}$. As it will appear from the explicit calculation in Sec.~\ref{sec:Stat}, fixing $\mu_M=x$ and $u_M=u$ leads to a modification of the joint distribution of the remaining eigenvalues $\grafe{\mu_i}_{i=1}^{M-1}$. In particular, the resulting joint distribution has the same form of the joint distribution of all the eigenvalues $\grafe{\mu_\alpha}_{\alpha=1}^{M}$ of the matrix \eqref{eq:BasicMf}, but with modified parameters $\tilde{\theta}, \tilde{\beta}$ defined by:
\begin{equation}\label{eq:NewPar}
\begin{split}
&\tilde{\theta}=\frac{\theta(1-u)(1+\beta)^4- x u (1-u) \beta^2 (2+\beta)^2}{[1+ \beta (1-u)(2+ \beta)]^2}, \quad \quad (1+ \tilde{\beta})^2=1+ \beta (1-u)(2+ \beta).
\end{split}
\end{equation}
This is equivalent to mapping $C_3 \to C_3(1-u)$ and $C_2 \to C_4(x,u)(1-u)$. 
One can easily check by substitution that the function $F(x,u)$ in \eqref{eq:Varie} can be written in terms of these parameters as:
\begin{equation}\label{eq:EffeDue}
F(x,u)=\frac{1}{\sigma^2 G_{\tilde{\sigma}}(\tilde{\theta})},
\end{equation}
where 
\begin{equation}
\tilde{\sigma}^2= \sigma^2 \quadre{\frac{\tilde{\beta} (\tilde{\beta}+2)}{(1+ \tilde{\beta})^{2}}}\leq \sigma^2.
\end{equation}
A comparison with \eqref{eq:BBP2} shows that the two regimes of \eqref{eq:LDP1} correspond to the regimes in which the typical value of the second- smallest eigenvalue sticks to the boundary of the semicircle $\rho_\sigma(\lambda)$ (when  $\sigma^2 F(x,u) \geq - \sigma$) or is smaller that $-2 \sigma$ (when  $\sigma^2 F(x,u) < - \sigma$). In the latter case, the typical value of the second-smallest eigenvalue takes precisely the form:
\begin{equation}\label{eq:TypSec}
\begin{split}
\mu_{M-1}^{\rm typ}(x,u)&= \mu_1(x,u)= G_\sigma^{-1}\tonde{G_{\tilde{\sigma}}(\tilde{\theta})}= G_\sigma^{-1} \tonde{\frac{1}{\sigma^2 \, F(x,u)}}\\
&= m^+_\sigma[C_4(x,u)(1-u), C_3(1-u)].
\end{split}
\end{equation}
Notice that the argument of $G^{-1}_\sigma$ is larger than $-1/\sigma$, as it should. Therefore, $\mathcal{L}_{\theta, \beta}(x,u)$ is proportional to $\mathcal{L}^{(a)}_{\theta, \beta}(x,u)$ whenever the parameters $x,u$ are chosen  in such a way that $x \leq \mu_{M-1}^{\rm typ}(x,u)$, and it is proportional to  $\mathcal{L}^{(b)}_{\theta, \beta}(x)$ otherwise. As it will follow from Section \ref{sec:OptimizationU}, this last regime is relevant only whenever $u$ is taken to be different from its typical value $u_{\rm typ}(x)$ defined in \eqref{eq:DefUTyp}: when the overlap is allowed to adjust itself to its typical value, one naturally finds that $x \leq \mu_{M-1}^{\rm typ}(x,u_{\rm typ}(x))$.

Case A can be analogously interpreted in terms of the effective parameters \eqref{eq:NewPar}. Indeed, we find that \eqref{eq:kappa} can be re-written as:
\begin{equation}
\kappa_{\theta, \beta}(x,u)= \frac{4 \tilde{\sigma}^2}{\tilde{\theta}^2},
\end{equation}
and therefore Case A corresponds to the regime in which $-2 \tilde{\sigma} < \tilde{\theta}<0$. In this regime, the functions $m^+_\sigma[C_4(x,u)(1-u), C_3(1-u)]$ are complex (and exactly equal at the boundary value $\tilde{\theta}=-2 \tilde{\sigma}$).

When interpreted in terms of the second-smallest eigenvalue, the large deviation function $\mathcal{L}_{\theta, \beta}(x,u)$ displays the same three regimes that will appear in Sec.~\ref{sec:OptimizedU}, with the substitutions $\theta \to \tilde{\theta}$, $\beta \to \tilde{\beta}$ and $\sigma' \to \tilde{\sigma}$.

\subsection{Large deviation function optimized over $u$ at fixed $\theta$}\label{sec:OptimizedU}
\begin{figure}[ht]
 \centering
    \includegraphics[width=.6\linewidth]{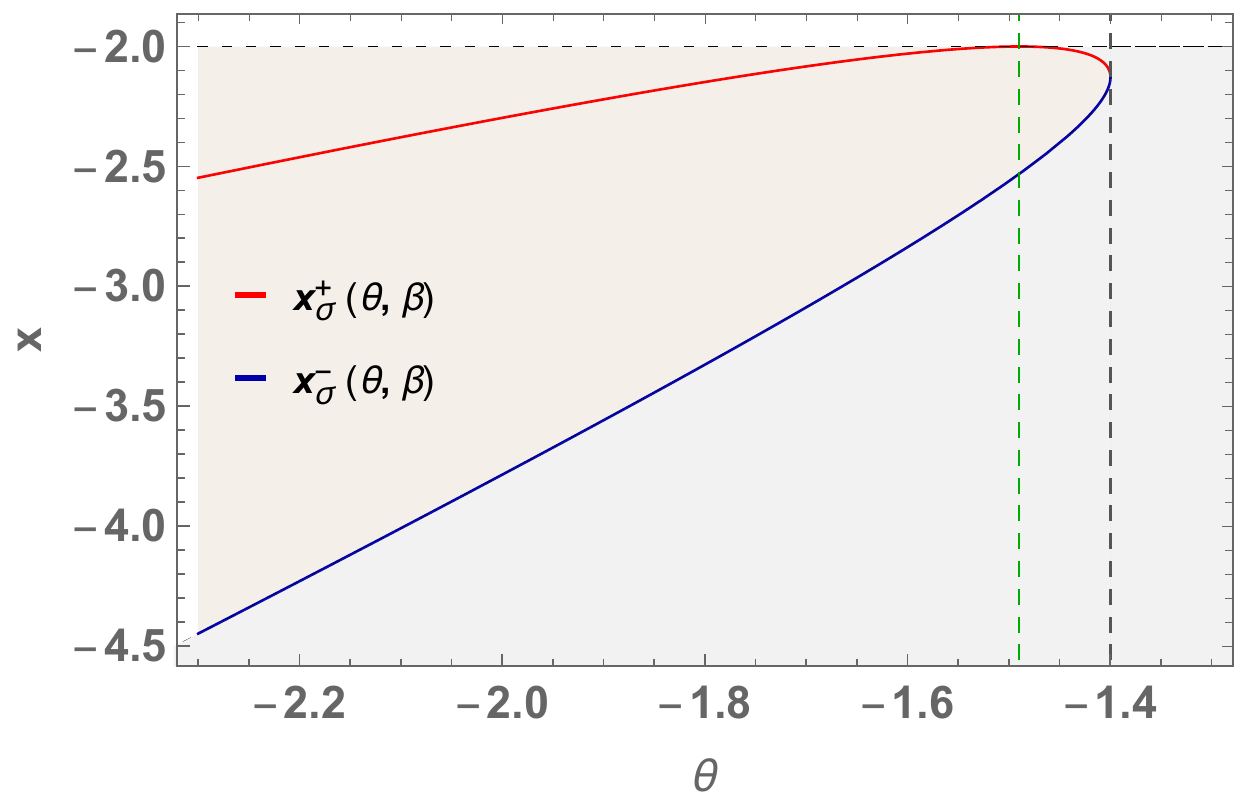} 
\caption{\small Plot of the functions $x^{\pm}_\sigma(\theta, \beta)$ for $\beta=.4$ and $\sigma=1$. The different colors correspond to the regions of parameters where the large deviation function $\overline{\mathcal{L}}_{\theta, \beta}(x)$ equals either to $\mathcal{G}_{\theta,\beta}(x)$ (red) or to $\mathcal{G}_{0}(x)$ (blue). The dashed vertical lines denote $\theta=\theta_{\rm c}$ (green) and $\theta= -2 \sigma'$ (black). The plot shows three regimes: (i) Regimes B.2, for $\theta<\theta_{\rm c}$: the function $x^+_\sigma(\theta, \beta)$ equals to the typical value for the smallest eigenvalue of the matrix, i.e., $x^+_\sigma(\theta, \beta)= \mu_{0}(\theta, \beta)$. At  $\theta=\theta_{\rm c}$ it becomes equal to $-2 \sigma$, signaling that the smallest eigenvalue is reabsorbed into the bulk; (ii) Regime B.1:  the smallest eigenvalue is typically not out of the bulk, $x^+_\sigma(\theta, \beta)$ gives the analytic continuation of the isolated eigenvalue into the second Riemann sheet in the complex plane. This ends at $\theta= -2 \sigma'$, when $x^+_\sigma(\theta, \beta)=x^-_\sigma(\theta, \beta)$; (iii) Regime A:  for $\theta> -2 \sigma'$, both functions are complex.  }\label{fig:MuPMRegimes}
  \end{figure}

We now state the form of the large deviation function \eqref{eq:LargeDDue}, obtained by optimizing \eqref{eq:LargeDUno} over the overlap $u$, at fixed $x$.
The behavior of the resulting function $\overline{\mathcal{L}}_{\theta,\beta}(x)$ depends on whether the parameters $\theta, \beta$ are such that $\mu_M^{\rm typ}$ is typically out of the bulk of the semicircle of not, and whether the parameter $x$ is taken to be larger or smaller than the following two thresholds:
\begin{equation}\label{eq:ExPIuMinus}
x^\pm_\sigma(\theta, \beta)=\frac{(1+\beta) [1+2 \beta  (\beta +2)] \theta \pm \sqrt{(1+\beta)^2 \theta ^2-4 \beta  (\beta
   +2) \sigma ^2}}{2 \beta  (\beta +1) (\beta +2)}=m^\pm_\sigma[C_2, C_3].
\end{equation}
For fixed $\beta$ and as a function of $\theta$, these thresholds have three regimes, see caption in Fig: \ref{fig:MuPMRegimes}, that correspond to three different regimes for the large deviation function $\overline{\mathcal{L}}_{\theta,\beta}(x)$:
\begin{itemize}
\item  Regime A: $-2 \sigma' \leq  \theta$: in this regime the functions $x^{\pm}_\sigma(\theta, \beta)$ in \eqref{eq:ExPIuMinus} are complex;
\item Regime B.1: $\theta_{\rm c} \leq \theta \leq -2 \sigma'$  or equivalently $\sigma^2 G_{\sigma'}(\theta)< - \sigma$ and $\theta \leq -2 \sigma'$: in this regime $\mu_M^{\rm typ}=-2 \sigma$;
\item Regime B.2: $\theta\leq \theta_{\rm c}=-\sigma [1+ (\sigma')^2/\sigma^2]$  or equivalently  $\sigma^2 G_{\sigma'}(\theta)\geq - \sigma$. In this regime the typical value $\mu_M^{\rm typ}$ of the smallest eigenvalue is out of  bulk and equals to $\mu_0(\theta, \beta)$, see \eqref{eq:BBP} and \eqref{eq:Lambda02}.

\end{itemize}

\begin{figure}[ht]
    \includegraphics[width=.48\linewidth]{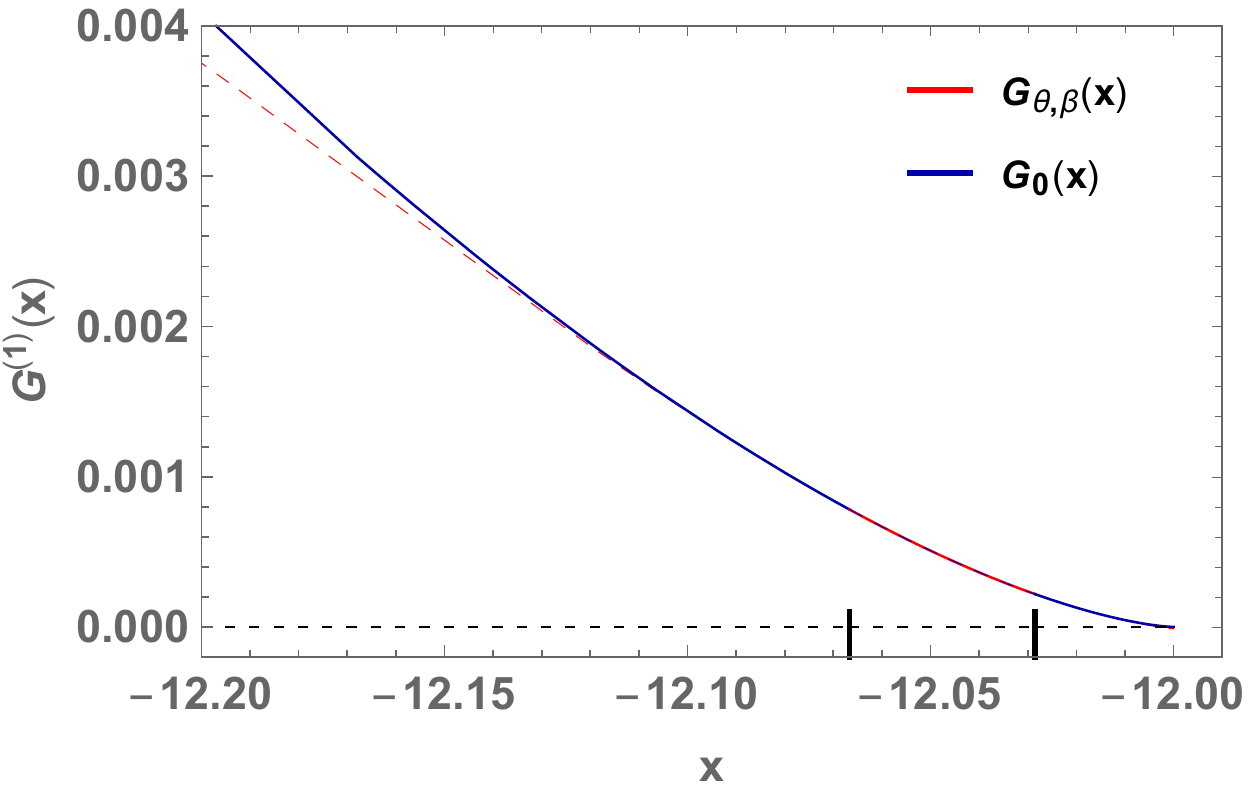} 
        \includegraphics[width=.46\linewidth]{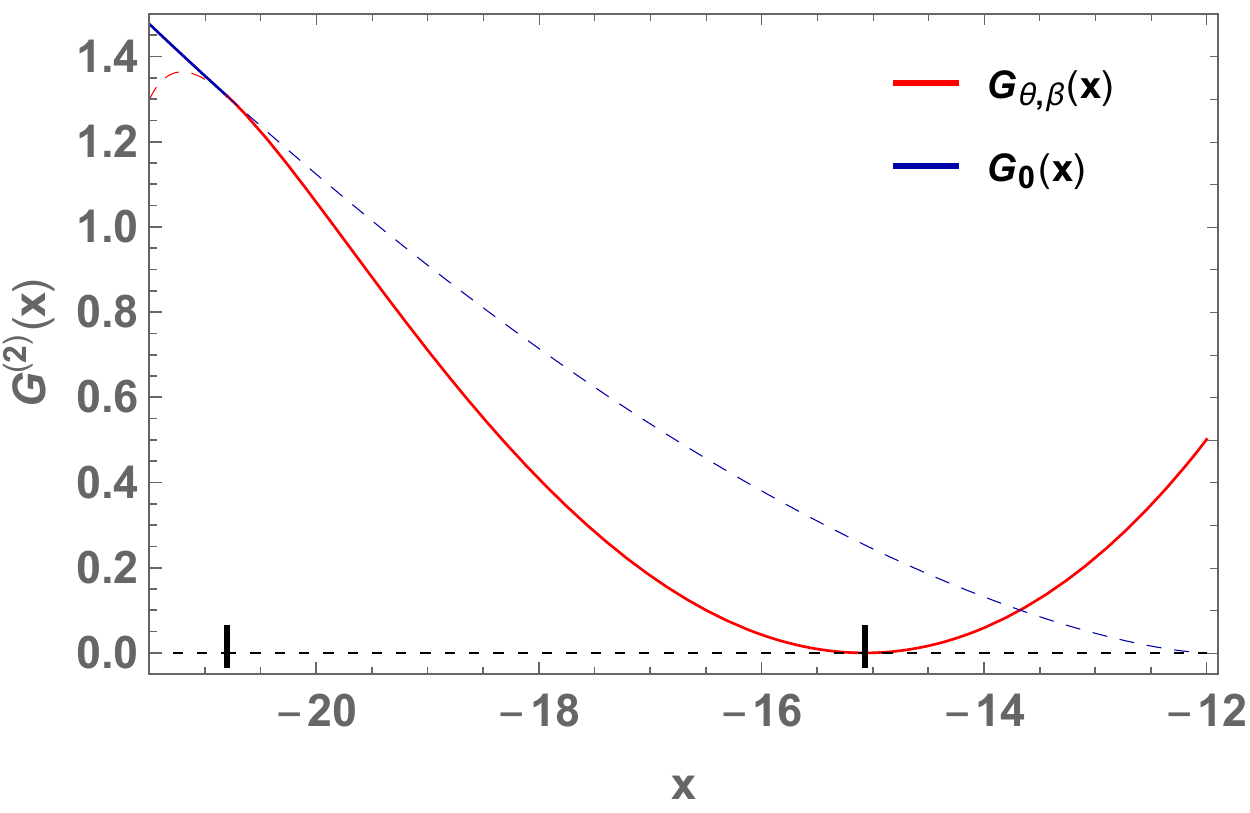} 
\caption{\small Plots of the large deviations function $\overline{\mathcal{L}}_{\theta, \beta}(x)$ for the smallest eigenvalue (solid lines), for values of parameters for which its typical value is at the boundary of the semicircle (\emph{left}), or it is out of the bulk (\emph{right}). The black ticks mark the values $x^{\pm}_\sigma(\theta, \beta)$. When the large deviation function coincides with $\mathcal{G}_{\theta, \beta}(x)$ (red curves), the eigenvector corresponding to the minimal eigenvalue has a typical projection along the special direction that is $u_{\rm typ}(x)>0$. }\label{fig:LDF}
  \end{figure}

Given the functions:
{
\medmuskip=0mu
\thinmuskip=0mu
\thickmuskip=0mu
\begin{equation}\label{eq:FundamentalRates}
\begin{split}
\mathcal{G}_{0}(x)&=\int_{x}^{-2 \sigma} \frac{\sqrt{z^2-4 \sigma^2}}{2 \sigma^2} dz=\frac{x^2}{4 \sigma^2}- \mathcal{I}(x)- \frac{1}{2}+ \log \sigma,\\
\mathcal{G}_{\theta,\beta}(x)&=\frac{[1+\beta(\beta+2)]^2 }{4 \sigma^2} x^2- \frac{\mathcal{I}(x)}{2}+\frac{(1+\beta)^4 \theta ^2-2 \theta x}{4 \sigma^2}+\frac{1}{2} \log [\beta  (\beta +2) x-(1+\beta)^2 \theta ],
\end{split}
\end{equation}
}
 it holds: 
\begin{equation}\label{eq:OverlineL}
\overline{\mathcal{L}}_{\theta, \beta}(x)=
\begin{cases}
 \mathcal{G}_{0}(x)  & \text{ in  Regime A}\\
\mathcal{G}^{(1)}_{\theta, \beta}(x) &\text{ in  Regime B.1}\\
\mathcal{G}^{(2)}_{\theta, \beta}(x) &\text{ in  Regime B.2},
\end{cases}
\end{equation}
where in the Regime B.1 one has:
\begin{equation}
 \mathcal{G}^{(1)}_{\theta, \beta}(x) = 
\begin{cases}
  \mathcal{G}_{0}(x)&\text{ if  }   x<x^-_\sigma(\theta, \beta) \text {  or  }x^+_\sigma(\theta, \beta)<x<-2 \sigma \\
     \mathcal{G}_{\theta, \beta}(x) &\text{ if  } x^-_\sigma(\theta, \beta)<x<x^+_\sigma(\theta, \beta),
     \end{cases}
\end{equation}
while in Regime B.2 one has:
\begin{equation}
 \mathcal{G}_{\theta, \beta}^{(2)}(x) = 
\begin{cases}
 \mathcal{G}_{\theta, \beta}(x) &\text{ if  }   x^-_\sigma(\theta, \beta)<x<- 2 \sigma\\
 \mathcal{G}_{0}(x)  &\text{ if  }  x<x^-_\sigma(\theta, \beta).
\end{cases}
\end{equation}
The function $ \mathcal{G}^{(1)}_{\theta, \beta}(x)$ has a minimum at $x_{\rm typ}=-2 \sigma$, while $ \mathcal{G}_{\theta, \beta}^{(2)}(x)$ vanishes at:
\begin{equation}\label{eq:xTyp}
x_{\rm typ}=G^{-1}_\sigma(G_{\sigma'}(\theta))=\mu_{0}(\theta, \beta),
\end{equation}
that is indeed the typical value of $\mu_M$ in this regime of parameters. 
Both large deviation functions are continuous at $x^{\pm}_\sigma(\theta, \beta)$. 
We notice that the explicit expressions of $\mu_{0}(\theta, \beta)$ and of $x^+_\sigma(\theta, \beta)$ coincide, even though $\mu_{0}(\theta, \beta)$ is well defined only in the regime $\theta \leq \theta_{\rm c}$ (otherwise the resolvent in \eqref{eq:xTyp} would not be invertible), while $x^+_\sigma(\theta, \beta)$ is defined in the opposite regime of parameters. The coincidence of the two expressions  follows from a symmetry of the GOE resolvent evaluated on the real axis, as we discuss more precisely in Sec. \ref{sec:SPinD} . Plots of the large deviation function $\overline{\mathcal{L}}_{\theta, \beta}(x)$ are given in Fig. \ref{fig:LDF}.

For what concerns the typical overlaps, $\mathcal{G}_{\theta, \beta}(x)$ corresponds to a non-trivial typical overlap $u_{\rm typ}(x)>0$ with the special direction, while $\mathcal{G}_{0}(x)$ corresponds to zero overlap, \emph{i.e.},  in Regime B.1 we have
\begin{equation}
u_{\text{typ}}(x)=
\begin{cases}
0 &\text{   if  } x^+_\sigma(\theta, \beta)<x<-2 \sigma \\
u^+_{\theta, \beta}(x) &\text{   if  } x^-_\sigma(\theta, \beta)<x<x^+_\sigma(\theta, \beta)\\
0 &\text{   if  } x<x^-_\sigma(\theta, \beta),
\end{cases}
\end{equation}
while in Regime  B.2 it holds:
\begin{equation}
u_{\text{typ}}(x)=
\begin{cases}
u^+_{\theta, \beta}(x) &\text{   if  } x^-_\sigma(\theta, \beta)<x<-2 \sigma\\
0 &\text{   if  } x<x^-_\sigma(\theta, \beta).
\end{cases},
\end{equation} 
The expression for $u^+_{\theta, \beta}(x)$ is given explicitly in \eqref{eq:Usol}. Notice that when the eigenvector associated to the smallest eigenvalue is uncorrelated with the special direction ($u_{\rm typ}(x)=0$), the large deviation function $\mathcal{G}_{0}(x)$ coincides with the one in absence of perturbations \cite{LargeDevGOE}. In the limit of a purely additive perturbation $\beta \to 0$, the Regime A disappears as $\sigma' \to 0$. Moreover, one finds $x^+_\sigma \to \theta + \sigma^2/\theta$ and $x^-_\sigma \to - \infty$. The typical value of the overlap, when positive, reduces to $u_{\rm typ}(x) \to1- [x \theta - \sqrt{\theta^2 (x^2 - 4 \sigma^2)}]/(2 \theta^2)$. The known results are recovered \cite{BiroliGuinnet}.

\subsection{Large deviations for fluctuating  $\theta$}\label{sec:FlucThetaLDP}
We finally state the expression for the large deviation function $\mathcal{F}_{\overline{\theta}, \sigma_\theta, \beta}(x)$ in \eqref{eq:FinalLDF}, obtained optimizing over the Gaussian fluctuations of $\theta$. In Regime A the optimization is trivial. In Regime B we shall show that all the inequalities in the previous section survive with the substitution $\theta \to \overline{\theta}$, meaning that we can identify once more the three regimes:
\begin{itemize}
\item  Regime A: $-2 \sigma' \leq  \overline{\theta}$ (the functions $x^{\pm}_\sigma(\overline{\theta}, \beta)$ are complex);
\item Regime B.1: $\sigma^2 G_{\sigma'}(\overline{\theta})< - \sigma$, meaning  $-\theta_{\rm c}\leq \overline{\theta} \leq -2 \sigma'$. In this regime $\mu_M^{\rm typ}(\overline{\theta}, \beta)=-2 \sigma$;
\item Regime B.2: $\sigma^2 G_{\sigma'}(\overline{\theta})\geq - \sigma$, meaning  $\overline{\theta}<\theta_{\rm c}$. In this regime   $\mu_M^{\rm typ}(\overline{\theta}, \beta)=\mu_0(\theta, \beta)$, see \eqref{eq:BBP}, \eqref{eq:Lambda02}.
\end{itemize}
The large deviation function for fluctuating $\theta$ reads:
\begin{equation}\label{eq:CurveF}
\mathcal{F}_{\overline{\theta}, \sigma_\theta, \beta}(x)=
\begin{cases}
 \mathcal{G}_{0}(x)  & \text{ in  Regime A}\\
\mathcal{F}^{(1)}_{\overline{\theta}, \sigma_\theta, \beta}(x)  &\text{ in  Regime B.1}\\
\mathcal{F}^{(2)}_{\theta, \beta}(x) &\text{ in  Regime B.2},
\end{cases}
\end{equation}
where
\begin{equation}\label{eq:Ultimo2}
\mathcal{F}^{(1)}_{\overline{\theta}, \sigma_\theta, \beta}(x)  = 
\begin{cases}
\mathcal{G}_0(x) &\text{ if  }  x<x^-_\sigma(\overline{\theta}, \beta) \text {  or  }x^+_\sigma(\overline{\theta}, \beta)<x<-2 \sigma \\
\mathcal{G}_{\theta^*,\beta}(x) &\text{ if  } x^-_\sigma(\overline{\theta}, \beta)<x<x^+_\sigma(\overline{\theta}, \beta).
     \end{cases}
\end{equation}
and
\begin{equation}\label{eq:Ultimo1}
\mathcal{F}^{(2)}_{\overline{\theta}, \sigma_\theta, \beta}(x) = 
\begin{cases}
\mathcal{G}_{\theta^*,\beta}(x)  &\text{ if  }   x^-_\sigma(\overline{\theta}, \beta)<x<- 2 \sigma\\
\mathcal{G}_0(x) &\text{ if  }  x<x^-_\sigma(\overline{\theta}, \beta).
\end{cases}
\end{equation}
Therefore the large deviation function has the same form as in the previous section with $\theta \to \overline{\theta}$, except for $\mathcal{G}_{\theta, \beta}$ which has to be computed at the shifted point $\theta \to \theta^*=\theta^*_0(x)$ whose explicit expression is given in \eqref{eq:NewSP}. Notice that, as it should, 
\begin{equation}
\theta^*_0(x) \stackrel{\sigma_\theta \to 0}{\longrightarrow} \overline{\theta}.
\end{equation}

\subsection{Derivation of the large deviations: the idea of the calculation}\label{sec:SummaryConto}
In this section we summarize the skeleton of the derivation of the large deviation functions, whose details are presented in the following. The starting point is the derivation of the joint density of the eigenvalues $\mu_\alpha$ of the matrix $\mathcal{Y}$ given in \eqref{eq:BasicMf}, and of the corresponding eigenvector components along ${\bm e}_M$. We set  $\mu_M \leq \mu_{M-1} \leq \cdots \mu_1$ and let ${\bf v}_\alpha$ be the matrix eigenvectors, and $u_\alpha= |{\bf v}_\alpha \cdot {\bm e}_M|^2 \in \quadre{0,1}$. As we derive in the following, the joint probability density of $\mu_\alpha, u_\alpha$ reads:
{
\medmuskip=0mu
\thinmuskip=0mu
\thickmuskip=0mu
\begin{equation}\label{eq:FinJointS}
 P_{\theta, \beta}(\mu_\alpha, u_\alpha)= \frac{e^{-M V(\mu_\alpha, u_\alpha)}}{\mathcal{Z}_M[\theta, \beta]}  \prod_{\gamma<\alpha} \tonde{\mu_\gamma- \mu_\alpha} \prod_{\alpha=1}^M \theta \tonde{\mu_{\alpha}- \mu_{\alpha+1}} \;   \delta\tonde{\sum_{\alpha=1}^M u_\alpha-1} \prod_\alpha \frac{1}{u_\alpha^{1/2}},
 \end{equation}
}
with $\mathcal{Z}_M[\theta, \beta]$ a normalization and
\begin{equation}\label{eq:PotentialS}
V(\mu_\alpha, u_\alpha) = \frac{1}{4 \sigma^2} \quadre{\sum_{\alpha} \mu^2_\alpha + \frac{C_3^2}{4}  \tonde{\sum_\alpha \mu_\alpha u_\alpha}^2 + C_2 \sum_\alpha \mu_\alpha u_\alpha+ C_3 \sum_\alpha \mu_\alpha^2 u_\alpha},
\end{equation}
with the constants given in \eqref{eq:Constants}. Therefore, the effect of the additive and multiplicative perturbations is to introduce a coupling between the $\mu_\alpha$ and $u_\alpha$ through the confinement potential $V(\mu_\alpha, u_\alpha)$.  The joint probability density of $\mu_M=x$ and $u_M=u$ has to be obtained integrating over all the other eigenvalues and eigenvector projections, as:
{
\medmuskip=0mu
\thinmuskip=0mu
\thickmuskip=0mu
\begin{equation}\label{eq:p1S}
\frac{ \mathcal{P}_{\theta, \beta}(x,u)}{p(u)}= \frac{e^{-\frac{M}{4 \sigma^2} \quadre{x^2+ C_2 x u + \frac{C_3^2}{4} \,x^2 u^2+ C_3 x^2 u}}}{\mathcal{Z}^*_M[\theta, \beta]} \int  \prod_{\alpha=1}^{M-1} d \mu_\alpha [(\mu_\alpha-x) \theta(\mu_\alpha-x) ]   F \tonde{\vec{\mu}}  I_{x,u}(\vec{\mu}).
\end{equation}
}
In this formula $p(u)$ is the distribution of a single eigenvector component for a GOE matrix, $F\tonde{\vec{\mu}} $ is the measure on the remaining $M-1$ eigenvalues:
\begin{equation}
 F \tonde{\vec{\mu}} = \prod_{\alpha >\gamma=1}^{M-1} (\mu_\gamma-\mu_\alpha) \theta(\mu_\gamma-\mu_\alpha) e^{-\frac{M}{4 \sigma^2} \quadre{\sum_{\alpha=1}^{M-1} \mu^2_\alpha }},
\end{equation}
 $I_{x,u}(\vec{\mu})$ is the integral over the remaining eigenvectors components, and the normalization is rescaled as $\mathcal{Z}^*_M[\theta, \beta]=\mathcal{Z}_M \Gamma(M/2)/ \pi^{M/2}$. From the explicit expression:  
\begin{equation}\label{eq:Comp2S}
\begin{split}
 I_{x,u}(\vec{\mu})=&\int_{-\infty}^{\infty} \prod_{\alpha=1}^{M-1} d e_\alpha \frac{\Gamma \tonde{\frac{M-1}{2} }}{\pi^{\frac{M-1}{2}}}  \delta \tonde{\sum_{\alpha=1}^{M-1} e^2_\alpha-1} \times \\
 &\times  e^{-\frac{M}{4 \sigma^2} \quadre{C_4 (x,u) (1-u) \sum_{\alpha=1}^{M-1} \mu_\alpha e_\alpha^2 + \frac{[C_3 (1-u)]^2}{4} \tonde{\sum_{\alpha=1}^{M-1} \mu_\alpha e_\alpha^2}^2+ C_3 (1-u)\sum_{\alpha=1}^{M-1} \mu_\alpha^2 e_\alpha^2}}
 \end{split}
\end{equation}
one sees that \eqref{eq:Comp2S}, up to normalization constants, has the same structure as \eqref{eq:FinJointS} with $C_3 \to C_3(1-u)$ and $C_2 \to C_4(x,u)(1-u)$. Therefore, at fixed $x$ and $u$ the distribution of the remaining eigenvalues is the one of a GOE matrix perturbed exactly as the original one, with modified parameters given in~\eqref{eq:NewPar}. We made use of this observation in the interpretation of the large deviation function.

Given \eqref{eq:p1S}, the core of the calculation is the computation of the integrals over the matrix eigenvalues and eigenvectors. This is done in  three steps: (i) introducing two auxiliary fields $y, \lambda$ the integration over the $e_\alpha$ becomes Gaussian and can be performed; (ii) the integration over the $\mu_\alpha$ is performed solving a variational problem for the eigenvalue density, both for its continuous part and for the isolated eigenvalue generated by the perturbations; (iii) the auxiliary parameters $y, \lambda$ are fixed with a saddle point calculation.

More precisely, the integration over the eigenvector components and over the continuous part of the eigenvalue density leads to the following expression for the joint probability: 
\begin{equation}\label{eq:p2S}
\begin{split}
 \mathcal{P}_{\theta, \beta}(x,u)\sim& \mathcal{A}_M\, e^{-M  \Psi_0(x,u)} \int_{\xi \geq x} d \xi e^{-M \quadre{\frac{\xi^2}{4 \sigma^2}- \mathcal{I}(\xi)}} \int_{\mathcal{D}(\xi)} dy d \lambda \, e^{M \phi(y, \lambda)},
 \end{split}
\end{equation}
where the remaining integrals are over the auxiliary parameters (with $\phi(y, \lambda)$ their action) and over the variable $\xi$, which represents the value of the second smaller eigenvalue $\mu_{M-1}$ of the matrix. The integration over this eigenvalue has to be done separately, since for certain values of parameters the effective perturbations \eqref{eq:NewPar} give rise to an outlier in the spectrum, that corresponds to its smaller eigenvalue $\mu_{M-1}$. The two integrals are coupled by the fact that $(\lambda, y)$ belong to a domain $\mathcal{D}(\xi)$ that depends explicitly on the value of $\xi$. All the integration can be performed with a saddle point approximation.
Depending on the values of $\xi$, the solutions $(\lambda^*, y^*)$ of the minimization problem for the action  $\phi(y, \lambda)$ are either within the domain, or outside the domain; in that case, the $\xi$-dependent boundary values have to be taken. Once the optimization over the auxiliary parameters is performed, performing the integral over $\xi$ with a saddle point approximation we are left with:
\begin{equation}\label{eq:LD2ndS}
\mathcal{P}_{\beta, \theta}(x,u) = \mathcal{A}_M e^{-M \quadre{\Psi_0(x,u)+ \inf_{-2 \sigma \geq \xi \geq x} \Psi_1(x,u, \xi)}},
\end{equation}
where $ \Psi_1(x,u, \xi)$ is (up to additive terms that are constant in $\xi$) the large deviation function for the smallest eigenvalue of a matrix  perturbed according to~\eqref{eq:NewPar}. The optimization over $\xi$ depends on wether $x$ is larger or smaller than the typical value $\mu^{\rm typ}_{M-1}$ of this eigenvalue: the different cases correspond to the different regimes of \eqref{eq:LDP1}. In particular, when $x,u$ are such that $\mu^{\rm typ}_{M-1}=-2 \sigma$ (meaning that $\sigma^2 F(x,u) \geq -\sigma $), the optimum of \eqref{eq:LD2ndS} is attained at $\xi^*=-2 \sigma$. When instead $\mu^{\rm typ}_{M-1} \equiv \mu_1(x,u)< -2 \sigma$  (meaning that $\sigma^2 F(x,u) < -\sigma $), the optimum is at $\xi^*=\mu_1(x,u)$ if $x< \mu_1(x,u)$, or at the boundary value $\xi^*=x$ otherwise. 

The other large deviation functions follow straightforwardly from an optimization over the overlap $u$ and over the additive perturbation $\theta$.

\section{Part III: Detailed derivation of large deviation functions}\label{sec:PartThree}
In this part of the paper, we present the derivation of the results summarized above. In particular, in Section \ref{sec:Stat} we show how the joint distribution of eigenvalues $\mu_\alpha$ and eigenvectors components $u_\alpha$ is modified by adding a combination of additive and multiplicative perturbations. In Section \ref{sec:EffPAr} we re-write the joint distribution $\mathcal{P}_{\theta, \beta}(x,u)$ as the integral of an action depending on the configuration of the second-smallest eigenvalue $\xi$, and over two additional auxiliary parameters $\lambda, y$. In Sections \ref{sec:SPinD} and \ref{sec:OD} we solve the saddle point equations for the auxiliary parameters $\lambda, y$, and in Section \ref{sec:VariationalSecond} we optimize over the value of the second-smallest eigenvalue. Finally, in Section \ref{sec:OptimizationU} we determine the optimal value of the overlap $u_{\rm typ}(x)$, and in Section \ref{sec:OptTheta} we optimize over the fluctuations of the additive perturbation $\theta$. Additional details on the calculation are given in the Appendices.

\subsection{The joint density of the smallest eigenvalue and eigenvector projection}\label{sec:Stat}
Let $\mu_\alpha$ be the eigenvalues of the matrix $\mathcal{Y}$ given in \eqref{eq:BasicMf}, with $\mu_M \leq \mu_{M-1} \leq \cdots \mu_1$. Let ${\bf v}_\alpha$ be the corresponding eigenvalues and $u_\alpha= |{\bf v}_\alpha \cdot {\bm e}_M|^2 \in \quadre{0,1}$. We consider $\theta$ to be fixed. We first argue that the joint probability density of $\mu_\alpha, u_\alpha$ reads:
{
\medmuskip=0mu
\thinmuskip=0mu
\thickmuskip=0mu
\begin{equation}\label{eq:FinJoint}
 P_{\theta, \beta}(\mu_\alpha, u_\alpha)= \frac{ e^{-M V(\mu_\alpha, u_\alpha)}}{\mathcal{Z}_M[\theta, \beta]} \prod_{\gamma<\alpha} \tonde{\mu_\gamma- \mu_\alpha} \prod_{\alpha=1}^M \theta \tonde{\mu_{\alpha}- \mu_{\alpha+1}} \;   \delta\tonde{\sum_{\alpha=1}^M u_\alpha-1} \prod_\alpha \frac{1}{u_\alpha^{1/2}},
 \end{equation}
}
with $\mathcal{Z}_M[\theta, \beta]$ a normalization and
\begin{equation}\label{eq:Potential}
V(\mu_\alpha, u_\alpha) = \frac{1}{4 \sigma^2} \quadre{\sum_{\alpha} \mu^2_\alpha + \frac{C_3^2}{4}  \tonde{\sum_\alpha \mu_\alpha u_\alpha}^2 + C_2 \sum_\alpha \mu_\alpha u_\alpha+ C_3 \sum_\alpha \mu_\alpha^2 u_\alpha},
\end{equation}
with the constants given in \eqref{eq:Constants}. As a matter of fact,
for the GOE matrix $\mathcal{X}$ the joint density of the ordered eigenvalues $\lambda_\alpha$ and of the eigenvectors squared components $z_\alpha= |{\bf e} \cdot {\bf w}_\alpha|^2$ along an arbitrary direction ${\bf e}$ is factorized, and reads:
\begin{equation}\label{eq:JointGOE}
\begin{split}
 p_{ \rm GOE}(\lambda_\alpha, z_\alpha)= &\frac{M! }{Z_M(\sigma)}e^{-M \sum_{\alpha=1}^M \frac{\lambda_\alpha^2}{4 \sigma^2}}  \prod_{\alpha<\gamma} |\lambda_\gamma- \lambda_\alpha| \prod_\alpha \theta(\lambda_{\alpha}- \lambda_{\alpha+1}) \times\\
 & \times \frac {\Gamma(M/2)}{(\Gamma(1/2))^M} \,  \delta\tonde{\sum_{\alpha=1}^M z_\alpha-1} \prod_\alpha \frac{1}{z_\alpha^{1/2}}
 \end{split}
\end{equation}
with $Z_M(\sigma)$ a normalization. 
The distribution \eqref{eq:FinJoint} is obtained through the change of variable:
\begin{equation}
 \mathcal{X}= F_\beta^{-1} \tonde{\mathcal{Y}-\theta {\bf e}_{M} {\bf e}^T_{M}} F_\beta^{-1}=  F_\beta^{-1} \mathcal{Y} F_\beta^{-1} -\theta(1+ \beta)^2 {\bf e}_{M} {\bf e}^T_{M},
\end{equation}
where 
\begin{equation}
 F_\beta^{-1}= \mathbb{1} + \beta\, {\bf e}_M {\bf e}^T_M.
\end{equation}
 The confinement potential is modified, since 
\begin{equation}
 \text{Tr} \mathcal{X}^2=  \text{Tr} \tonde{F_\beta^{-1}\mathcal{Y} F_\beta^{-1}}^2 + \theta^2(1+ \beta)^4-2 \theta (1+ \beta)^2 \, \text{Tr} \tonde{F_\beta^{-1}\mathcal{Y}F_\beta^{-1} {\bf e}_M {\bf e}^T_M}
\end{equation}
Using that:
\begin{equation}
\begin{split}
& \text{Tr}(F_\beta^{-1} \mathcal{Y} F_\beta^{-1} {\bf e}_M {\bf e}^T_M)= (1 + \beta)^2 \, {\bf e}_M \cdot \mathcal{Y} \cdot  {\bf e}_M \\
 &\text{Tr} \tonde{F_\beta^{-1} \mathcal{Y} F_\beta^{-1}}^2=  \text{Tr}{\mathcal{Y}}^2 + (4 \beta^2 + 4 \beta^3 + \beta^4) \tonde{ {\bf e}_M \cdot \mathcal{Y} \cdot  {\bf e}_M}^2 + (4 \beta + 2 \beta^2) {\bf e}_M \cdot \mathcal{Y}^2 \cdot {\bf e}_M,
 \end{split}
 \end{equation}
one finds
\begin{equation}
 \text{Tr} \mathcal{X}^2=\text{Tr}\mathcal{Y}^2 + \frac{C_3^2}{2} \tonde{ {\bf e}_M \cdot \mathcal{Y} \cdot  {\bf e}_M}^2  + C_2  {\bf e}_M \cdot \mathcal{Y} \cdot  {\bf e}_M+ C_3  {\bf e}_M \cdot \mathcal{Y}^2 \cdot {\bf e}_M+ \theta^2(1+ \beta)^4
\end{equation}
with the constants given in \eqref{eq:Constants}. The confinement potential for the eigenvalues of $\mathcal{Y}$ is therefore given by \eqref{eq:Potential} and depends explicitly on their eigenvector components $u_\alpha$ (the constant term $\theta^2(1+ \beta)^4$ is absorbed in the normalization). On the other hand, it can be easily argued that the joint measure of the eigenvector components and of the eigenvalues is left invariant by the change of variables (for an additive rank-1 perturbation, this was shown in \cite{Bogomolny} following \cite{Aleiner}). 
As a consequence, the only effect of the additive and multiplicative perturbations is to introduce a coupling between the $\mu_\alpha$ and $u_\alpha$ through the confinement term. 
From \eqref{eq:FinJoint} we can then get that the joint density of $\mu_M=x, u_M=u$ reads:
{
\medmuskip=0mu
\thinmuskip=0mu
\thickmuskip=0mu
\begin{equation}\label{eq:p1}
 \frac{\mathcal{P}_{\theta, \beta}(x,u)}{p(u)}= \frac{e^{-\frac{M}{4 \sigma^2} \quadre{x^2+ C_2 x u + \frac{C_3^2}{4} \,x^2 u^2+ C_3 x^2 u}}}{\mathcal{Z}^*_M[\theta, \beta]} \int  \prod_{\alpha=1}^{M-1} d \mu_\alpha [(\mu_\alpha-x) \theta(\mu_\alpha-x) ]  F \tonde{\vec{\mu}}I_{x,u}(\vec{\mu}).
\end{equation}
}
In this formula $p(u)$ the distribution of a single eigenvector component:
\begin{equation}
\begin{split}
 p(u)&=\frac{\Gamma(M/2)}{\sqrt{\pi} \Gamma \tonde{\frac{M-1}{2}}} \frac{(1-u)^{\frac{M-3}{2}}}{\sqrt{u}} \sim (1-u)^{\frac{M}{2}},
 \end{split}
\end{equation}
$\mathcal{Z}^*_M[\theta, \beta]=\mathcal{Z}_M \Gamma(M/2)/ \pi^{M/2}$ is a rescaled normalization,
 $F\tonde{\vec{\mu}} $ is the measure on the remaining $M-1$ eigenvalues:
\begin{equation}
 F \tonde{\vec{\mu}} = \prod_{\alpha >\gamma=1}^{M-1} (\mu_\gamma-\mu_\alpha) \theta(\mu_\gamma-\mu_\alpha) e^{-\frac{M}{4 \sigma^2} \quadre{\sum_{\alpha=1}^{M-1} \mu^2_\alpha }},
\end{equation}
while $ I_{x,u}(\vec{\mu})$ is an integral over the remaining $u_\alpha$:
{
\medmuskip=0mu
\thinmuskip=0mu
\thickmuskip=0mu
\begin{equation}\label{eq:IntI}
I_{x,u}(\vec{\mu})= \int_0^\infty  \prod_{\alpha=1}^{M-1} du_\alpha \;  p(\vec{u}|u) e^{-\frac{M}{4 \sigma^2} \quadre{C_4 (x,u) \sum_{\alpha=1}^{M-1} \mu_\alpha u_\alpha + \frac{C_3^2}{4}\tonde{\sum_{\alpha=1}^{M-1} \mu_\alpha u_\alpha}^2+ C_3 \sum_{\alpha=1}^{M-1} \mu_\alpha^2 u_\alpha}}.
\end{equation}
}
Here $ C_4(x,u)= C_2 + (C_3^2/2) \, x u$ , and $p(\vec{u}|u) $
is the uniform distribution on a sphere of radius $1-u$ in dimension $M-1$:
\begin{equation}
\begin{split}
 p(\vec{u}|u) &=  \frac{\Gamma \tonde{\frac{M-1}{2} }}{\pi^{\frac{M-1}{2}} (1-u)^{\frac{M-1}{2}-1}}  \, \prod_{\alpha=1}^{M-1}\frac{1}{u_\alpha^{1/2}}\, \delta \tonde{\sum_{\alpha=1}^{M-1} u_\alpha-(1-u)}. \end{split}
\end{equation}
Notice that $0 \leq u_\alpha \leq 1$, but the distribution \eqref{eq:FinJoint} can be integrated on the whole positive semi-axis because the delta enforces this constraint automatically. 
Explicitly, we can write:
\begin{equation}\label{eq:Comp2}
\begin{split}
 I_{x,u}(\vec{\mu})=&\int_{-\infty}^{\infty} \prod_{\alpha=1}^{M-1} d e_\alpha \frac{\Gamma \tonde{\frac{M-1}{2} }}{\pi^{\frac{M-1}{2}}}  \delta \tonde{\sum_{\alpha=1}^{M-1} e^2_\alpha-1} \times\\
 & \times e^{-\frac{M}{4 \sigma^2} \quadre{C_4 (x,u) (1-u) \sum_{\alpha=1}^{M-1} \mu_\alpha e_\alpha^2 + \frac{[C_3 (1-u)]^2}{4} \tonde{\sum_{\alpha=1}^{M-1} \mu_\alpha e_\alpha^2}^2+ C_3 (1-u)\sum_{\alpha=1}^{M-1} \mu_\alpha^2 e_\alpha^2}}.
 \end{split}
\end{equation}

As anticipated, the distribution \eqref{eq:Comp2}, up to normalization constants, has the same structure as \eqref{eq:FinJoint} but with modified constants $C_3 \to C_3(1-u)$ and $C_2 \to C_4(x,u)(1-u)$. This implies that, fixing $x$ and $u$, the distribution of the remaining eigenvalues is the one of a GOE matrix perturbed exactly as the original one, with modified parameters given in~\eqref{eq:NewPar}.

\subsection{Integration over the remaining eigenvectors and eigenvalues}\label{sec:EffPAr}
As we show in Appendix \ref{app:Auxiliary}, \eqref{eq:Comp2} can be re-written in the following more convenient form:
{
\medmuskip=0mu
\thinmuskip=0mu
\thickmuskip=0mu
\begin{equation}\label{eq:Rapp1}
I_{x,u}(\vec{\mu})=- \frac{\Gamma \tonde{\frac{M-1}{2} }}{\pi^{\frac{M-1}{2}}} \sqrt{\frac{M^3 \,4 \sigma^2}{\pi C_3^2 (1-u)^2}}   \quadre{\frac{2 \sigma^2}{C_3 (1-u)}}^{\frac{M-3}{2}} \int \int_{-i \infty}^{i \infty} dy\, d \lambda e^{M \tonde{\frac{y^2}{\sigma^2}-\lambda}}  I_2
\end{equation}
}
with
\begin{equation}\label{eq:GaussE}
I_2(y, \lambda, \vec{\mu})=    
\int_{-\infty}^\infty \prod_{\alpha=1}^{M-1} d e_\alpha e^{-\frac{M}{2} \sum_{\alpha=1}^{M-1} e_\alpha^2\quadre{ \mu_\alpha^2 +\tonde{\frac{C_4 (x,u)}{C_3} -2 y}\mu_\alpha -2 \lambda  \frac{2 \sigma^2}{C_3 (1-u)}} }.
\end{equation}
The parameters $y$ and $\lambda$ in \eqref{eq:Rapp1} are auxiliary fields that will be fixed through a saddle point calculation, while the integrals \eqref{eq:GaussE} are decoupled Gaussian integrals whose convergence imposes some constraints on the domain of $y,\lambda$. In particular, given the functions
\begin{equation}
\mu^{\pm }_{x,u}(y,\lambda)= -\frac{1}{2} \tonde{\frac{C_4(x,u)}{C_3}- 2 y} \pm \frac{1}{2} \sqrt{8 \lambda \frac{2 \sigma^2}{C_3(1-u)} + \tonde{\frac{C_4(x,u)}{C_3}- 2 y}^2},
\end{equation}
the condition for the convergence of the integrals in \eqref{eq:GaussE} reads (assuming that $\lambda, y$ are real):
{
\medmuskip=0mu
\thinmuskip=0mu
\thickmuskip=0mu
\begin{equation}\label{eq:QuadForm}
\mu_\alpha^2 +\tonde{\frac{C_4 (x,u)}{C_3} -2  y}\mu_\alpha -2  \lambda  \frac{2 \sigma^2}{C_3 (1-u)}= [\mu_\alpha - \mu^{+ }_{x,u}(y,\lambda)] [\mu_\alpha - \mu^{- }_{x,u}(y,\lambda)]>0 \quad \forall \alpha.
\end{equation}
}
For a given configuration of eigenvalues $	\mu_\alpha$, we denote with $\mathcal{D}[{\mu_\alpha}]$ the domain of $\lambda, y$ for which \eqref{eq:QuadForm} is satisfied. Performing the Gaussian integration, \eqref{eq:p1} becomes equal to:
{
\medmuskip=0mu
\thinmuskip=0mu
\thickmuskip=0mu
\begin{equation}\label{eq:BasisDiscuss}
\begin{split}
\frac{ \mathcal{P}_{\theta, \beta}(x,u)}{p(u)}=& \frac{\alpha_M(u) }{\mathcal{Z}_M \,\Gamma \tonde{\frac{M}{2}} }\quadre{\frac{2 \pi \sigma^2}{e \,C_3(1-u)}}^{\frac{M}{2}}e^{-\frac{M}{4 \sigma^2} \quadre{x^2+ C_2 x u + \frac{C_3^2}{4} \,x^2 u^2+ C_3 x^2 u}} \mathcal{J}_{\theta, \beta}(x,u)
\end{split}
\end{equation}
}
where 
\begin{equation}
\alpha_M(u) =-\frac{M^2}{2 \sigma^2} \, \Gamma \tonde{\frac{M-1}{2} } \tonde{\frac{2 e}{M}}^{\frac{M}{2}}  \sqrt{\frac{ C_3(1-u)}{\pi}}
\end{equation}
scales less than exponentially with $M$, 
{
\medmuskip=0mu
\thinmuskip=0mu
\thickmuskip=0mu
\begin{equation}
\mathcal{J}_{\theta, \beta}(x,u)=\int \prod_{\alpha=1}^{M-1} d\mu_\alpha   \, 
{\bm 1}_{x<\mu_{M-1}<\cdots \leq \mu_1}   \int  d y \, d \lambda\,  {\bm 1}_{\lambda, y \in \mathcal{D}[\mu_\alpha]} \, e^{-M^2 \tilde{S}_1[\vec{\mu}]- M \tilde{S}_0[y, \lambda, \vec{\mu}]} 
\end{equation}
}
where ${\bm 1}$ is the indicator function,
and the actions have the following expression:
{
\medmuskip=0mu
\thinmuskip=0mu
\thickmuskip=0mu
\begin{equation}
\begin{split}
&\tilde{S}_1[\vec{\mu}]=\frac{1}{4 \sigma^2} \frac{1}{M}\sum_{\alpha=1}^{M-1} \mu^2_\alpha -\frac{1}{M^2} \sum_{\alpha>\gamma=1}^{M-1} \log (\mu_\gamma-\mu_\alpha),\\
&\tilde{S}_0[y, \lambda,\vec{\mu}]=\lambda- \frac{y^2}{\sigma^2}+\frac{1}{2M} \sum_{\alpha=1}^{M-1} \log [(\mu_\alpha- \mu^+_{x,u})(\mu_\alpha - \mu^-_{x,u})]- \frac{1}{M}\sum_{\alpha=1}^{M-1} \log(\mu_\alpha-x).
\end{split}
\end{equation}
}
Notice that the action $\tilde{S}_1[\vec{\mu}]$ is the one corresponding to the joint distribution of the eigenvalues of an unperturbed GOE matrix, and is given by one-point functions of the eigenvalues. These actions can be expressed in terms of the eigenvalue density $\nu(\mu)= \sum_{\alpha=1}^M \delta(\mu- \mu_\alpha)$, performing the change of variable $\vec{\mu} \to \nu(\mu)$. Naturally, the density $\nu(\mu)$ can have both a continuous part and some poles, corresponding to the isolated eigenvalues. The dominating term of $\tilde{S}_1$  depends only on the continuous part of $\nu(\mu)$, and reproduces exactly then term that one would get from an unperturbed GOE; therefore, the corresponding action is zero at the typical density $\nu^{\rm typ}_{\rm cont}(\mu)= \rho_\sigma(\mu)$ corresponding to the semicircle law \eqref{eq:UsualGOE}. Any contribution to $\nu(\mu)$ coming from isolated poles is  of $O(1/M)$, and gives rise to sub-leading contributions to $\tilde{S}_1$ that have to be added to the linear term in $M$ of the exponent in \eqref{eq:BasisDiscuss}.  

To proceed with the calculation, we assume that \emph{only one} of these poles can be present, corresponding to the second-smallest eigenvalue $\mu_{M-1}$ of the matrix. We show that, under this assumption, the saddle-point equations obtained by minimizing the linear term in $M$ of the resulting action fix this eigenvalue to its typical value $\mu^{\rm typ}_{M-1}(x,u)$ at fixed $x$ and $u$, which is either $-2 \sigma$ when $\sigma^2 F(x,u)\geq -\sigma$, or \eqref{eq:TypSec} otherwise, consistently with the results in Sec. \ref{sec:InterpretationSmaller}. 
We subsequently need to check that the hypothesis is consistent, meaning that whenever the second eigenvalue is fixed to its typical value, the third-smallest eigenvalue typically sticks to the boundary of the semicircle $\mu_{M-2}^{\rm typ}=-2 \sigma$. We discuss this check in Appendix \ref{app:OnlyOneEvalue}.

We therefore assume that the only eigenvalue that can take values smaller that $-2 \sigma$ is $\mu_{M-1}$ and integrate over the remaining ones, getting:
\begin{equation}
\begin{split}
&\mathcal{J}_{\theta, \beta}(x,u)=A_M \int_{\xi \geq x}    d\xi \,h(\xi,x)  \int  d y \, d \lambda\,  {\bm 1}_{\lambda, y \in \mathcal{D}[\xi]} \times \\
& \quad \times e^{-M \quadre{\frac{\xi^2}{4 \sigma^2}- \int d\mu \log[(\mu-\xi)(\mu-x)] \rho_\sigma(\mu)+ \lambda-\frac{y^2}{\sigma^2}+ \frac{1}{2} \int d\mu \log[(\mu-\mu^+_{x,u})(\mu-\mu^-_{x,u})] \rho_\sigma(\mu)}},
\end{split}
\end{equation}
where $h(\xi,x)= (\xi-x)/[(\xi-\mu^+_{x,u})(\xi -\mu^-_{x,u})]^{1/2}$, and $A_M$ contains constant terms coming from the change of variables $\vec{\mu} \to \nu(\mu)$. 
Combining everything, asymptotically at the exponential scale in $M$ we find:
\begin{equation}\label{eq:p2}
\begin{split}
 \mathcal{P}_{\theta, \beta}(x,u)\sim& \mathcal{A}_M\, e^{-M  \Psi_0(x,u)} \int_{\xi \geq x} d \xi e^{-M \quadre{\frac{\xi^2}{4 \sigma^2}- \mathcal{I}(\xi)}} \int_{\mathcal{D}(\xi)} dy d \lambda \, e^{M \phi(y, \lambda)},
 \end{split}
\end{equation}
with 
\begin{equation}\label{eq:Phi}
\begin{split}
&\Psi_0(x,u)= \frac{1}{4 \sigma^2} \tonde{x^2+ C_2 x u + \frac{C_3^2}{4} x^2 u^2+ C_3 x^2 u} - \frac{1}{2} \log \tonde{\frac{2 \sigma^2}{C_3 }}-\mathcal{I}(x)+ \frac{1}{2},\\
&\phi(y, \lambda)= \frac{y^2}{\sigma^2}- \lambda - \frac{1}{2} \int d\mu \rho_\sigma(\mu) \log \quadre{(\mu- \mu^-_{x,u}(y,\lambda))(\mu- \mu^+_{x,u}(y, \lambda))}.
\end{split}
\end{equation}
The expression for $\mathcal{I}(z)$ is given in \eqref{eq:GOEI} (and we are using that $x \leq - 2 \sigma$), and we made use of the identity:
\begin{equation}
\frac{\sqrt{z^2-4 \sigma^2}}{2}= \sigma^2 G(z)-\frac{z}{2} \quad \quad \text{for} \quad z<- 2 \sigma.
\end{equation}
The constant $\mathcal{A}_M$ in \eqref{eq:p2} contains exponential contributions that have to be determined from the condition:
\begin{equation}\label{eq:NormTyp}
 \mathcal{P}_{\theta, \beta}(x_{\rm typ},u_{\rm typ}) \sim O(1),
\end{equation}
where $x_{\rm typ},u_{\rm typ}$ are the typical values of $\mu_M$ and $u_M$ at fixed $\theta, \beta$.

Finally, we comment on the domain $\mathcal{D}$.  The latter changes depending on whether the roots $\mu^{\pm}_{x,u}$ are real or complex. We can distinguish the following two cases :
\begin{itemize}
\item Case A: the roots $ \mu^{\pm}_{x,u}(y,\lambda)$ are complex: this happens whenever the discriminant is negative, corresponding to
\begin{equation}\label{eq:NewCond}
 \lambda< -\frac{1}{8} \tonde{\frac{C_4 (x,u)}{C_3} -2  y}^2 \frac{C_3 (1-u)}{2 \sigma^2} \leq 0.
\end{equation}
In this case the condition \eqref{eq:QuadForm} is always met, and one can set 
\begin{equation}
\mathcal{D}=\grafe{(y, \lambda):  \,  \lambda \leq 0 \quad \text{  and  } \quad y \in \mathbb{R}}.
\end{equation}

\item Case B: The roots $\mu^-_{x,u}(y,\lambda)\leq \mu^+_{x,u}(y,\lambda)$ are real; for $\theta<0$, one can self-consistently check that  $\mu^+_{x,u}(y,\lambda)<0$. A necessary condition for \eqref{eq:QuadForm} to hold true is that $ \mu^+_{x,u}(y,\lambda) \leq -2 \sigma$, meaning that the support of the continuous part of the eigenvalue distribution lies to the right of $ \mu^+_{x,u}$. Additionally, we have to impose that the eigenvalues that do not belong to the continuous part of the eigenvalue density satisfy the condition. This implies that either $\xi <\mu^-_{x,u}(y,\lambda)$ or $\mu^+_{x,u}(y,\lambda) < \xi$, meaning:  
\begin{equation}\label{eq:Domain}
\mathcal{D}(\xi)= \grafe{(y, \lambda):  \,  \xi \leq \mu^-_{x,u}(y, \lambda) \text{   or  } \mu^+_{x,u}(y, \lambda) \leq \xi \,\, \text{  and  } \mu^+_{x,u}(y, \lambda) <- 2 \sigma}.
\end{equation}	
\end{itemize}

\subsection{Saddle point equations for the auxiliary fields I: inside the domain}\label{sec:SPinD}
In this section we discuss the saddle point equations for $\phi(y, \lambda)$ in \eqref{eq:Phi}, at fixed values of $\xi$. To simplify the notation, we denote $\mu^\pm_{x,u}$ simply with $\mu^\pm$. 

The minimization of $\phi(y, \lambda)$ gives the following two equations:
\begin{equation}\label{eq:Rel1}
\begin{split}
 \frac{C_3(1-u)}{2 \sigma^2}  \tonde{\mu^+-\mu^-}&=G_\sigma(\mu^-)- G_\sigma(\mu^+) \\
  \frac{4}{ \sigma^2} y + \frac{C_3(1-u)}{2 \sigma^2}  \tonde{\mu^+ + \mu^-}&= G_\sigma(\mu^-)+ G_\sigma(\mu^+).
 \end{split}
\end{equation}
Summing and subtracting these equations we get the relations:
\begin{equation}\label{eq:Fund}
\begin{split}
 \frac{C_3(1-u)}{2 \sigma^2}  \mu^+ +   \frac{2}{ \sigma^2} y &=G_\sigma(\mu^-)\\
 \frac{C_3(1-u)}{2 \sigma^2}  \mu^- +   \frac{2}{  \sigma^2} y &=G_\sigma(\mu^+).
 \end{split}
\end{equation}
Assuming that $G_\sigma$ can be inverted, these can be re-written as:
\begin{equation}\label{eq:Root}
\begin{split}
\mu^+(\lambda,y) &= G^{-1}_\sigma \tonde{ \frac{C_3(1-u)}{2 \sigma^2}  \mu^-(\lambda,y) +   \frac{2}{ \sigma^2} y}\\
\mu^-(\lambda,y) &= G^{-1}_\sigma \tonde{ \frac{C_3(1-u)}{2 \sigma^2}  \mu^+(\lambda,y) +   \frac{2}{ \sigma^2} y}.
\end{split}
\end{equation}
As we show in Appendix \ref{app:AuxFieldsSol}, regardless of whether $\mu^\pm_{x,u}$ are complex or real, the solutions of these equations is given by:
\begin{equation}\label{eq:ySaddle}
\begin{split}
y^*&=\frac{C_4(x,u) C_3(1-u)^2}{ 2 \quadre{2+ C_3 (1-u)}^2}, \quad \quad
\lambda^* =  -\frac{ \quadre{\sigma ^2 (C_3 (1-u)+2)^2+C_4^2   (1-u)^2}}{\sigma^2 (C_3 (1-u)+2)^3}.
\end{split}
\end{equation}
When $\mu^\pm$ are computed at the saddle point solutions $y^*, \lambda^*$, the correspondent action is given by:
\begin{equation}\label{eq:ActionSaddle}
\phi(\lambda^*, y^*)= \frac{1}{2}-\frac{1}{2}\log \quadre{\sigma^2 \tonde{1+ \frac{2}{C_3(1-u)}}}+ \frac{C_4^2(x,u) (1-u)^2}{4 \sigma^2 \quadre{2 + C_3(1-u)}^2}.
\end{equation}

We now discuss the conditions under which the GOE resolvent can be inverted, and the saddle point solutions lie in the right domain $\mathcal{D}[\xi]$. If Case A holds, the equation are always invertible given that the resolvent is never singular. When $\mu^{\pm}_{x,u}$ are real, \emph{i.e.} when Case B holds, to write \eqref{eq:Root} it must hold:
\begin{equation}\label{eq:Riemann}
\begin{split}
 \Big| \frac{C_3(1-u)}{2 \sigma^2}  \mu^{\pm} +   \frac{2}{ \sigma^2} y \Big| \leq \frac{1}{\sigma}.
 \end{split}
\end{equation}
Since for $\mu^\pm <0$ and thus $G(\mu^\pm)<0$, these conditions become 
\begin{equation}
\begin{split}
 F^\pm \equiv \frac{C_3(1-u)}{2 \sigma^2}  \mu^{\pm}( \lambda^*, y^*) +   \frac{2}{ \sigma^2} y^*  \geq  -\frac{1}{\sigma},
 \end{split}
\end{equation}
and given that $F^+>F^-$ one has to impose that
{
\medmuskip=0mu
\thinmuskip=0mu
\thickmuskip=0mu
\begin{equation}\label{eq:CritUse}
F(x,u) \equiv F^-=-\frac{C_4(1-u)+ \sqrt{C_4^2(1-u)^2- \sigma^2 C_3(1-u) \quadre{2+ C_3(1-u)}^3}}{ \sigma^2 \quadre{2 + C_3(1-u)}^2} \geq - \frac{1}{\sigma}.
\end{equation}
}
As we anticipated in \eqref{eq:EffeDue}, this condition is equivalent to $ [G_{\tilde{\sigma}}(\tilde{\theta})]^{-1}\geq -\sigma$,
which is the condition under which the typical value of the second-smaller eigenvalue $\mu_{M-1}^{\rm typ}$ \emph{is not} out of the bulk. In this case we find:
\begin{equation}
\xi^{\pm}_\sigma(x,u) \equiv \mu^{\pm}_{x,u}(\lambda^*, y^*) = m^{\pm}_\sigma[C_4(x,u)(1-u), C_3(1-u)],
\end{equation}
where $m^{\pm}_\sigma$ are given in \eqref{eq:Emme}. In this regime of parameters, the saddle point solution $(y^*, \lambda^*)$ lies within $\mathcal{D}[\xi]$ iff
\begin{equation}\label{eq:Cond2}
\xi \geq \mu^+(y^*, \lambda^*)=G^{-1}_\sigma \tonde{F(x,u)} = G^{-1}_\sigma \tonde{\frac{1}{\sigma^2 G_{\tilde{\sigma}}(\tilde{\theta})}} \quad \text{           or        } \quad \xi  \leq \mu^-(y^*, \lambda^*),
\end{equation}
and has to be discarded otherwise.
 
When \eqref{eq:Riemann} is not met and  $F^+ \geq -1/\sigma > F^-$, the equation for $\mu^+(y,\lambda)$ still admits a solution $\mu^+(y^*,\lambda^*)$, which nevertheless belongs to the second Riemann sheet in the complex plane. This is due to the fact that the quadratic equation for the resolvent of a GOE matrix $\sigma^2 G^2_\sigma(z)-z G_\sigma(z)+1=0$  admits another solution 
\begin{equation}\label{eq:ResolventII}
 G^{(II)}_\sigma(z)=\frac{1}{2 \sigma^2} \tonde{z+ \text{sign}(z) \sqrt{z^2- 4 \sigma^2}}
\end{equation}
for $z$ real, which is obtained from $G_\sigma(z)$ changing the sign in front of the square root. This function is  defined on the second Riemann sheet, and it takes values in $|z|>1/\sigma$. Its inverse is again given by $G_\sigma^{-1}(z)= z^{-1}+ \sigma^2 z$, but now evaluated in this domain $|z|>1/\sigma$. 

When $F^+ \geq -1/\sigma > F^-$, $\mu^+(y^*,\lambda^*)$ 
solves the second of Eqs. \eqref{eq:Fund} with $G_\sigma \to G_\sigma^{II}$. In this case, the saddle point solution $(y^*, \lambda^*)$ can still be considered, and it lies within the integration domain $\mathcal{D}[\xi]$ iff $\xi < \mu^-(y^*, \lambda^*)$. Notice that this is the regime of parameters in which the typical value of the second smallest eigenvalue \emph{is} out of the bulk. Using the results of Sec. \ref{sec:InterpretationSmaller} we know that the latter can be written as: 
\begin{equation}\label{eq:typ22}
\mu_1(x,u) \equiv \mu^{\rm typ}_{M-1}= G_\sigma^{-1} \tonde{G_{\tilde{\sigma}}(\tilde{\theta})}= G_\sigma^{-1} \tonde{\frac{1}{\sigma^2 \, F(x,u)}},
\end{equation}
where here the argument of $G^{-1}_\sigma$ is larger than $-1/\sigma$. 

We notice that the explicit expression of the typical value \eqref{eq:typ22} is exactly the same as the one of $ \xi^+_\sigma(x,u) =\mu^+_{x,u}(\lambda^*, y^*)=G_\sigma^{-1} \tonde{F(x,u)}$. The two expressions coincide due to the following symmetry of the function $G_\sigma^{-1}$ on the real axis:
\begin{equation}\label{eq:Symmetry}
G_\sigma^{-1}(x)= G_\sigma^{-1} \tonde{\frac{1}{\sigma^2 x}}.
\end{equation}
Therefore, the threshold value $\xi^+_\sigma(x,u) $ can be though of as the analytic continuation of the expression for $\mu_1(x,u)$, extended to a regime of parameters for which typically the eigenvalue \emph{is not} out of the bulk of the semicircle \footnote{Notice that when $\beta \to 0$, we correctly recover $F \to  \mu (1-u)/\sigma^2$ and thus \eqref{eq:CritUse} reduces to  $\mu(1-u) > - \sigma$. The other threshold value $ \xi^{-}_\sigma(x,u)$ diverges to $-\infty$ in this limit.}. The difference between the two quantities is that in this regime, while $\mu_1(x,u)$ lies in the first Riemann sheet, $\xi^+_\sigma(x,u) $ lies in the second.

Finally, the case $F^+ < -1/\sigma$ needs not to be considered, since $F^+$ becomes complex before reaching the threshold value $F^+ = -1/\sigma$
\footnote{Indeed, exactly at the point when $F^\pm$ develop a complex part and we transition to Case A, the functions take the value:
\begin{equation}
F^\pm(x,u)= - \frac{C_4(1-u)}{\sigma^2 [2+ C_3(1-u)]^3}= -\frac{1}{\sigma} \sqrt{\frac{C_3(1-u)}{2 + C_3(1-u)}} \geq -\frac{1}{\sigma}.
\end{equation}}. In summary, the saddle point solutions is acceptable whenever the resulting $\mu^\pm(y^*,\lambda^*)$ are either complex (case A), or when they satisfy any of the two conditions \eqref{eq:Cond2}.

\subsection{Saddle point equations for the auxiliary fields II: boundary of domain}\label{sec:OD}
When Case B holds but \eqref{eq:Cond2} is not met,  $y^*, \lambda^*$ do not belong to the domain $\mathcal{D}[\xi]$, and the rate function $\phi$ has to be computed at boundary manifold, where one of the two following equalities hold:
\begin{equation}\label{eq:Mup}
\xi= -\frac{1}{2} \tonde{\frac{C_4}{C_3}- 2 y} \pm \frac{1}{2} \sqrt{8 \lambda \frac{2 \sigma^2}{C_3(1-u)} + \tonde{\frac{C_4}{C_3}- 2 y}^2} = \mu^\pm_{x,u}(y, \lambda).
\end{equation}
This is an equation relating $y, \lambda$. Assuming that \eqref{eq:Mup} holds for some $\lambda=\lambda_{\rm ext}(\xi)$ and $y=y_{\rm ext}(\xi)$,  taking its square we get the relations:
\begin{equation}\label{eq:GenOut}
\begin{split}
\lambda_{\rm ext}(y; \xi )&=\frac{C_3(1-u)}{4 \sigma^2} \quadre{\xi^2 + \tonde{\frac{C_4(x,u)}{C_3}- 2 y}\xi},\\
y_{\rm ext}(\lambda; \xi)&=\frac{1}{2}\quadre{\frac{C_4(x,u)}{C_3}- \frac{4 \sigma^2 \lambda}{C_3(1-u) \xi}+ \xi}.
\end{split}
\end{equation}
Substituting the first of these equations into $\phi(\lambda, y)$ and minimizing over $y$ we get:
\begin{equation}\label{eq:Eq1Ex}
\frac{2 y}{\sigma^2} + \frac{C_3(1-u)}{2 \sigma^2} \xi- G \tonde{- \xi -\frac{C_4}{C_3}+2 y }=0.
\end{equation}
The two equations are solved by:
\begin{equation}\label{eq:SolExt}
\begin{split}
y_{ \rm ext}(\xi)&=- \,\frac{C_3 \sigma ^2}{C_3  \xi (C_3 (1-u)+2)+2 C_4}-\frac{1}{4}
   C_3  (1-u)  \xi,\\ 
        \lambda_{ \rm ext}(\xi)&=\frac{C_3 \xi  (1-u)}{4 \sigma ^2} G^{-1}\tonde{\frac{2 C_3}{2 C_4 + C_3 \xi \quadre{2+ C_3(1-u)}}}
  \end{split}
    \end{equation}
If the second equation in \eqref{eq:GenOut} is used we get an equivalent result, see Appendix \ref{app:AuxFieldsSol} for the details. 
The rate function $\phi(y, \lambda)$ computed at \eqref{eq:SolExt} reads
{
\medmuskip=0mu
\thinmuskip=0mu
\thickmuskip=0mu
\begin{equation}\label{eq:ActExt}
\phi(y_{\rm ext}, \lambda_{\rm ext})=- \frac{(1-u) \xi \quadre{4 C_4 + C_3 \xi (4 + C_3(1-u))}}{16 \sigma^2} - \frac{ \mathcal{I}(\xi)}{2}+ \frac{1}{2}\log \quadre{\frac{2 C_3}{C_3 \xi (C_3 (1-u)+2)+ 2 C_4}},
\end{equation}
}
as we derive in the same Appendix.

\subsection{The variational problem for the second smaller eigenvalue}\label{sec:VariationalSecond}
Combining \eqref{eq:p2} with the results of the last two section, we find that:
{
\medmuskip=0mu
\thinmuskip=0mu
\thickmuskip=0mu
\begin{equation}
\mathcal{P}_{\beta, \theta}(x,u) = \mathcal{A}_M e^{-M \quadre{\Psi_0(x,u)+ \substack{\inf}_{-2 \sigma \geq \xi \geq x} \Psi_1(x,u, \xi)}}= e^{-M \quadre{\Psi_0(x,u)+ \inf_{-2 \sigma \geq \xi \geq x} \Psi_1(x,u, \xi) -l(\theta, \beta)}}
\end{equation}
}
with $l(\theta, \beta)$ defined by $\mathcal{A}_M=\text{exp} \tonde{M l(\theta, \beta)+ o(M)}$, and
\begin{equation}\label{eq:Psi1}
\begin{split}
&\Psi_1(x,u, \xi)=  \frac{1}{4 \sigma^2} \xi^2 - \int d\mu \rho_\sigma(\mu) \log(\mu-\xi) - \Phi(x, u; \xi).
\end{split}
\end{equation}
The function $ \Phi(x, u; \xi)$ is given by:
\begin{equation}
\Phi(x,u; \xi)=\begin{cases}
 \phi_1(x,u)  & \text{ if Case A, Cond 1 or Cond 4 }\\
 \phi_2(x,u; \xi)   & \text{ if Cond 2 or Cond 3 }
\end{cases}
\end{equation}
where
{
\medmuskip=0mu
\thinmuskip=0mu
\thickmuskip=0mu
\begin{equation}
\begin{split}
&\phi_1  \equiv \frac{1}{2}-\frac{1}{2}\log \quadre{\sigma^2 \tonde{1+ \frac{2}{C_3(1-u)}}}+ \frac{C_4^2(x,u) (1-u)^2}{4 \sigma^2 \quadre{2 + C_3(1-u)}^2}\\
&\phi_2 \equiv \frac{1}{2}\log \quadre{\frac{2 C_3}{C_3 \xi (C_3 (1-u)+2)+ 2 C_4(x,u)}}- \frac{(1-u) \xi \quadre{4 C_4 + C_3 \xi (4 + C_3(1-u))}}{16 \sigma^2} - \frac{ \mathcal{I}(\xi)}{2}
\end{split}
\end{equation}
}
where the conditions are:
\begin{equation}\label{eq:Conditions}
\begin{split}
&\text{Cond 1:}  \quad \sigma^2 F(x,u) \geq - \sigma  \text{ and } \xi \geq  \xi^{+}_\sigma(x,u) \text{   or   } \xi \leq \xi^{-}_\sigma(x,u)\\
 &\text{Cond 2:} \quad \sigma^2 F(x,u) \geq - \sigma  \text{ and } \xi^{-}_\sigma(x,u)<\xi < \xi^{+}_\sigma(x,u)\\
  &\text{Cond 3:} \quad \sigma^2 F(x,u) < - \sigma  \text{ and } \xi^{-}_\sigma(x,u)<\xi   \\
    &\text{Cond 4:} \quad \sigma^2 F(x,u) < - \sigma  \text{ and } \xi^{-}_\sigma(x,u) \geq \xi.
\end{split}
\end{equation}

The function $\Psi_1$ in \eqref{eq:Psi1} is (up to constants) the large deviation function for the second eigenvalue, than we need to optimize in the domain $\quadre{x, -2 \sigma}$. we can distinguish the following two cases:

\begin{itemize}

\item If $\sigma^2 F(x,u) \geq - \sigma$, typically the second smallest eigenvalue \emph{is not} out of the bulk. In this case the large deviation function has \emph{three} regimes:
\begin{equation}\label{eq:LDFsecond}
\Psi_1(x,u,\xi)= 
\begin{cases}
 \frac{1}{4 \sigma^2} \xi^2 - \mathcal{I}(\xi)- \phi_1(x,u)  &\text{  if  } \xi^{+}_\sigma \leq \xi \leq -2 \sigma\\
  \frac{1}{4 \sigma^2} \xi^2 - \mathcal{I}(\xi)- \phi_2(x,u, \xi) &\text{  if  }  \xi^{-}_\sigma < \xi <  \xi^{+}_\sigma\\
  \frac{1}{4 \sigma^2} \xi^2 - \mathcal{I}(\xi)- \phi_1(x,u)  &\text{  if  }  \xi \leq \xi^{-}_\sigma\\
\end{cases}
\end{equation}
and it is always minimal at $\xi=-2 \sigma$, meaning that:
\begin{equation}
\inf_{- 2 \sigma \geq \xi \geq x} \Psi_1(x,u,\xi)=  1 - \mathcal{I}(-2 \sigma)- \phi_1(x,u)= \frac{1}{2} - \log \sigma- \phi_1(x,u).
\end{equation}

\item
If $\sigma^2 F(x,u) < - \sigma$, typically the second smallest eigenvalue \emph{is} out of the bulk, and takes value $\mu_1(x,u)$. In this case for any $\xi \in \quadre{\xi^-_\sigma, -2 \sigma}$ it holds:
{
\medmuskip=0mu
\thinmuskip=0mu
\thickmuskip=0mu
\begin{equation}\label{eq:PSi1Cite}
\Psi_1= \frac{ \xi^2}{4 \sigma^2} -\frac{\mathcal{I}(\xi)}{2} +\frac{(1-u) \xi \quadre{4 C_4 + C_3 \xi (4 + C_3(1-u))}}{16 \sigma^2} - \frac{1}{2}\log \quadre{\frac{2 C_3}{C_3 \xi (C_3 (1-u)+2)+ 2 C_4}},
\end{equation}
}
which has a minimum at $\mu_1(x,u)$; indeed
the derivative of $\Psi_1$ is proportional to:
{
\medmuskip=0mu
\thinmuskip=0mu
\thickmuskip=0mu
\begin{equation}\label{eq:Deriv}
\frac{2 \sigma^2 C_3 (C_3 (1-u)+2)}{C_3 \xi 
   (C_3 (1-u)+2)+2 C_4}+\frac{\xi C_3(1-u)  (C_3 (1-u)+4)}{2 }+C_4 (1-u)-\xi +\sqrt{\xi ^2-4 \sigma ^2}=0.
\end{equation}
}
Among the solutions to this equation, the only one that does not diverge in the limit $\beta\to 0$ is precisely given by  $\mu_1(x,u)$. Depending on the position of $x$ with respect to $\mu_1(x,u)$, the infimum is either attained at the minimum or at the boundary, meaning:
\begin{equation}\label{eq:Accendi}
\inf_{- 2 \sigma \geq \xi \geq x} \Psi_1(x,u,\xi)=
\begin{cases}
\Psi_1(x, u, \xi=x)  &\text{  if  } \mu_1(x,u) \leq x\\
\Psi_1(x, u, \mu_1(x,u)) &\text{  if  } x < \mu_1(x,u).
\end{cases}
\end{equation}

\end{itemize}

In Appendix \ref{sec:AppCheckSecond}, we comment on the consistence between the large deviation function for the second-smallest eigenvalue $\Psi_1(x,u; \xi)$ and known results in the literature \cite{Maida} valid in the limit $\beta \to 0$. 
To conclude this section, we simplify the resulting expressions by noticing that $\Psi_1$ in \eqref{eq:PSi1Cite} one has the identity:
\begin{equation}\label{eq:Idy1}
\begin{split}
 \Psi_1(x,u; y\to x)  =&\frac{x^2}{4 \sigma^2} - \frac{1}{2} \mathcal{I}(x)- \frac{1}{2}\log \quadre{\frac{2 C_3}{C_3^2 x+ 2 C_3 x+ 2 C_2} } \\
 &+ \frac{(1-u)x}{4 \sigma^2}\tonde{ C_4(x,u)+  C_3 x+ \frac{C_3^2}{4} x(1-u)},
 \end{split}
\end{equation}
implying that in the relevant regime, $\Psi_0(x,u)+  \Psi_1(x,u; y\to x) = \mathcal{L}^{(b)}_{\theta, \beta}(x)$ given in \eqref{eq:Case11}. Similarly, as we show in the same Appendix it holds:
{
\medmuskip=0mu
\thinmuskip=0mu
\thickmuskip=0mu
\begin{equation}\label{eq:Idy2}
\Psi_1(x,u; y \to \xi^+_\sigma)  =- \frac{1}{2} \log \tonde{\frac{C_3 (1-u)}{C_3(1-u)+2 }}-\frac{C_4^2(x,u) (1-u)^2}{4 \sigma^2 \quadre{2 + C_3(1-u)}^2}=\frac{1}{2}-\log \sigma -\phi_1(x,u),
\end{equation}
}
and thus in all other regimes the sum $\Psi_0(x,u)+  \Psi_1$ equals to $\mathcal{L}^{(a)}_{\theta, \beta}(x,u)$ given again in \eqref{eq:Case11}. Combining all this we recover the results stated in Sec. \ref{sec:LDFfixedXU}, up to the constant  $l(\theta, \beta)$. 
In the following subsection, we determine the typical value of the overlap parameter $u$ at fixed $x$, and compute the constant $l(\theta, \beta)$.

\subsection{Optimization over the overlap $u$}\label{sec:OptimizationU}
\begin{figure}[ht]
    \includegraphics[width=.48\linewidth]{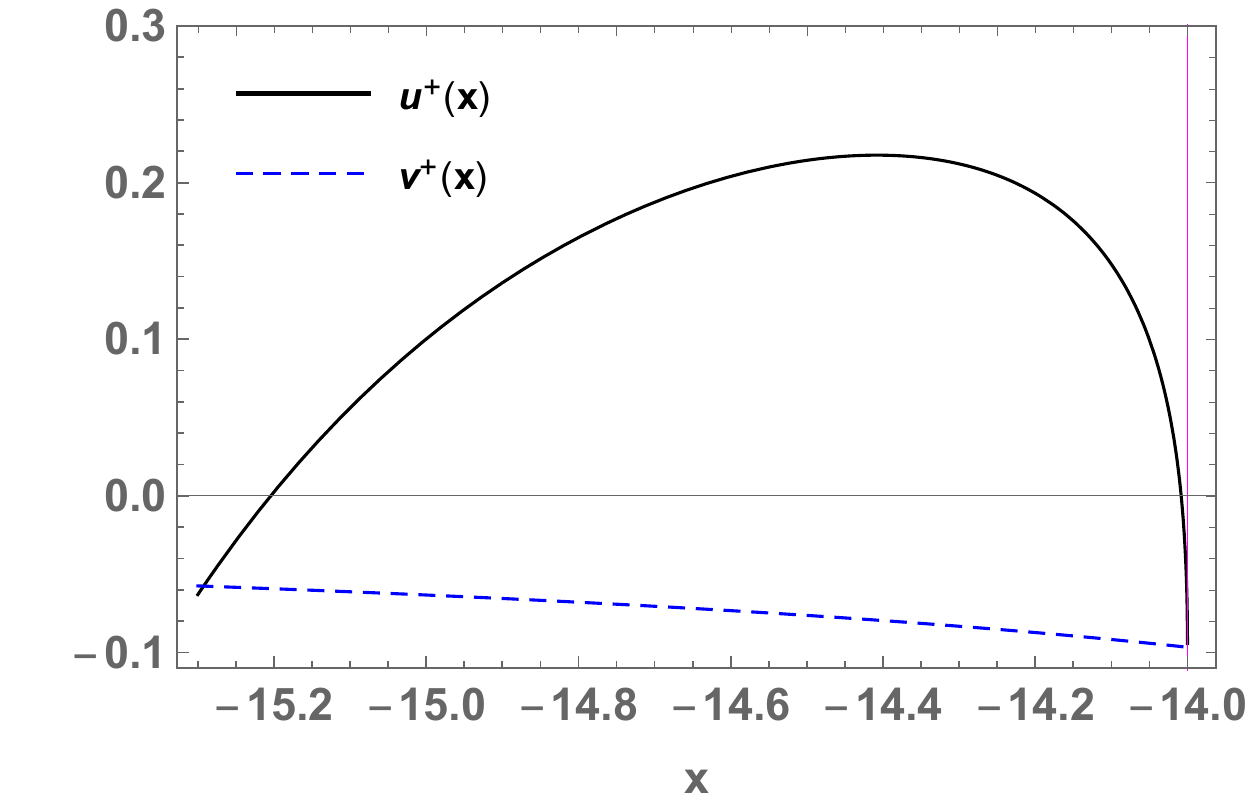} 
        \includegraphics[width=.48\linewidth]{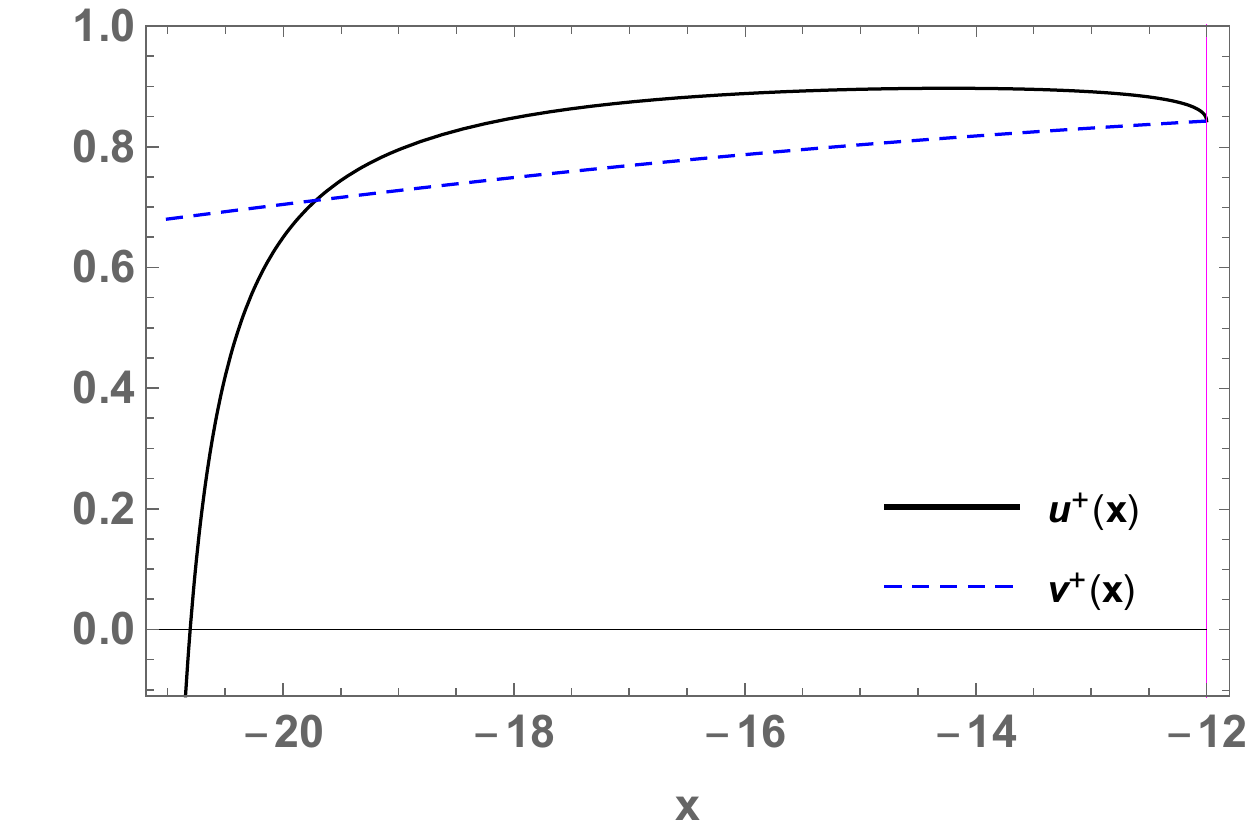} 
\caption{ \small Plots of $u^+_{\theta, \beta}(x)$ for values of parameters in which the smallest eigenvalue is typically at the boundary of the semicircle (\emph{left}) or out of the bulk (\emph{right}). The dashed blue curve denotes $v^+_{\theta, \beta}(x)$, see the discussion in Appendix \ref{app:SelfConsTyp}. The region where $u^+_{\theta, \beta}(x) \geq v^+_{\theta, \beta}(x)$ corresponds to the regime of parameters in which $\sigma^2 F(x,u)+ \sigma>0$.}\label{fig:UTyp}
  \end{figure}

We now discuss the optimization of the large deviation function $\mathcal{L}_{\theta, \beta}(x,u)$ over the overlap $u \in \quadre{0,1}$. The functions to be optimize change across the different regimes. Since $\mathcal{L}^{(b)}_{\theta, \beta}(x)$ is independent on $u$, the integral over $u$ in this case does not give any exponential contribution. We therefore focus on $\mathcal{L}^{(a)}_{\theta, \beta}(x,u)$ and identify the solutions of 
\begin{equation}
\frac{\partial \mathcal{L}^{(a)}_{\theta, \beta}(x,u)}{\partial u}=0
\end{equation}
that lie within the unit interval. This variational equation is quadratic in $u$, with two solutions 
\begin{equation}\label{eq:Usol}
\begin{split}
u^{\pm}_{\theta, \beta}(x)&=\frac{4 C_2^2+4 C_2 [C_3 (C_3+3)+1] x+C_3
   (C_3+2) [(C_3 (C_3+4)+2) x^2+4 \sigma ^2]}{4
   C_2^2+4 C_2 C_3 (C_3+3) x+C_3^2 \left\{(C_3+2)
   (C_3+4) x^2+4 \sigma ^2\right\}}\\
  & \pm \frac{2 \sqrt{\left(x^2-4 \sigma ^2\right) (2 C_2+C_3
   (C_3+2) x)^2}}{4
   C_2^2+4 C_3 C_3 (C_3+3) x+C_3^2 \left\{(C_3+2)
   (C_3+4) x^2+4 \sigma ^2\right\}},
   \end{split}
\end{equation}
of which the relevant one satisfying $0 \leq u \leq 1$ for at least some values of $x$ is  $u^{+}_{\theta, \beta}(x)$, which corresponds to a minimum of $\mathcal{L}^{(a)}_{\theta, \beta}(x,u)$. Notice that in the limit $C_3 \to 0$ (equivalently, $\beta \to 0$) corresponding to a purely additive perturbation, this reduces to (using $\theta<0$):
\begin{equation}
u^{+}_{\theta, \beta}(x) \to 1- \frac{x + \sqrt{x^2-4 \sigma^2}}{2 \theta},
\end{equation} 
which agrees with the known results \cite{BiroliGuinnet}.
When $u^{+}_{\theta, \beta}(x)$ is non-negative,  we always find $u^{+}_{\theta, \beta}(x)<1$. Therefore we can set:
\begin{equation}\label{eq:utypa}
u_{\rm typ}^{(a)}(x) \equiv \max \grafe{0, u^{+}_{\theta, \beta}(x)}.
\end{equation}
Here the superscript denotes that $u_{\rm typ}^{(a)}(x)$ is obtained assuming $\mathcal{L}_{\theta, \beta}(x,u) \propto \mathcal{L}^{(a)}_{\theta, \beta}(x,u)$. In order to discuss the form of $u_{\rm typ}^{(a)}(x)$, we find it convenient to separate the three following regimes of the parameters $\theta, \beta$:
 
\begin{itemize}
\item Regime A: When $- 2 \sigma'< \theta <0$, we find that $u^{+}_{\theta, \beta}(x)<0$ and thus $u_{\rm typ}^{(a)}(x)=0$. 

\item Regime B1: When $\theta_{\rm c} < \theta < -2 \sigma'$ with 
$   \theta_{\rm c}=- \sigma [1+ 4 \beta + 2 \beta^2]/(1+ \beta)^2$,
 given in \eqref{eq:BBP}, we find that the function $u^{+}_{\theta, \beta}(x)$ behaves as in Fig. \ref{fig:UTyp} (\emph{left}): it is non-monotonic in $x$ \footnote{This is due to the fact that the coefficient of the quadratic term in the equation for $u$ depends on $x$, and vanishes at a value of $x$ which corresponds to the poles of \eqref{eq:Usol}. So at this value of $x$ one has a divergence of the solution for $u$ to $-\infty$ [the pole diverges to $- \infty$ when $C_3 \to 0$]. The divergence gives the non-monotonicity.}, and at $x= -2 \sigma$ it takes the value
 \begin{equation}\label{eq:v2s}
u^{+}_{\theta, \beta}(- 2 \sigma)=  \frac{(1+\beta)^4 \theta +\sigma (\beta +1)^2  [1+2 \beta  (2+\beta)] }{(1+\beta)^4 \theta +\sigma \beta 
   (\beta +2) [3+2 \beta  (2+\beta)]},
 \end{equation}  
which is always negative. Indeed, $\theta \leq -2 \sigma'$ implies that the denominator is always negative, while the numerator changes sign exactly at $\theta= \theta_{\rm c}$. The function $u^{+}_{\theta, \beta}(x)$ vanishes exactly at the points $x^\pm_\sigma(\theta, \beta) $ given in \eqref{eq:ExPIuMinus}, and it positive in the regime $x^-_\sigma(\theta, \beta) < x< x^+_\sigma(\theta, \beta)$. Therefore in this regime the optimization of $\mathcal{L}^{(a)}_{\theta, \beta}(x,u)$ subject to the constraint $u \in \quadre{0,1}$  the gives:
\begin{equation}\label{eq:utypa}
u_{\rm typ}^{(a)}(x)=\begin{cases}
0 &\text{   if  } x^+_\sigma(\theta, \beta)<x<-2 \sigma \\
u^+_{\theta, \beta}(x) &\text{   if  } x^-_\sigma(\theta, \beta)<x<x^+_\sigma(\theta, \beta)\\
0 &\text{   if  } x<x^-_\sigma(\theta, \beta).
\end{cases}
\end{equation}
Notice that these values of $\theta$ coincide with the regime in which typically the smallest eigenvalue of the perturbed matrix is at the boundary of the semicircle. 

\item Regime B2: When $\theta \leq \theta_{\rm c}$  the function $u^{+}_{\theta, \beta}(x)$ behaves as in Fig. \ref{fig:UTyp} (\emph{right}): it is again non-monotonic in $x$, but it is positive at $x=-2 \sigma$, with only one zero at $x=x^-_\sigma(\theta, \beta)$. Therefore in this case:
\begin{equation}\label{eq:utypa2}
u_{\rm typ}^{(a)}(x)=\begin{cases}
u^+_{\theta, \beta}(x) &\text{   if  } x^-_\sigma(\theta, \beta)<x<-2 \sigma\\
0 &\text{   if  } x<x^-_\sigma(\theta, \beta).
\end{cases}
\end{equation}
Notice that these values of $\theta$ coincide with the regime in which typically the smallest eigenvalue of the perturbed matrix is smaller that $-2 \sigma$, and equals to $\mu_0(\theta, \beta)$.
\end{itemize}
In order for $u_{\rm typ}^{(a)}(x)$ to be the correct solution for the optimal overlap, we have to check self-consistently that the conditions that imply $\mathcal{L}_{\theta, \beta}(x,u)\propto \mathcal{L}^{(a)}_{\theta, \beta}(x,u)$ are satisfied when $u \to u_{\rm typ}^{(a)}$. In Appendix \ref{app:SelfConsTyp} we perform this self-consistent check, showing that when the optimization over $u$ is performed, the relevant rate function is always  $\mathcal{L}^{(a)}_{\theta, \beta}$: when the overlap $u$ is allowed to take its typical value, one always finds that the typical value of the second-smallest eigenvalue is out of the bulk and \emph{larger} than $x$, which is the large-deviation value of the smallest one, as it is natural to expect.
Using the above expressions, we find:
\begin{equation}
\begin{split}
\mathcal{L}^{(a)}_{\theta, \beta}(x, u_{\theta, \beta}^+(x))&= 1-\log \sigma^2+ \frac{1}{2}\log \frac{C_3}{2} +
\frac{x^2}{4 \sigma^2}- \mathcal{I}(x)-\tilde{\phi}_2(x) ,\\
\mathcal{L}^{(a)}_{\theta, \beta}(x, 0)&= 1-\log \sigma^2+ \frac{1}{2}\log \frac{C_3}{2} +
\frac{x^2}{4 \sigma^2}- \mathcal{I}(x)-\tilde{\phi}_1(x) 
\end{split}
\end{equation}
with:
\begin{equation}
\begin{split}
\tilde{\phi}_1&=\frac{1}{2}-\frac{1}{2}\log \tonde{\frac{\sigma^2 (C_3+2)}{C_3}}+ \frac{C_2^2}{4 \sigma^2 (2+ C_3)^2}\\
\tilde{\phi}_2(x)&=-\frac{x}{16 \sigma^2} \quadre{4 C_2+ C_3 x (4+ C_3)}+ \frac{1}{2} \log \tonde{\frac{2 C_3}{2 C_2 + C_3 x (2+ C_3)}}- \frac{\mathcal{I}(x)}{2},
\end{split}
\end{equation}
implying that for $\theta_{\rm c} < \theta < -2 \sigma'$ we have: 
\begin{equation}\label{eq:SecAin}
\begin{split}
&\mathcal{L}^{(a)}_{\theta, \beta}(x, u_{\text{typ}}(x))- \tonde{1-\log \sigma^2+ \frac{1}{2}\log \frac{C_3}{2}} =\\
&
\begin{cases}
\frac{x^2}{4 \sigma^2}- \mathcal{I}(x)- \tilde{\phi}_1&\text{   if  } x^+_\sigma(\mu, \beta)<x<-2 \sigma \\
\frac{x^2}{4 \sigma^2}- \frac{\mathcal{I}(x)}{2}-\tonde{\tilde{\phi}_2(x) + \frac{\mathcal{I}(x)}{2}} &\text{   if  } x^-_\sigma(\mu, \beta)<x<x^+_\sigma(\mu, \beta)\\
\frac{x^2}{4 \sigma^2}- \mathcal{I}(x)- \tilde{\phi}_1 &\text{   if  } x<x^-_\sigma(\mu, \beta),
\end{cases}
\end{split}
\end{equation}
while for $\theta<\theta_{\rm c}$:
\begin{equation}\label{eq:SecAout}
\begin{split}
&\mathcal{L}^{(a)}_{\theta, \beta}(x, u_{\text{typ}}^{(a)}(x))- \tonde{1-\log \sigma^2+ \frac{1}{2}\log \frac{C_3}{2}} =\\
&
\begin{cases}
\frac{x^2}{4 \sigma^2}- \frac{\mathcal{I}(x)}{2}-\tonde{\tilde{\phi}_2 + \frac{\mathcal{I}(x)}{2}} &\text{   if  } x^-_\sigma(\mu, \beta)<x<- 2 \sigma\\
\frac{x^2}{4 \sigma^2}- \mathcal{I}(x)- \tilde{\phi}_1&\text{   if  } x<x^-_\sigma(\mu, \beta).
\end{cases}
\end{split}
\end{equation}

The expression \eqref{eq:SecAin} agrees -up to a constant- with the one for the large deviation function of the second smallest eigenvalue that appears in the calculation (see Eq. \eqref{eq:LDFsecond}), provided one keeps in mind the substitution $C_4(1-u) \to C_2$ and $C_3(1-u) \to C_2$, see Sec. \ref{sec:InterpretationSmaller}. Similarly, \eqref{eq:SecAout} is consistent with Eq. \eqref{eq:PSi1Cite}.

  To conclude this section, we determine the constant $l(\theta, \beta)=\mathcal{L}^{(a)}_{\theta, \beta}(x_{\rm typ }, u_{\text{typ}})$. When  $\theta_{\rm c} < \theta$, the typical value of the smallest eigenvalue is $x_{\rm typ }= -2 \sigma$ and $u_{\rm typ}=0$ leading to:
\begin{equation}
\mathcal{L}^{(a)}_{\theta, \beta}(-2 \sigma, 0)=1-\frac{1}{2}\log \tonde{\frac{2 \sigma^4}{C_3+2}} -\frac{C_2^2}{4 \sigma^2 (C_3+2)^2}= 1-\log \tonde{\frac{\sigma^2}{1+ \beta}}-\frac{\theta^2}{2 \sigma^2 [1+\beta]^2}.
\end{equation}
When  $\theta <\theta_{\rm c}$ instead  we have 
\begin{equation}
x_{\rm typ}=\mu_0(\theta, \beta)= G^{-1}_\sigma \tonde{G_{\sigma'}(\theta)}, \quad G_{\sigma'}(\theta)= \frac{\sqrt{C_2^2-C_3 (C_3+2)^3 \sigma ^2}-C_2}{C_3
   (C_3+2) \sigma ^2}
\end{equation}
and $u_{\rm typ}=u^{(a)}_{\rm typ}(x_{\rm typ})$. Using that $G_{\sigma'}(\theta) [2 C_2+ C_3 x^+_\sigma(2+ C_3)]= 2 (C_3+2)$, we obtain that also in this regime:
\begin{equation}
\mathcal{L}^{(a)}_{\theta, \beta}(x_{\rm typ}, u_{\rm typ})= 1-\frac{1}{2}\log \tonde{\frac{2 \sigma^4}{C_3+2}} -\frac{C_2^2}{4 \sigma^2 (C_3+2)^2}=1-\log \tonde{\frac{\sigma^2}{1+ \beta}}-\frac{\theta^2}{2 \sigma^2 [1+\beta]^2}.
\end{equation}
Thus, we recover \eqref{eq:Constf}. The final expression for the function $\overline{\mathcal{L}}_{\theta, \beta}(x)$ in \eqref{eq:OverlineL} is obtained as $\overline{\mathcal{L}}_{\theta, \beta}(x)= \mathcal{L}^{(a)}_{\theta, \beta}(x, u_{\rm typ}(x))- l(\theta, \beta)$, substituting the expressions above.

\subsection{Optimization over the Gaussian fluctuations of $\theta$}\label{sec:OptTheta}
\begin{figure}[ht]
    \includegraphics[width=.48\linewidth]{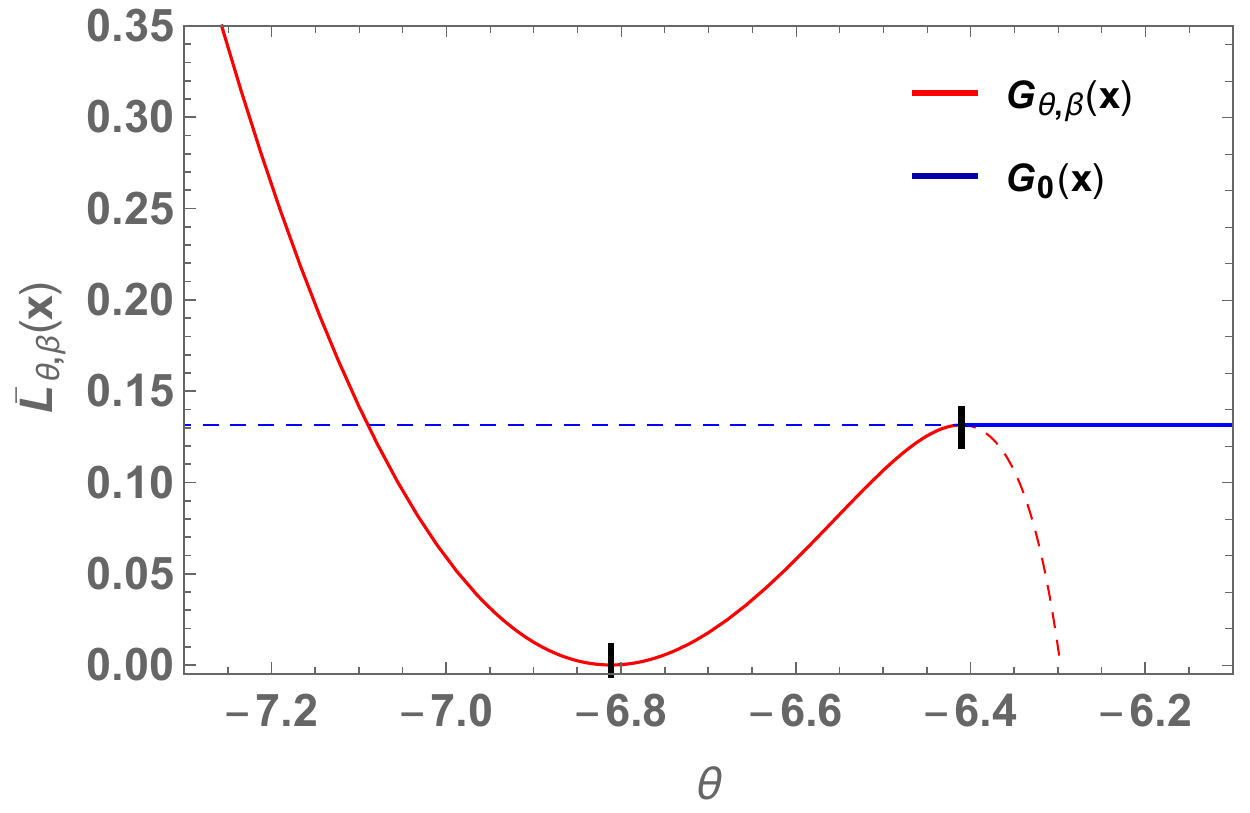} 
        \includegraphics[width=.46\linewidth]{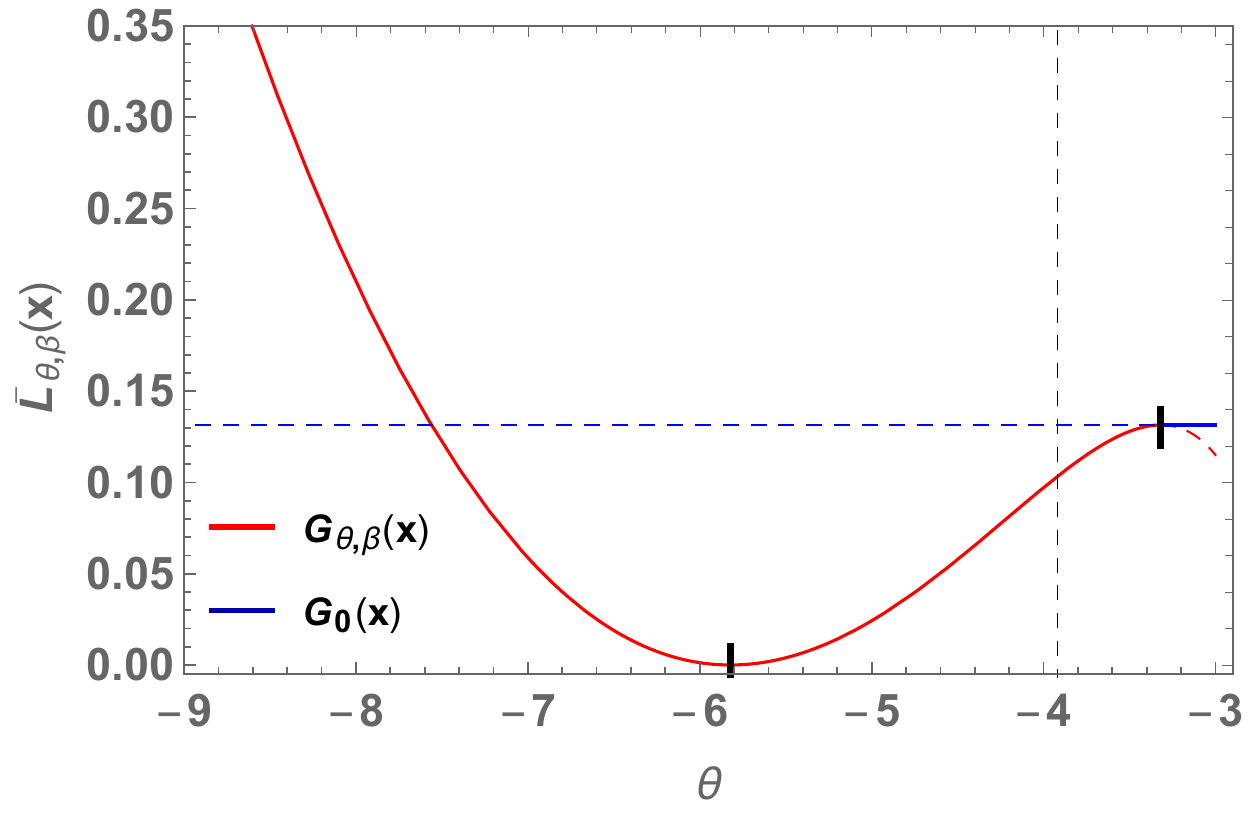} 
\caption{ \small \emph{Left. }Large deviation function $\overline{\mathcal{L}}_{\theta, \beta}(x)$ as a function of $\theta$ for $\beta = 2, \sigma = 3$ and $x =-7<x^*_\sigma(\beta)$. The ticks correspond to the local minimum and maximum attained at  $\theta^*_-$ and $\theta^*_+$, respectively. In this case the local maximum $\theta^*_+<\theta_{\rm c}$. \emph{Right. } Large deviation function $\overline{\mathcal{L}}_{\theta, \beta}(x)$ for $\beta = 0.2, \sigma = 3$ and $x =-7>x^*_\sigma(\beta)$. The dashed vertical lines marks  $\theta_{\rm c}$, which is smaller than $\theta^*_+$ in this case. }\label{fig:LargeMu}
  \end{figure}

The above calculations are performed for fixed $\theta<0$. In this section, we allow for fluctuations of $\theta$ and determine the rate function in \eqref{eq:FinalLDF}:
\begin{equation}\label{eq:NewToMin}
\mathcal{F}_{\overline{\theta}, \sigma_\theta, \beta}(x)= \min_{\theta} \quadre{\frac{(\theta- \overline{\theta})^2}{2 \sigma_\theta^2} + \overline{\mathcal{L}}_{\theta,\beta}(x)},
\end{equation}
focusing on Regime B. Viewed as a function of $\theta$ and at fixed $x$, the rate
function $\overline{\mathcal{L}}_{\theta, \beta}(x)$ in \eqref{eq:OverlineL} takes different forms depending on whether $\theta$ is such that $x^{\pm}_\sigma(\theta, \beta)$ are smaller or larger than $x$. More precisely, we find that 
\begin{equation}
\begin{split}
&x \leq x^{-}_\sigma(\theta, \beta)  \longrightarrow \theta \geq \theta^*_+(x)= \frac{x+2 \beta  (\beta +2) x + \sqrt{x^2-4 \sigma ^2}}{2 (1+\beta)^2}\\
&x \geq x^{+}_\sigma(\theta, \beta)  \longrightarrow \theta \leq \theta^*_-(x)= \frac{x+2 \beta  (\beta +2) x - \sqrt{x^2-4 \sigma ^2}}{2 (1+\beta)^2}\\
\end{split}
\end{equation}
Notice that $\theta^*_\pm(x)$ are also the stationary points satisfying
\begin{equation}
\frac{\partial}{\partial \theta} \quadre{ \mathcal{G}_{\theta, \beta}(x)}=0.
\end{equation}
In particular, $\theta^*_-(x)$ is a local minimum of $\mathcal{G}_{\theta, \beta}$: when the additive perturbation equals to $\theta^*_-(x)$, then $x$ is precisely the typical value of the smallest eigenvalue, \emph{i.e.}, $x=G^{-1}_\sigma \tonde{G_{\sigma'}(\theta_-^*(x))}$, see \eqref{eq:TypFinal}. The point $\theta^*_+(x)$ is a local maximum of $\mathcal{G}_{\theta, \beta}$. For $x<-2 \sigma$ it holds $\theta^*_-(x)<\theta^*_+(x)$ and $\theta^*_-(x)<\theta_{\rm c}$. The position of the local maximum $\theta^*_+(x)$ with respect to $\theta_{\rm c}$ depends instead on $x$: $\theta^*_+(x)<\theta_{\rm c}$ for $x<x^*_\sigma(\beta)$ while  $\theta^*_+(x)>-\theta_{\rm c}$ for $x^*_\sigma(\beta)<x<-2 \sigma$, with $x^*_\sigma(\beta)=-2 \sigma - \sigma [\beta(1+ \beta)^2(2+ \beta)]^{-1}$. Therefore, viewed as a function of $\theta$ the rate $\overline{\mathcal{L}}_{\theta, \beta}(x)$ in Regime B reads, see Fig. \ref{fig:LargeMu}:
\begin{equation}
\overline{\mathcal{L}}_{\theta, \beta}(x)= 
\begin{cases}
\mathcal{G}_{\theta, \beta}(x)  &\text{  if  } \theta \leq \theta^*_+(x)\\
\mathcal{G}_{0}(x)  &\text{  if  } \theta > \theta^*_+(x).
\end{cases}
\end{equation}

\begin{figure}[ht]
    \includegraphics[width=.48\linewidth]{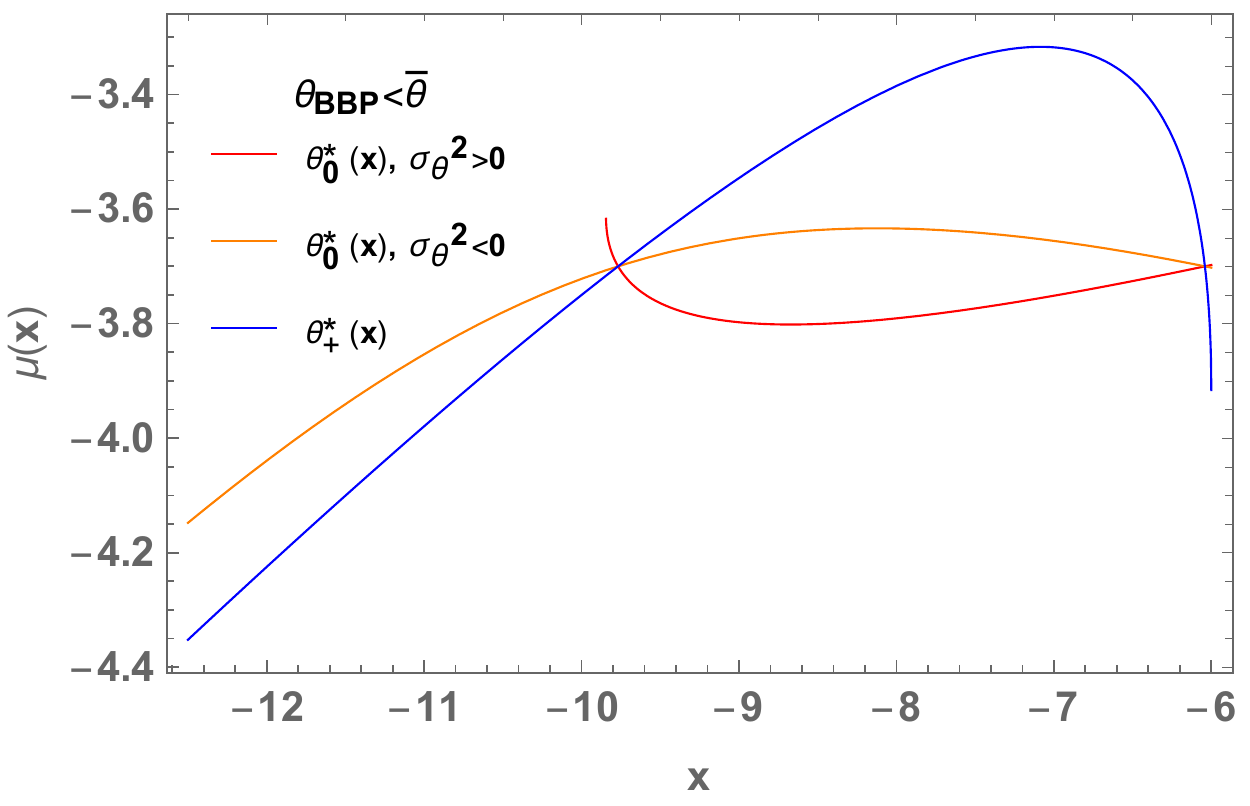} 
       \includegraphics[width=.48\linewidth]{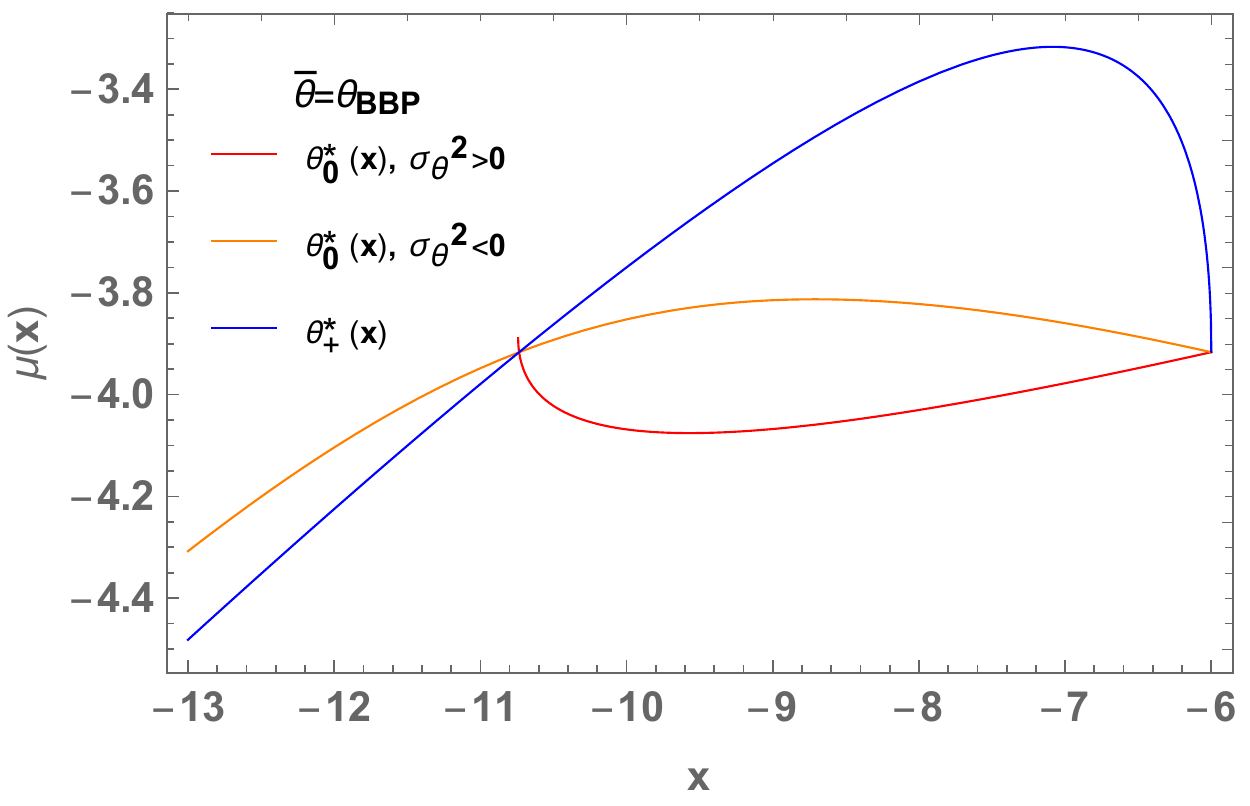} 
        \includegraphics[width=.48\linewidth]{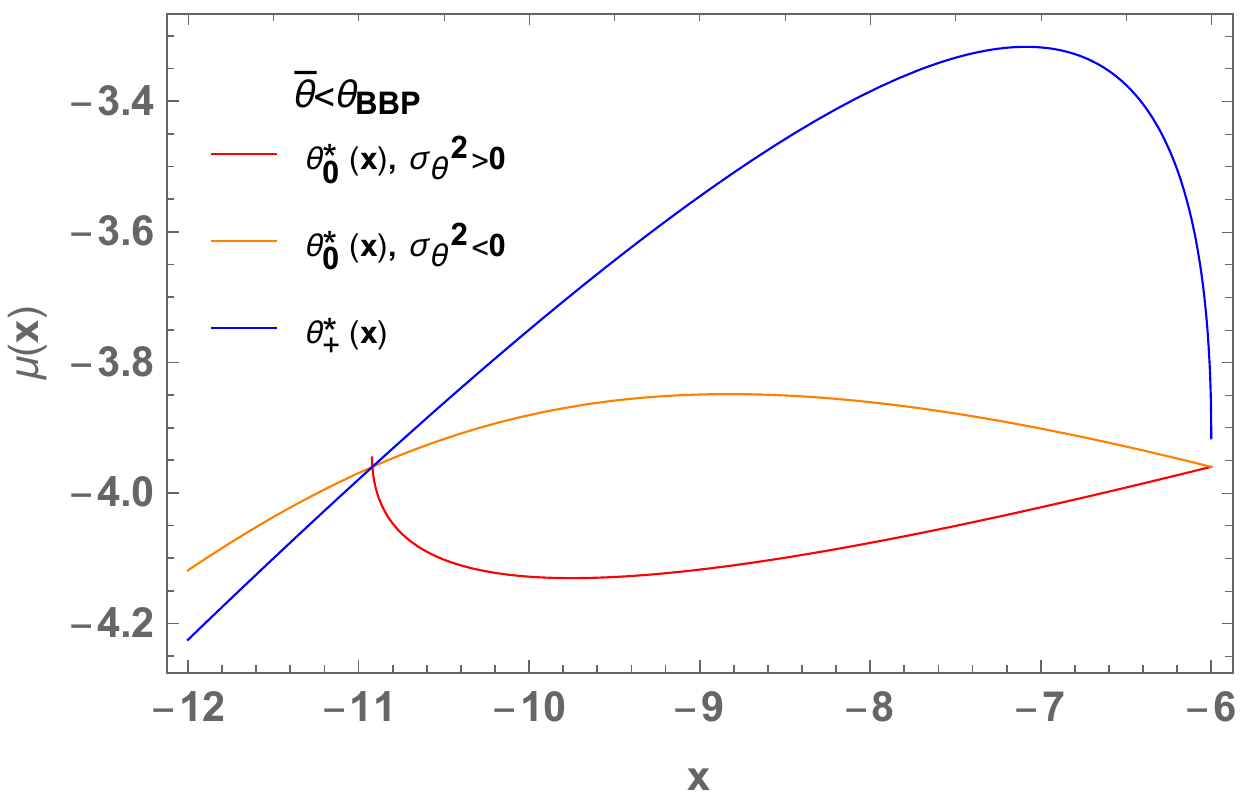} 
         \includegraphics[width=.48\linewidth]{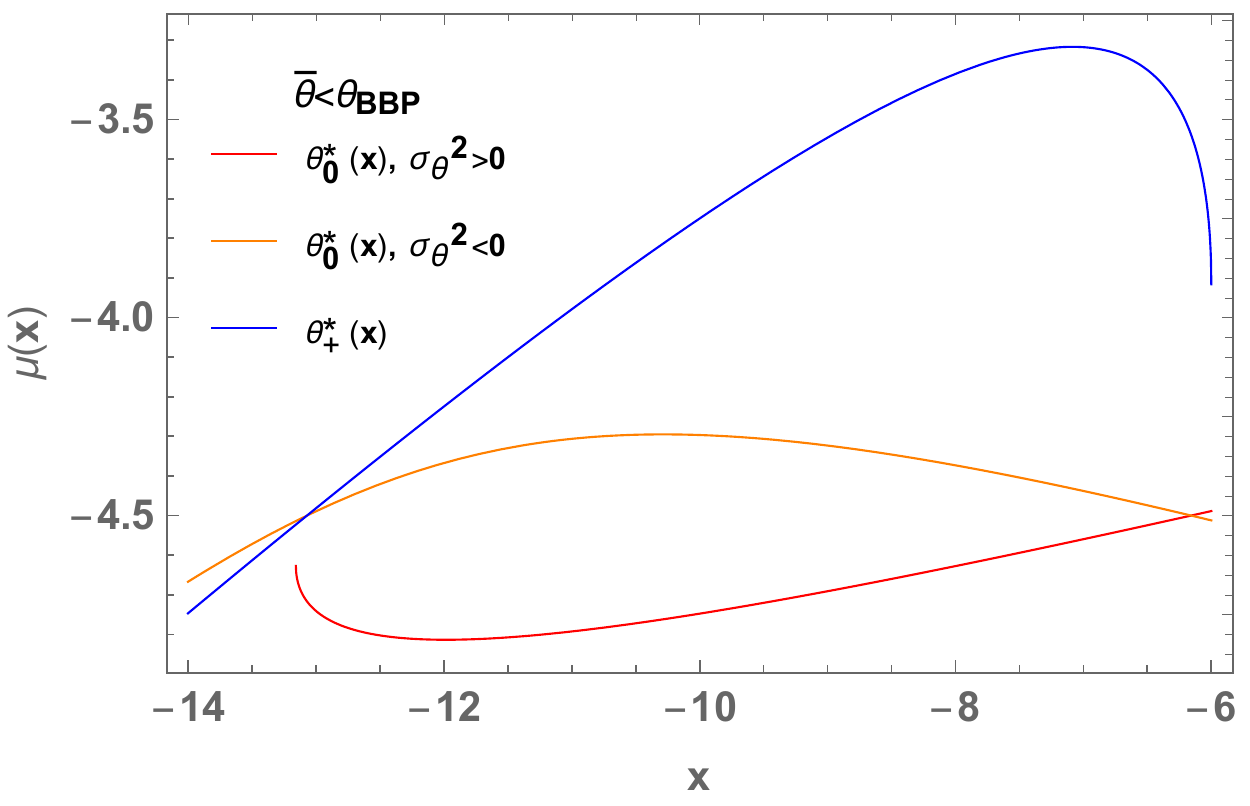} 
\caption{ \small Comparison between the function $\theta^*_+(x)$ and $\theta^*_0(x)$ for either positive and negative values of $\sigma^2_\theta=\pm 0.8 $ and $\beta = .2, \sigma= 3$ (giving $\theta_{\rm c}= -3.92$) and $\overline{\theta}=-3.7$ (\emph{Top Left}), $\overline{\theta}=\theta_{\rm c}$ (\emph{Top Right}), $\overline{\theta}=-3.96$ (\emph{Bottom Left}) and $\overline{\theta}=-4.5$ (\emph{Bottom Right}) }\label{fig:Casistica}
  \end{figure}

The Gaussian weight in \eqref{eq:NewToMin} shifts the local minimum from $\theta^*_-(x)$ to
\begin{equation}\label{eq:NewSP}
\begin{split}
\theta^*_0(x| \sigma, \beta, \overline{\theta}, \sigma^2_\theta) \equiv \frac{2\overline{\theta}\sigma ^2+2 \beta  (\beta +2) \sigma ^2 (\overline{\theta}+x)+[2 \beta  (2+\beta )+1] (\beta
   +1)^4 x \sigma_\theta^2- \sqrt{T}}{4 (1+\beta)^2 \sigma ^2+2 (1+\beta)^6 \sigma_\theta^2}
  \end{split}
\end{equation}
with
\begin{equation}
\begin{split}
T= &4\sigma ^4 (1+\beta)^4  [ \overline{\theta}^2-2  \sigma_\theta^2]+4 \sigma ^2 (1+\beta)^2 \overline{\theta} x
   [(1+\beta)^4  \sigma_\theta^2-2 \beta  (2+\beta) \sigma ^2]+\\
   +&[(1+\beta)^4 x  \sigma_\theta^2-2
   \beta  (\beta +2) \sigma ^2 x]^2-4 (1+\beta)^8 \sigma ^2  \sigma_\theta^4.
   \end{split}
\end{equation}
Henceforth we denote $\theta^*_0(x| \sigma, \beta, \overline{\theta}, \sigma^2_\theta)$ simply with  $\theta^*_0(x)$.
This point lies in the correct domain provided that: 
\begin{equation} \label{eq:COndTheta}
\theta^*_0(x) \leq \theta^*_+(x).
\end{equation}
We find that, irrespectively of the value of the variance $\sigma_\theta$, the two curves in \eqref{eq:COndTheta} meet at \emph{at most} two values of $x$, see Fig. \ref{fig:Casistica}, that are given precisely by:
\begin{equation}
x=x^\pm_\sigma(\overline{\theta}, \beta),
\end{equation}
where $x^\pm_\sigma$ are as in \eqref{eq:ExPIuMinus} and we are assuming that $\theta^*_0(x)$ is real. When the curve meet, they equal to:
\begin{equation}
\theta^*_0(x^\pm_\sigma(\overline{\theta}, \beta)) = \overline{\theta}.
\end{equation}
More precisely, as it appears from Fig. \ref{fig:Casistica}, we find that:
\begin{itemize}
\item When $\theta_{\rm c} \leq \overline{\theta}$, the two function in \eqref{eq:COndTheta} cross at both $x^{\pm}_\sigma$, and the solution $\theta^*_0(x)$ is to be retained for $x \in \quadre{x^-_\sigma, x^+_\sigma}$; at the boundary of the interval one has $\theta^*_0=\overline{\theta}$, that is the solution to be kept for all $x$ outside the interval;
\item When $ \overline{\theta}<\theta_{\rm c}$ the solution cross only at $x^-_\sigma$ for $\sigma_\theta^2<0$ (orange curves), and the solution $\theta^*_0(x)$ is to be retained for $x > x^-_\sigma$; for $\sigma_\theta^2<0$ a transition occurs: if $\sigma_\theta$ becomes large enough the solution  $\theta^*_0(x)$ becomes complex before crossing at $x^-_\sigma$  (as it follows from Sec. \ref{sec:statHess}, this regime of positive variance is not of direct interest for applications to the $p$-spin landscape).
\end{itemize}
Evaluating the rate functions at the correct value of $\theta$, we recover \eqref{eq:Ultimo1} and \eqref{eq:Ultimo2}. Notice that the fact that the conditions are unaltered provided one performs the substitution  $\theta \to \overline{\theta}$ is consistent with the observation that  $x^+_\sigma$ is related to the typical value of the smallest eigenvalue, that should not be shifted by fluctuations of order $1/\sqrt{M}$ of the $MM$ element of the matrix. For the purely additive case, this is proved in \cite{Benaych-Georges}, see the Remark 2.16.

\section{Summary and conclusions}\label{sec:Conclusions}
Characterizing the geometry of high-dimensional landscape in terms of the distribution of their stationary points is a fundamental step to understand quantitatively the dynamical exploration of the landscape. This is particularly true when the landscape is rugged with plenty of energy barriers, and the dynamics is expected to be dominated by activated processes. In this work we have considered a prototypical energy landscape, that of the $p$-spin model, and we have determined the statistics of the index-1 saddles surrounding an arbitrary local minimum, as a function of its energy. In particular, we have identified the range of energy densities and overlaps in which an exponentially large population of saddles is found, and computed their complexity. This completes the analysis initiated in \cite{RBCBarriers}, where only the saddles at shorter distance from the reference minimum were obtained. We have characterized a transition occurring in the population of dominant saddles, separating a regime in which they are geometrically connected to the local minimum and a regime in which they are not, meaning that the corresponding downhill direction in the landscape points in a random direction in configuration space that is not correlated to the direction connecting the saddle to the local minimum. 

A relevant question to address once the saddles are identified concerns the properties (typical energy and overlap) of the minima that are connected to the reference one through a given index-1 saddle. For the saddles that are closer to the reference minimum, these properties are determined in \cite{RBCProgress}, where it is shown that the closest saddles connect the reference minimum to minima that are quite close to it in configuration space. Therefore, escaping through these saddles the system is likely unable to decorrelate from the first trapping minimum. It is an interesting open question whether the same holds true also for the saddles at larger distance from the minimum, whose statistics is determined in this work. An alternative possibility (which is not ruled out by known results, see the discussion in Sec.~\ref{sec:DiscussionDynamics}) is that the marginal saddles found in this work allow the system to decorrelate, \emph{i.e.}, to reach regions of configuration space that are orthogonal to the reference minimum. This would open interesting scenarios for the activated dynamics in this model, allowing the system to decorrelate from the trapping minimum while staying at energies that lie \emph{below} the threshold value.  How to validate or rule out this scenario through numerical simulations \cite{Stariolo,Marco} and how to embed this type of processes within simple phenomenological models \cite{BouchaudTrap,Dyre} are open direction to explore. 

On the technical side, the landscape analysis performed in this work required to extend the large deviation principles derived in \cite{BiroliGuinnet} to the case of a GOE matrix deformed with both an additive and a multiplicative finite-rank perturbation. The resulting large deviation functions display features similar to the ones obtained in case of a purely additive perturbation: in particular, we find that the different regimes displayed by these functions have an interpretation in terms of a \emph{BBP-like} transition of the second-smallest eigenvalue of the perturbed matrix, as it happens in the purely additive case \cite{BiroliGuinnet}. Some new feature emerge nonetheless as a consequence of the multiplicative part of the perturbation: for instance, when the smallest eigenvalue is fixed to values of $x$ for which the second-smallest eigenvalue is not an outlier but lies within the bulk of the eigenvalue density (see Fig. \ref{fig:LDF} \emph{left}), the large deviations are affected by the finite rank perturbation only in an intermediate regime $x \in [x^-_\sigma, x^+_\sigma]$, while they coincide with the unperturbed GOE large deviation for both small-enough and large-enough $x$. Correspondingly, the correlation of the smallest eigenvector with the direction of the perturbation (measured by $u_{\rm typ}(x)$) displays a non-monotonic behavior in $x$. The scale $x^-_\sigma$ appears only in presence of a multiplicative perturbation, and diverges to $x^-_\sigma \to -\infty$ in the limit of a purely additive perturbation. 

Obtaining a rigorous proof of these results, and more generally of the fact that the annealed \emph{constrained} complexities of saddles are exact for the $p$-spin model, are also interesting open problems.

\section*{Acknowledgments}
I thank G. Biroli for the many insightful discussions on this problem and on related topics, and for the useful feedback on the manuscript. This work is supported by the Simons Foundation collaboration Cracking the Glass Problem (No. 454935 to G. Biroli).

\section{Appendices}
\subsection{The statistics of conditioned Hessian}\label{app:ConstantsHessian}
In this appendix, we recall the explicit expressions of the functions $\Delta, \tilde{\Delta}$ and $\mu$ defining the statistics of the Hessian matrices discussed in Sec. \ref{sec:statHess}. We recall that $\sigma^2=p(p-1)$.  The variances of the elements $m_{iM}$ ($i \neq M$) of the matrix $\mathcal{M}$ are given by:
\begin{equation}
\Delta^2(q)=p(p-1) \quadre{1-\frac{(p-1)(1-q^2) q^{2p-4}}{1-q^{2p-2}}} \leq \sigma^2.
\end{equation}

The element $m_{MM}$ has a different variance given by:
\begin{equation}
 \Tilde{\Delta}^2(q)=p(p-1)  \,  \frac{b_1(q)}{b_2(q)},
\end{equation}
with 
{
\medmuskip=0mu
\thinmuskip=0mu
\thickmuskip=0mu
\begin{equation}
\begin{split}
b_1(q)=&p(p-1) q^{4 p}-(p-1)(p-2)^2  q^{2 p+2}+(p-1)^2(p-2)  q^{2 p+8}+2 q^8-p \left(3 p^2-13 p+14\right) q^{2 p+6}+\\
  &(p-2)(p-3) q^{4 p+4}+\left(3 p^3-14 p^2+17
   p-6\right) q^{2 p+4} -2 (p-1)(p-2) q^{4 p+2}\\
b_2(q)=&q^4 \left[q^4-(p-1)^2 q^{2 p}+q^{4 p}+2 p(p-2)  q^{2
   p+2}-(p-1)^2 q^{2 p+4}\right]
\end{split}
\end{equation}
}
and we find that in general  $ \Tilde{\Delta}^2 (q)<\Delta^2(q)$.  For $p=3$, in particular, one finds $ \Tilde{\Delta}^2 (q)=0$. 
Finally, the element $m_{MM}$ has a non-zero average given by:
\begin{equation}\label{eq:Mu}
 \begin{split}
 \mu(q, \epsilon, \epsilon_0) \equiv\frac{\sqrt{2} (p-1) p \left(1-q^2\right) \left(a_0(q) \epsilon_0-a_1(q) \epsilon \right)}{q^{6-p}+q^{3 p+2}- q^{p+2}\left((p-1)^2 (q^4+1)-2 (p-2) p q^2\right)}
 \end{split}
\end{equation} 
with 
\begin{equation}
\begin{split}
 a_1&=q^{3 p}+ q^{p+2}\left(p-2-(p-1) q^2\right)\\
 a_0&= q^4+ q^{2 p}\left(1-p +(p-2)q^2\right).
   \end{split}
\end{equation}

\subsection{Computing the expectation value of the Hessian determinant}\label{app:determinantHessian}
In Sec. \ref{sec:ConstrComp1} we use the fact that the expectation value of the Hessian determinant in the Kac-Rice formula, conditioned to the values of the smallest eigenvalue $\lambda_{\rm min }= \lambda$ and of $u_{\rm min}=u$, to leading order in $N$ is independent of this conditioning. To show this, we first notice that the diagonal shift in 
\eqref{eq:Shift} is independent of the conditioning, which only affects the matrix $\mathcal{M}$. We let $\mu_\alpha$ be the eigenvalues of $\mathcal{M}$, ordered as $\mu_M \leq \mu_{M-1} \leq \cdots \leq \mu_1$. Setting $x=  \lambda + \sqrt{2} \, p\, \epsilon$, we condition $\mathcal{M}$ to the event $\mu_M= x$ and to the overlap $u$, and denote with $P_{\epsilon, q, \epsilon_0}\tonde{\grafe{\mu_\alpha}_{\alpha=1}^{M-1} | x,u}$ the joint distribution of the remaining eigenvalues. We can therefore write:
\begin{equation}\label{eq:Det1}
\begin{split}
&  \Big\langle  \left| \text{det} \mathcal{H}[{\bm \sigma}]\right| \Big|  \grafe{ \begin{subarray}{l}
 {\bf g}[{\bm \sigma}^0]=0, {\bf g}[{\bm \sigma}]=0\\
  h[{\bm \sigma}^0]=\sqrt{2 N} \epsilon_0,   h[{\bm \sigma}]=\sqrt{2 N} \epsilon\\
  \lambda_{\rm min}=\lambda , \, \,  u_{\rm min}=u
  \end{subarray}} \Big\rangle= |\lambda| \, \int \prod_{\alpha=1}^{M-1} d \lambda_\alpha \,  |\lambda_\alpha|\times \\
 &\times P_{\epsilon, q, \epsilon_0}\tonde{\grafe{\lambda_\alpha+ \sqrt{2}\, p\, \epsilon} \Big| \lambda+ \sqrt{2}\, p\, \epsilon, u} .
  \end{split}
\end{equation}
As we derive in more generality in Sec. \ref{sec:Stat}, the joint distribution $P_{\epsilon, q, \epsilon_0}\tonde{\grafe{\mu_\alpha}_{\alpha=1}^{M-1} | x,u}$ has the same structure as the one of the eigenvalues of the \emph{unconditioned} matrix $\mathcal{M}$, {\it i.e.}, it equals to the joint distribution of eigenvalues of a matrix perturbed with both an additive and a multiplicative rank-1 perturbation along the same direction in configuration space. The values of the additive and multiplicative perturbations depend explicitly on the parameters $x$ and $u$, see Sec. \ref{sec:InterpretationSmaller}. 
These perturbations do not modify the typical eigenvalue density of the matrix $\mathcal{M}$ to leading order in $N$, which remains a GOE semicircle of the form $\rho_\sigma(\mu)= \sqrt{4 \sigma^2-\mu^2}/2 \pi \sigma^2$: their only effect is to generate (for certain values of parameters) isolated eigenvalues, that correspond to sub-leading corrections of order $1/N$ to the eigenvalue density. Nevertheless, these perturbation do not matter when computing  \eqref{eq:Det1} to leading exponential order in $N$, as only the bulk of the density of states does. In particular, using the fact that the determinant is a 1-point function of the eigenvalues, and computing  \eqref{eq:Det1} with a saddle point in the space of eigenvalue densities we get:
\begin{equation}
\Big\langle  \left| \text{det} \mathcal{H}[{\bm \sigma}]\right| \Big|  \grafe{ \begin{subarray}{l}
 {\bf g}[{\bm \sigma}^0]=0, {\bf g}[{\bm \sigma}]=0\\
  h[{\bm \sigma}^0]=\sqrt{2 N} \epsilon_0,   h[{\bm \sigma}]=\sqrt{2 N} \epsilon\\
  \lambda_{\rm min}=\lambda,   \, \,  u_{\rm min}=u
  \end{subarray}} \Big\rangle=  e^{\quadre{\frac{M}{2} \log M + \int d\lambda \, \rho_\sigma(\lambda+\sqrt{2}\, p\, \epsilon) \log |\lambda| + o(N) }},
\end{equation}
which is exactly the same contribution that we would obtain from the unconstrained Hessian. Notice that this contribution does not depend neither on the geometrical conditioning on $q$, nor on the conditioning to the value of the smallest eigenvalue.

\subsection{Generalized Kac-Rice formula for the quenched complexity}\label{app:QuenchedKR}
The general expression of the higher moments appearing in \eqref{eq:RepTrick} is given by:
\begin{equation}\label{eq:moments1}
\begin{split}
 \Big\langle \mathcal{N}_{{\bm \sigma}^0}^n(\epsilon, q, \lambda, u| \epsilon_0)\Big\rangle_0&=  \int \prod_{a=1}^n d{\bm \sigma}^{(a)} \, \delta\hspace{-0.05 cm}\tonde{{\bm \sigma}^{(a)} \hspace{-0.06 cm}\cdot\hspace{-0.06 cm} {\bm \sigma}^0\hspace{-0.06 cm}-\hspace{-0.06 cm} q\hspace{-0.02 cm}} 
  \times  p_{\vec{\bm \sigma}|{\bm \sigma}^0}({\bf 0}, \epsilon) \mathbb{G}^{(n)}_{\vec{{\bm \sigma}}| {\bm \sigma}^0} \tonde{\vec{\lambda},\vec{u}} \times\\
  &\times \Big\langle \prod_{a=1}^n  \left| \text{det} \mathcal{H}[{\bm \sigma}^{(a)}]\right| \Big|  \grafe{ \begin{subarray}{l}
 {\bf g}[{\bm \sigma}^0]=0, {\bf g}[{\bm \sigma}^{(a)}]=0\\
  h[{\bm \sigma}^0]=\sqrt{2 N} \epsilon_0,   h[{\bm \sigma}^{(a)}]=\sqrt{2 N} \epsilon\\
  \lambda^{(a)}_{\rm min}=\lambda,   \, \,  u^{(a)}_{\rm min}=u
  \end{subarray}} \Big\rangle
\end{split}
 \end{equation}
where $p_{\vec{\bm \sigma}|{\bm \sigma}^0}$ now denotes the \emph{joint} distribution of all gradients ${\bf g}[{\bm \sigma}^{(a)}]$ and all energy fields $ h[{\bm \sigma}^{(a)}]$, each Hessian $\mathcal{H}[{\bm \sigma}^{(a)}]$ in the expectation value is conditioned to gradients, energy fields and smallest Hessian eigenvalues at all the other points ${\bm \sigma}^{(b)}$, and $\mathbb{G}^{(n)}_{\vec{{\bm \sigma}}| {\bm \sigma}^0} \tonde{\vec{\lambda},\vec{u}}$ is the \emph{joint} probability distribution of the smallest eigenvalues and of the correspondent eigenvector components of these conditioned Hessians. Following the reasoning elucidated in Refs.~\cite{SpikedRepKacRice, RBCBarriers} one can show that, as a consequence of the isotropy of the correlations of the random energy field, all these statistical distributions depend on the points ${\bm \sigma}^{(a)}$ only through their mutual overlaps $q_{ab} \equiv N({\bm \sigma}^{(a)} \cdot {\bm \sigma}^{(b)})$.  Introducing an $n \times n$ symmetric overlap matrix $\hat{Q}$ with components $Q_{ab}= \delta_{ab} + (1-\delta_{ab}) q_{ab}$ we can parametrize the above integral as:
\begin{equation}\label{eq:moments2}
 \Big\langle \mathcal{N}_{{\bm \sigma}^0}^n(\epsilon, q, \lambda, u| \epsilon_0)\Big\rangle_0=  \int \prod_{a<b=1}^n d q_{ab} \, \text{exp}\quadre{N S_n (\epsilon, q, \hat{Q} | \epsilon_0) + o(N n)} \, \mathbb{G}^{(n)}_{\epsilon, q, \hat{Q}| \epsilon_0} \tonde{\vec{\lambda},\vec{u}}.
 \end{equation}
This integral can now be computed with a saddle-point approximation, optimizing over the matrix $\hat{Q}$. The total constrained complexity is contributed by stationary points for which $\lambda_{\rm \min}$ and $u_{\rm min}$ take their typical values, implying that the joint distribution $\mathbb{G}^{(n)}_{\epsilon, q, \hat{Q}| \epsilon_0} \tonde{\vec{\lambda},\vec{u}}$ does not scale exponentially with $N$ but it is of $O)(1)$. In that case the saddle point of the remaining action is attained at $q_{ab} \equiv q_1=q^2$ \cite{RBCBarriers}. In presence of the conditioning, to compute \eqref{eq:moments2} one has to determine the large deviations of the smallest eigenvalues and eigenvectors of all the $n$ Hessian matrices. This will in general depend on the parameters $q_{ab}$: to prove that the annealed calculation is correct, one has to show that this dependence is such that the saddle point value $q_{ab} \equiv q_1=q^2$ is not shifted by additional contributions coming from this large deviation function, that are exponentially large in $N$. Notice that for all values of $q_{ab} \neq 0$ the Hessian matrices are coupled with each others: therefore, determining the joint distribution $\mathbb{G}^{(n)}_{\epsilon, q, \hat{Q}| \epsilon_0} \tonde{\vec{\lambda},\vec{u}}$ to linear exponential order in $N$, and its generic dependence on the parameters $q_{ab}$, is a highly non-trivial task.

\subsection{Large deviations at fixed $\theta,u$: limiting cases}\label{app:LimitingCases}
From the above expressions, we can easily recover the limiting cases of the large deviations for an unperturbed GOE~\cite{LargeDevGOE} and for a purely additive perturbation~\cite{BiroliGuinnet}. 
In the case in which all the perturbations vanish, $\mu_1(x,u)$ diverges and $F(x,u) \to 0$. The function $\mathcal{L}^{(a)}_{\theta, \beta}(x,u)$  tends to:
{
\medmuskip=0mu
\thinmuskip=0mu
\thickmuskip=0mu
\begin{equation}
\begin{split}
\mathcal{L}^{(a)}_{\theta, \beta}(x,u)- l(\theta, \beta) \stackrel{\theta, \beta \to 0}{\longrightarrow}  -\frac{x}{4 \sigma^2} \sqrt{x^2-4\sigma^2}- \log \tonde{-\frac{x}{2}+ \frac{1}{2}\sqrt{x^2-4 \sigma^2}}+ \log \sigma,
\end{split}
\end{equation}
}
which for $u=0$ coincides exactly with the large deviations for the smallest eigenvalue of an orthogonal matrix with variance $\sigma^2$, given by:
\begin{equation}\label{eq:UnPert}
  \mathcal{G}_{0}(x)=\int_{x}^{-2 \sigma} \frac{\sqrt{z^2-4 \sigma^2}}{2 \sigma^2} dz= \tonde{\frac{x^4}{4 \sigma^2}- \mathcal{I}(x)+\frac{1}{2}}+ \log \sigma- 1,
\end{equation}
where $\mathcal{I}(z)$ is defined in \eqref{eq:GOEI}.
This function vanishes at $x=-2 \sigma$, which is indeed the typical value of the smallest eigenvalue.

In the case of a purely additive perturbation $\beta=0$, the relevant case is Case B. For a negative perturbation $\theta <0$, it holds 
$\sigma^2 F(x,u) \to \theta (1-u)$, and the typical value of the second-smaller eigenvalue, when smaller than $-2 \sigma$, becomes:
\begin{equation}\label{eq:Typ2Add}
\mu_1(x,u) \stackrel{\beta \to 0}{\longrightarrow} \theta(1-u)+ \frac{\sigma^2}{\theta (1-u)} \equiv \mu_1(u),
\end{equation}
consistently with the fact that in this case the effective perturbation induced by fixing $x$ is an additive perturbation with strength $\tilde{\theta}=\theta (1-u)$, see Eq. \eqref{eq:NewPar}. 
Therefore, for $\theta (1-u) \geq - \sigma$
{
\medmuskip=0mu
\thinmuskip=0mu
\thickmuskip=0mu
\begin{equation}\label{eq:Case1}
\begin{split}
 \mathcal{L}^{(a)}_{\theta, \beta}(x,u)  \stackrel{\beta \to 0}{\longrightarrow} \frac{x^2}{4 \sigma^2}- \frac{\theta x u}{2 \sigma^2} -\mathcal{I}(x) -\frac{1}{2}\log (1-u)- \quadre{-\frac{1}{2}+ \log \sigma + \frac{\theta^2 (1-u)^2}{4 \sigma^2}},
\end{split}
\end{equation}
}
which coincides \footnote{See the combination of Eq. (1) in \cite{BiroliGuinnet} and the beginning of Sec. 7; in particular $C'=-1/2 + \log \sigma$ and $\sigma=1$ in that work.} with what is found in \cite{BiroliGuinnet}. For $\theta(1-u) <  - \sigma$ we have instead:
\begin{equation}\label{eq:Case2}
\begin{split}
&\mathcal{L}_{\theta, \beta}(x,u) + l(\theta, \beta) \stackrel{\beta \to 0}{\longrightarrow}\\
&
\begin{cases}
 a(x,u)- \quadre{-\frac{x^2}{4 \sigma^2}+ \frac{\mathcal{I}(x)}{2}+ \frac{\theta (1-u) x}{2 \sigma^2}- \frac{1}{2} \log \tonde{\frac{\theta (1-u)}{\sigma^2}}- \frac{1}{2}}    &\text{ if  } x \geq \mu_1(u) \\
 a(x,u)- \quadre{-\frac{y^2}{4 \sigma^2}+ \frac{\mathcal{I}(y)}{2}+ \frac{\theta (1-u) y}{2 \sigma^2}- \frac{1}{2} \log \tonde{\frac{\theta (1-u)}{\sigma^2}}- \frac{1}{2}}  \Big|_{y=\xi^+_\sigma(u)}    &\text{ if  } x < \mu_1(u)
\end{cases}
\end{split}
\end{equation}
with 
\begin{equation}
a(x,u)= \frac{x^2}{4 \sigma^2} -\theta \frac{x u}{2 \sigma^2} - \mathcal{I}(x)- \frac{1}{2}\log (1-u),
\end{equation}
which again coincides with the result in \cite{BiroliGuinnet}.

\subsection{Introduction of the auxiliary fields $y, \lambda$}\label{app:Auxiliary}
In this Appendix we show how the representation \eqref{eq:Rapp1} is obtained. 
First, using the Hubbard-Stratonovich transformation we set:
{
\medmuskip=0mu
\thinmuskip=0mu
\thickmuskip=0mu
\begin{equation}
 e^{-\frac{M}{2} \frac{C_3^2 (1-u)^2}{8 \sigma^2} \tonde{\sum_{\alpha=1}^{M-1} \mu_\alpha e^2_\alpha}^2}= \sqrt{\frac{4 \sigma^2}{\pi C_3^2 (1-u)^2}} \int_{-\infty}^{\infty} dy \, e^{-\frac{4 \sigma^2 y^2}{ C_3^2 (1-u)^2} +i \sqrt{M} y \tonde{\sum_{\alpha=1}^{M-1} \mu_\alpha e_\alpha^2}}
\end{equation}
}
so that the integral \eqref{eq:Comp2} can be re-written as:
\begin{equation}
\begin{split}
&I_{x,u}(\vec{\mu})= \frac{\Gamma \tonde{\frac{M-1}{2} }}{\pi^{\frac{M-1}{2}}} \sqrt{\frac{4 M \,\sigma^2}{\pi C_3^2 (1-u)^2}} \int_{-\infty}^{\infty} dy \, e^{-\frac{ 4 M \, \sigma^2 y^2}{ C_3^2 (1-u)^2}}\times \\
&\times \quadre{   
\int \prod_{\alpha=1}^{M-1} d e_\alpha \delta \tonde{\sum_{\alpha=1}^{M-1} e^2_\alpha-1}  e^{-\frac{M}{2} \quadre{\frac{C_4 (x,u) (1-u)}{2 \sigma^2} -2 i y}\sum_{\alpha=1}^{M-1} \mu_\alpha e_\alpha^2 + \frac{C_3 (1-u)}{2 \sigma^2} \sum_{\alpha=1}^{M-1} \mu_\alpha^2 e_\alpha^2} }
\end{split}
\end{equation}
Exponentiating the constraint, we can re-write the quantity in square brackets as:
\begin{equation}\label{eq:Sq}
\begin{split}
\quadre{\cdot}
&= -i M  \tonde{\frac{2 \sigma^2}{C_3 (1-u)}}^{\frac{M-1}{2}}\int_{-i \infty}^{i \infty} d \lambda e^{- M \lambda}     
\times\\
&\times \int \prod_{\alpha=1}^{M-1} d e_\alpha e^{-\frac{M}{2} \sum_{\alpha=1}^{M-1} e_\alpha^2\quadre{ \mu_\alpha^2 +\tonde{\frac{C_4 (x,u)}{C_3} -2 i y \frac{2 \sigma^2}{C_3 (1-u)}}\mu_\alpha -2 \lambda  \frac{2 \sigma^2}{C_3 (1-u)}} } 
\end{split}
\end{equation}
The representation \eqref{eq:Rapp1} is obtained with 
the change of variable:
\begin{equation}
y'= i y \frac{2 \sigma^2}{C_3 (1-u)}.
\end{equation}

\subsection{Derivation of the solutions for the auxiliary fields $y, \lambda$}\label{app:AuxFieldsSol}
In this appendix, we report the derivation of the solutions \eqref{eq:ySaddle} of the saddle point equations for $y, \lambda$, as well as of \eqref{eq:SolExt}.
We begin with the derivation of \eqref{eq:ySaddle}, starting from Eq. \eqref{eq:Root}. We introduce the notation: \begin{equation}
\begin{split}
a=\frac{C_3(1-u)}{2 \sigma^2}, \quad \quad 
b=\frac{2}{ \sigma^2} 
\end{split}
\end{equation}
If $\mu^{\pm}$ are complex, they can be written as:
\begin{equation}
\mu^{\pm }= -\frac{1}{2} \tonde{\frac{C_4}{C_3}- 2 y} \pm \frac{i}{2} \sqrt{-8 \lambda \frac{2 \sigma^2}{C_3(1-u)} - \tonde{\frac{C_4}{C_3}- 2 y}^2} \equiv X \pm i Y,
\end{equation}
where now the quantity under the square root is positive. The two equations \eqref{eq:Root} are one the adjoint of the other, and read explicitly:
\begin{equation}
X+ i Y= \frac{a X + b y + i a Y }{(a X + b y)^2 + a^2 Y^2}+ \sigma^2 \tonde{a X + b y -i a Y},
\end{equation}
and equating real and imaginary parts (assuming $Y \neq 0$) gives:
\begin{equation}\label{eq:Im}
\frac{1}{(a X + b y)^2 + a^2 Y^2}=\frac{1+ a \sigma^2}{a} \longrightarrow (a X + b y) \tonde{a^{-1} + 2 \sigma^2}=X,
\end{equation}
that gives the solution for $y^*$. The first equation \eqref{eq:Im} allows then to solve for $\lambda$ as:
\begin{equation}\label{eq:LambdaSaddle}
8 \frac{\lambda^* 2 \sigma^2}{C_3(1-u)}=  -\frac{16 \quadre{\sigma ^2 (C_3 (1-u)+2)^2+C_4^2
   (1-u)^2}}{C_3 (1-u) (C_3 (1-u)+2)^3}.
\end{equation}
If on the other hand $\mu^{\pm}$ are real, they can be written as 
\begin{equation}
\mu^{\pm }= -\frac{1}{2} \tonde{\frac{C_4}{C_3}- 2 y} \pm \frac{1}{2} \sqrt{8 \lambda \frac{2 \sigma^2}{C_3(1-u)} + \tonde{\frac{C_4}{C_3}- 2 y}^2} \equiv X \pm \sqrt{Y},
\end{equation}
where the quantity under the square root is again positive. Eqs. \eqref{eq:Root} read in this case:
\begin{equation}
\tonde{X \pm \sqrt{Y}}\tonde{a X + b y \mp a \sqrt{Y}}= 1+ \sigma^2 \tonde{a X + b y \mp a \sqrt{Y}}^2,
\end{equation}
which are equivalent to:
\begin{equation}
a X^2 + b y X - a Y -1 -\sigma^2 (a X+ b y)^2- \sigma^2 a^2 Y= \mp \sqrt{Y}\tonde{b y + 2 a \sigma^2 (a X+ b y)}
\end{equation}
Summing and subtracting these two equations, we get two linear equations for $\lambda^*, y^*$:
\begin{equation}
\begin{split}
b y + 2 a \sigma^2 (a X+ b y)&=0\\
a X^2 + b y X - a Y -1 -\sigma^2 (a X+ b y)^2- \sigma^2 a^2 Y&=0,
\end{split}
\end{equation}
which are again solved by \eqref{eq:ySaddle}. Notice that $\lambda^*<0$, which implies that $\mu^+$ is the largest of the two real solutions, and it is negative. The action $\phi(\lambda^*, y^*)$ is obtained noticing that the saddle point equations imply:
\begin{equation}
\mathcal{I}(\mu^+)+\mathcal{I}(\mu^-)=\log \quadre{\sigma^2 \tonde{1+ \frac{2}{C_3(1-u)}}}-1 + a \mu^+ \mu^- + \frac{b y^*}{2} \tonde{\mu^+ + \mu^-}.
\end{equation}

We now come to the derivation of  \eqref{eq:SolExt}. First, taking $y$ as a free parameter we find that the expression for $\lambda_{\rm ext}(y; \xi)$ in \eqref{eq:GenOut} implies that: 
\begin{equation}
\begin{split}
& y < \frac{C_4(x,u)}{2 C_3}+ \xi \longrightarrow \xi=  \mu^+_{x,u}(y, \lambda_{\rm ext}(y; \xi))\\
& y > \frac{C_4(x,u)}{2 C_3}+ \xi  \longrightarrow \xi=  \mu^-_{x,u}(y, \lambda_{\rm ext}(y; \xi))
\end{split}
\end{equation}
and at the threshold value:
\begin{equation}\label{eq:Condy}
y = \frac{C_4(x,u)}{2 C_3}+\xi  \longrightarrow  \xi=  \mu^-_{x,u}(y, \lambda_{\rm ext}(y; \xi))=\mu^+_{x,u}(y, \lambda_{\rm ext}(y; \xi)).
\end{equation}
Using that in both cases:
\begin{equation}\label{eq:MumExt}
\begin{split}
&\mu^-_{\rm ext}= \mu^+ - \sqrt{8 \lambda \frac{2 \sigma^2}{C_3(1-u)} + \tonde{\frac{C_4}{C_3}- 2 y}^2}
=- \xi - \frac{C_4}{C_3}+2 y =-\frac{4 \sigma^2 \lambda}{C_3(1-u) \xi},\\
&\mu^+_{\rm ext}= \mu^- + \sqrt{8 \lambda \frac{2 \sigma^2}{C_3(1-u)} + \tonde{\frac{C_4}{C_3}- 2 y}^2}
=- \xi - \frac{C_4}{C_3}+2 y =-\frac{4 \sigma^2 \lambda}{C_3(1-u) \xi},
\end{split}
\end{equation}
for any value of $y$ we get that the action evaluated at $ \lambda_{\rm ext}(y; \xi)$ reduces to:
{
\medmuskip=0mu
\thinmuskip=0mu
\thickmuskip=0mu
\begin{equation}
\tilde{\phi}(y)= \frac{y^2}{ \sigma^2} - \frac{C_3(1-u)}{4 \sigma^2} \quadre{\xi^2 + \tonde{\frac{C_4}{C_3}- 2 y}\xi}-\frac{1}{2M}\sum_{\alpha=1}^{M-1} \log (\mu_\alpha-\xi)-\frac{1}{2M}\sum_{\alpha=1}^{M-1} \log (\mu_\alpha-\mu^\pm_{\rm ext}(y)).
\end{equation}
}
Equivalently, if  $ y_{\rm ext}(\lambda; \xi)$ is used one finds
{
\medmuskip=0mu
\thinmuskip=0mu
\thickmuskip=0mu
\begin{equation}
\tilde{\phi}(\lambda)= \frac{1}{ 4 \sigma^2}\tonde{\frac{C_4}{C_3}- \frac{4 \sigma^2 \lambda}{C_3(1-u) \xi}+ \xi}^2 -\lambda-\frac{1}{2M}\sum_{\alpha=1}^{M-1} \log (\mu_\alpha-\xi)-\frac{1}{2M}\sum_{\alpha=1}^{M-1} \log (\mu_\alpha-\mu^\pm_{\rm ext}(\lambda))
\end{equation}
}
These functions can be further optimized in $y$ or $\lambda$, by solving the equations:
\begin{equation}\label{eq:Eq1Ex}
\begin{split}
&\frac{2 y}{\sigma^2} + \frac{C_3(1-u)}{2 \sigma^2} \xi- G \tonde{- \xi -\frac{C_4}{C_3}+2 y }=0\\
&\frac{2 C_4}{C_3}+ 2 \xi+C_3(1-u) \xi- \lambda \frac{8 \sigma^2}{C_3(1-u) \xi}- 2 \sigma^2 G \tonde{-\frac{4 \sigma^2 \lambda}{C_3 (1-u)\xi}}=0
\end{split}
\end{equation}
Note that in the first equation the argument of the resolvent is positive, in the second equation it is negative because $\lambda<0$. 
Both these equations are linear for a GOE (the coefficients of the quadratic terms simplify, with solutions given in \eqref{eq:SolExt}. 
The threshold condition \eqref{eq:Condy} becomes equivalent to:
    \begin{equation}\label{eq:a1}
    \xi  (C_3 (1-u)+4)+\frac{4 C_3 \sigma
   ^2}{C_3\xi  (C_3 (1-u)+2)+2 C_4}+\frac{2 C_4}{C_3}=0.
    \end{equation}    
  To determine \eqref{eq:ActExt} we use that $ \mu^\pm_{\rm ext}=\xi$ (with $\pm$ chosen depending on the value of $y_{\rm ext}$), as well as \eqref{eq:MumExt} and the saddle point condition for $y_{\rm ext}$, we get
  {
\medmuskip=0mu
\thinmuskip=0mu
\thickmuskip=0mu
\begin{equation}
\begin{split}
\mathcal{I}(\mu^\mp_{\rm ext})&=\log \quadre{\tonde{\frac{C_3(1-u)}{2}+1}\xi+ \frac{C_4(x,u)}{C_3}}- \frac{1}{2}-\frac{\xi+ \frac{C_4}{C_3}- 2 y_{\rm ext}}{2} \quadre{\frac{2 y_{\rm ext}}{ \sigma^2 } + \frac{C_3(1-u)}{2 \sigma^2}\xi},
\end{split}
\end{equation}
}
and calling
\begin{equation}
H( \xi)=\frac{2 C_3}{C_3 \xi (C_3 (1-u)+2)+ 2 C_4}
\end{equation}
this is:
\begin{equation}
\begin{split}
\mathcal{I}(\mu^\mp_{\rm ext})
&=\log \tonde{\frac{1}{H( \xi )}}- \frac{1}{2}-\frac{H( \xi )}{2}G^{-1} \quadre{-H( \xi )} =\log \tonde{\frac{1}{H( \xi )}}+\frac{ \sigma^2 }{2}H^2( \xi)
\end{split}
\end{equation}
The expression  \eqref{eq:ActExt} is obtained using that:\begin{equation}
\frac{y_{\rm ext}^2}{\sigma^2}- \lambda_{\rm ext}=- \frac{(1-u) \xi}{16 \sigma^2} \quadre{4 C_4 + C_3 \xi (4 + C_3(1-u))}+ \frac{\sigma^2}{4}H^2.
\end{equation}

To conclude the Appendix, we remark that equation \eqref{eq:Idy2} follows from the general identity:
\begin{equation}
\mathcal{I}(x)=\log \tonde{-\frac{1}{G_\sigma(x)}}-\frac{1}{2}+ \frac{x}{2}G_\sigma(x),
\end{equation}
using that:
\begin{equation}
\xi^+_\sigma=G_\sigma^{-1}\tonde{\frac{1}{\sigma^2 F(x,u) }}, \;\; 2 C_4(x,u)+ \xi^+_\sigma C_3 (2+ C_3(1-u))= -\frac{2 \sigma^2 F(x,u) [2+ C_3 (1-u)]}{(1-u)}
\end{equation}
as well as:
\begin{equation}
\begin{split}
&\frac{(\xi^+_\sigma)^2}{4 \sigma^2}\tonde{1+\frac{(1-u) C_3}{4} [4+ C_3(1-u)]}+ \frac{\xi^+_\sigma}{4 \sigma^2} \tonde{C_4(1-u)- \frac{1}{F(x,u)}}=\\
&=-\frac{1}{4} - \frac{(1 - u)^2 (2 C_2 + C_3^2 \,u\, x)^2}{16 \sigma^2 [2 + C_3(1-u)]^2}.
\end{split}
\end{equation}

\subsection{Large deviations for the second smaller eigenvalue: the case of purely additive perturbation}\label{sec:AppCheckSecond}
In this Appendix we compare the large deviation function for the second smallest eigenvalue computed in Sec. \ref{sec:VariationalSecond} with the results given in \cite{Maida} for the large deviations in the case of a purely additive perturbation. Indeed, for $\beta=0$ and $x,u$ fixed, the eigenvalue $\mu_{M-1}$ is the smallest eigenvalue of a matrix subject to an additive rank-1 perturbation of strength $\tilde{\theta}= \theta(1-u)$, see Eq. \eqref{eq:NewPar}. In this limit, given that $\sigma^2 F(x,u) \to \theta(1-u)$ and that $\xi^-_\sigma \to -\infty$, the two cases discussed in Sec. \ref{sec:VariationalSecond} reduce to the following:
\begin{itemize}
\item
If $\theta(1-u) < - \sigma$, typically the second smallest eigenvalue \emph{is} out of the bulk and \begin{equation}
\Psi_1(x,u,\xi)\to \frac{1}{4 \sigma^2} \xi^2 -\frac{1}{2} \mathcal{I}(\xi)- \frac{\theta (1-u) \xi}{2 \sigma^2} +\frac{1}{2} \log (- 2 \theta)- \frac{1}{2}\log C_3,
\end{equation}
and the logarithmic divergence due to $C_3$ gets canceled by another term in $\Psi_0$. This function (up to constants that do not depend on $\xi$) matches with $L_\theta^\beta(\xi)$ in Th. 1.1 of \cite{Maida}. It has a minimum in $\xi^*=\theta(1-u)+\sigma^2/(\theta(1-u))$, that is indeed the typical value of the smallest eigenvalue of a GOE matrix subject to the additive perturbation of strength $\tilde{\theta}=\theta(1-u)$.

\item If $\theta (1-u)> - \sigma$, typically the second smallest eigenvalue \emph{is not} out of the bulk. In this case the large deviation function has only two regimes:
\begin{equation}
\Psi_1(x,u,\xi) \to
\begin{cases}
 \frac{1}{4 \sigma^2} \xi^2 - \mathcal{I}(\xi)- \phi_1(x,u)  &\text{  if  } \xi \geq \xi^*\\
 \frac{1}{4 \sigma^2} \xi^2 -\frac{1}{2} \mathcal{I}(\xi)- \theta \frac{(1-u) \xi}{2\sigma^2} + \frac{1}{2} \log (- 2 \theta)- \frac{1}{2}\log C_3 &\text{  if  } \xi < \xi^*,
\end{cases}
\end{equation}
which matches with the function $M_\theta^\beta(x)$ of \cite{Maida}  (up to constants that do not depend on $\xi$).
\end{itemize}
The difference in the constants comes from the fact that the large deviation function $\Psi_1(x,u,\xi) $ in Sec. \ref{sec:VariationalSecond} is not normalized to zero at the typical value.

\subsection{Self-consistent checks on $u_{\rm typ}(x)$}\label{app:SelfConsTyp}
    
In this Appendix we check under which conditions the rate functions to be optimized is $\mathcal{L}^{(a)}(\theta, \beta)$. If Case A holds (see \eqref{eq:Cases}), this is always the case. 
In Case B, in order to perform the check we need to determine the sign of the function:
\begin{equation}
\tilde{F}(x,v)=\sigma^2 F(x,v)+ \sigma,
\end{equation}
evaluated at  $v=u_{\rm typ}^{(a)}(x)$. This function is quadratic in $v$, with two roots given by:
 {
\medmuskip=0mu
\thinmuskip=0mu
\thickmuskip=0mu
\begin{equation}
v^{\pm}_{\theta, \beta}(x)=\frac{-2
   C_2+C_3^2 (4 \sigma +x)+6 C_3 \sigma \pm \sqrt{\left(2 C_2+C_3^2 x\right)^2-4 C_3 \sigma  (6
   C_2+C_3 (3 C_3+4) x)+4 C_3^2 \sigma ^2}}{2 C_3^2 (2 \sigma
   +x)}.
   \end{equation}
}
Again, it is convenient to consider the above regimes of $\theta, \beta$:

\begin{itemize}

\item Regime A: In this case $u_{\rm typ}^{(a)}=0$. Plugging $u=0$ into \eqref{eq:Cases}, it can be checked that the condition to be in Case A becomes:
\begin{equation}
\frac{4 \sigma^2 \beta(\beta+2)}{\theta^2 (1+ \beta)^2} >1,
\end{equation}
which is always satisfied for $- 2 \sigma'< \theta <0$. Therefore, in this regime Case A holds and $\mathcal{L}^{(a)}_{\theta, \beta}(x,u)$ is the right large deviation function to be optimized.

\item Regime B1: When $\theta_{\rm c} < \theta < -2 \sigma'$, we find that $v^-_{\theta, \beta}(x) \geq v^+_{\theta, \beta}(x)$; moreover, when real, $\tilde{F}(x,v) \geq 0$ for $v^{+}_{\theta, \beta}(x) \leq v \leq v^{-}_{\theta, \beta}(x)$. The function  $v^-_{\theta, \beta}(x)$ is a monotonically increasing function of $x$ which satisfies $v^-_{\theta, \beta}(x) \stackrel{x \to -\infty}{\rightarrow} 1$ and  $v^-_{\theta, \beta}(x) \stackrel{x \to -2 \sigma}{\rightarrow} \infty$. Similarly,  $v^+_{\theta, \beta}(x)$ is monotonic and satisfies $v^+_{\theta, \beta}(x) \stackrel{x \to -\infty}{\rightarrow} 0$, while $v^+_{\theta, \beta}(-2 \sigma)= u^{+}_{\theta, \beta}(- 2 \sigma)<0$, implying that $ v^+_{\theta, \beta}(x)<0$ for any $x$. Therefore, we always have $v^+_{\theta, \beta}(x) < u_{\text{typ}}^{(a)}(x) < v^-_{\theta, \beta}(x)$, which implies that $\tilde{F}(x,u_{\text{typ}}^{(a)}(x)) \geq 0$. Therefore, also in this regime the solution is self-consistent, meaning that the correct large-deviation function to optimize  is $\mathcal{L}^{(a)}_{\theta, \beta}(x,u)$. 

\item Regime B2: When $\theta<\theta_{\rm c}$, we find  $ v^+_{\theta, \beta}(- 2 \sigma)= u_{\theta, \beta}^+(-2 \sigma)>0$ and  $ v^+_{\theta, \beta}(x)>0$ for any $x$. The functions $ v^+_{\theta, \beta}(x)$ and $u_{\theta, \beta}^+(x) $ cross at a point $x^-_\sigma(\mu, \beta) <x^{**}< -2 \sigma$, where $\tilde{F}(x,u_{\theta, \beta}^+(x))$ becomes negative.  
It can be checked that $\mu_1(x,u_{typ}(x))-x \geq 0$ for  $x<x^{**}$, meaning that also in this regime the function to be optimized is again $\mathcal{L}^{(a)}_{\theta, \beta}$. 
\end{itemize}

\subsection{Self-consistent check: at most one isolated eigenvalue is generated}\label{app:OnlyOneEvalue}
In the derivation of the large deviation function, we made the assumption that for any value of the parameters $\theta, \beta$ and for any choice of $x$ and $u$, the $M-1$ eigenvalues $\mu_{M-1}, \cdots, \mu_1$ typically arrange themselves in such a way that \emph{at most one} of them, namely $\mu_{M-1}$, is found to be smaller than $-2 \sigma$ and isolated from the continuous part of the density of states. In order to validate this hypothesis self-consistently, we have to check that when $\mu_{M-1}$ takes its typical value, the third-smallest eigenvalue satisfies $\mu_{M-2}^{\rm typ}=-2 \sigma$. 
As we pointed out several times already, once the values of $\mu_M$ and $u_M$ are fixed to $x,u$ the distribution of the remaining $M-1$ eigenvalues is the one of a GOE matrix perturbed with both an additive and multiplicative perturbation along a given direction ,with parameters $\tilde{\theta}$ and $\tilde{\beta}$ (that depend explicitly on $x$ and $u$, see Eq. \ref{eq:NewPar}). Similarly, when the values of $\mu_{M-1}$ and $u_{M-1}$ are kept fixed,  the distribution of the remaining $M-2$ eigenvalues is again the one of a perturbed GOE matrix with modified parameters. Our goal is to argue that when fixing $\mu_{M-1}=\mu_{M-1}^{\rm typ}$ and  $u_{M-1}^{\rm typ}$, then $\mu_{M-2}^{\rm typ}=-2 \sigma$. This is totally equivalent to stating that, when  $\mu_M$ and $u_M$ are fixed to their typical value, then $\mu_{M-1}^{\rm typ}= - 2 \sigma$. 
This is trivially true when $\mu_M^{\rm typ}=- 2 \sigma$, \emph{i.e.}, when $\theta \geq \theta_{\rm c}$. In the regime $\theta < \theta_{\rm c}$, then $\mu_M^{\rm typ}=\mu_0(\theta, \beta)$ and $u_{\rm typ}= u^+_{\theta, \beta}(\mu_0)$. In order for the second eigenvalue to stick to the boundary of the semicircle, it must hold:
\begin{equation}\label{eq:IneCheck0}
\sigma^2 F(\mu_0, u^+_{\theta, \beta}(\mu_0))+ \sigma \geq 0,
\end{equation}
see Sec. \ref{sec:InterpretationSmaller}. From the discussion in Appendix \ref{app:SelfConsTyp} it follows that this is guaranteed if, for arbitrary values of $\sigma, \beta$ and for $\theta< \theta_{\rm c}$, we find:  
\begin{equation}\label{eq:IneCheck}
v^+_{\theta, \beta}(\mu_0) \leq u^+_{\theta, \beta}(\mu_0).
\end{equation}

 \begin{figure}[ht]
 \centering
    \includegraphics[width=.6\linewidth]{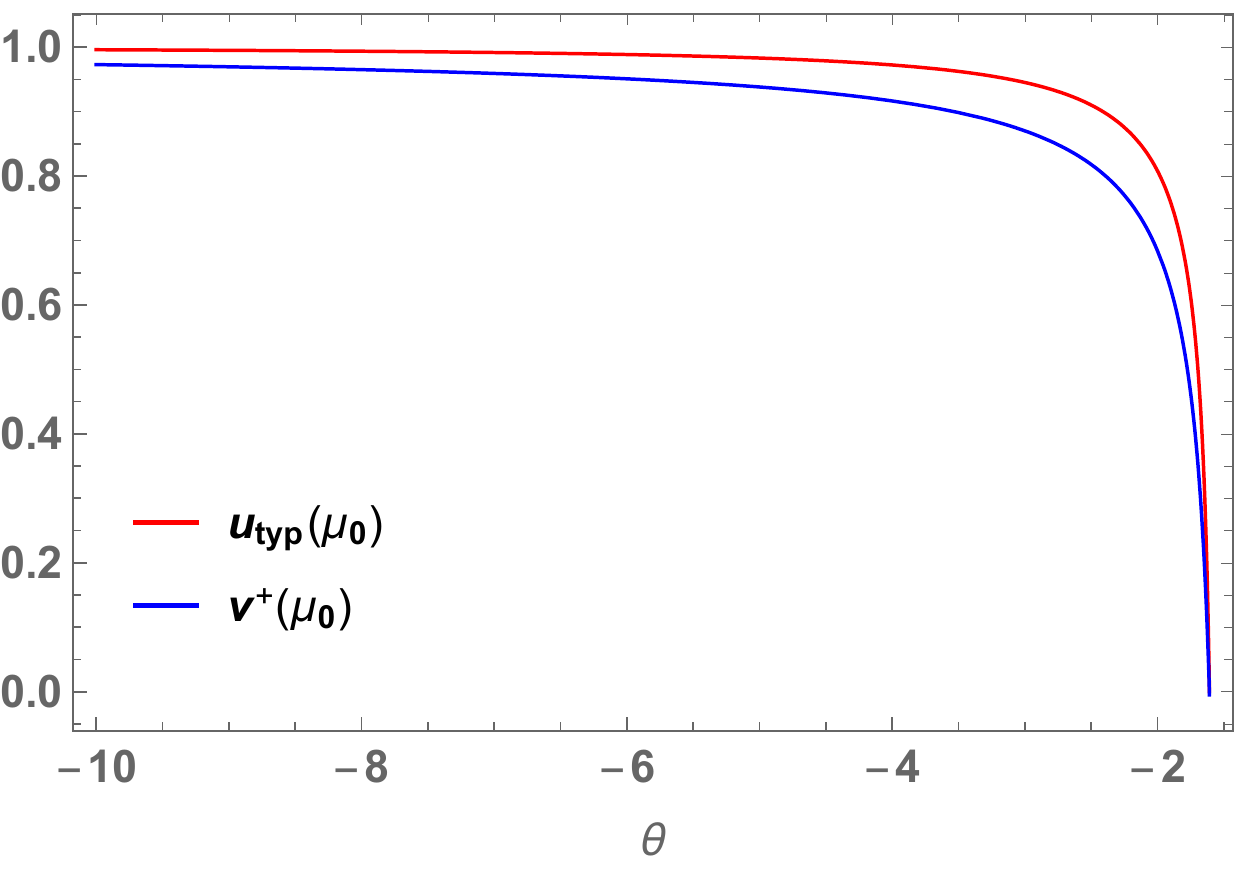} 
\caption{ \small Values of  $u^+_{\theta, \beta}(\mu_0)$ and $v^+_{\theta, \beta}(\mu_0)$ for $\sigma=1$, $\beta=0.6$ and $\theta< \theta_{\rm c}=-1.61$. The plot shows that the inequality \eqref{eq:IneCheck} is always satisfied, implying  \eqref{eq:IneCheck0}.}\label{fig:OnlyOneValue}
  \end{figure}

This inequality can be checked graphically: in Fig. \ref{fig:OnlyOneValue} we give an example for a fixed value of $\beta, \sigma$. Very similar results are obtained for different values of $\beta, \sigma$.

\end{document}